**Strategic differences between regional investments into graphene technology and how corporations and universities manage patent portfolios**

Ai Linh Nguyen,[1] Wenyuan Liu,[1] Khiam Aik Khor,[2] Andrea Nanetti,[3] Siew Ann Cheong[1]

[1]Division of Physics and Applied Physics, School of Physical and Mathematical Sciences, Nanyang Technological University, 21 Nanyang Link, Singapore 637371

[2]School of Mechanical & Aerospace Engineering, Nanyang Technological University, 50 Nanyang Avenue, Singapore 639798

[3]School of Art, Design and Media, Nanyang Technological University, 81 Nanyang Dr, Singapore 637458

## Abstract

Nowadays, patenting activities are essential in converting applied science to technology in the prevailing innovation model. To gain strategic advantages in the technological competitions between regions, nations need to leverage the investments of public and private funds to diversify over all technologies or specialize in a small number of technologies. In this paper, we investigated who the leaders are at the regional and assignee levels, how they attained their leadership positions, and whether they adopted diversification or specialization strategies, using a dataset of 176,193 patent records on graphene between 1986 and 2017 downloaded from Derwent Innovation. By applying a co-clustering method to the IPC subclasses in the patents and using a z-score method to extract keywords from their titles and abstracts, we identified seven graphene technology areas emerging in the sequence *synthesis – composites – sensors – devices – catalyst – batteries – water treatment*. We then examined the top regions in their investment preferences and their changes in rankings over time and found that they invested in all seven technology areas. In contrast, at the assignee level, some were diversified while others were specialized. To illustrate this difference, we classified assignees into different categories (corporations, universities, and others) and three quartile groups based on their accumulated patent credits. We also visualized the distributions of portfolio entropy and accumulated patent credits and how these evolved. We found significant differences between small entities and large entities, as well as between universities and corporations, in how they managed their patent portfolios. While large entities diversified their portfolios across multiple technology areas, small entities specialized around their core competencies. In addition, we found that universities had higher entropy values than corporations on average, leading us to the hypothesis that corporations file, buy, or sell patents to enable product development. In contrast, universities focus only on licensing their patents. We validated this hypothesis through an aggregate analysis of reassignment and licensing and a more detailed analysis of three case studies – SAMSUNG, RICE UNIVERSITY, and DYSON.

*Keywords*

Technological innovation; patent portfolio management; graphene; diversification/specialization; corporations/universities

## 1. Introduction

Patents form a critical stage in innovation, from pure scientific research to applied scientific research to technology disclosure to commercial products. Therefore, in the scientific and technological competition between nations, an aspiring region must cultivate all aspects of its scientific and technological ecosystem but must pay particular attention to its performance in generating patents. Indeed, there is an immense advantage in getting a head start in specific technology areas. To make this happen, public and private investments in scientific research and technology development must be made. Each nation must decide which fields of potential technologies to invest into, how much to invest, when to support, and when to stop investing. Large countries can invest in many more fields, whereas smaller countries must be more selective. Unfortunately, this area of public policy is not yet entirely evidence-based, even though it should be, given how much bibliometric data is available for science and technology.

Many hot fields today attract funding from many countries, such as Artificial Intelligence, Block Chain and Financial Technology, and Renewable Energies, among many others. In this paper, we focus on graphene, which is a material that promises diverse applications in electronic devices, sensors, energy storage, and water treatment. Since the experimental demonstration in 2004 that graphene can be fabricated (Novoselov et al., 2004), much research has gone into understanding the material's properties, and we can say that the science of graphene is now mature. At the same time, many applications have also been developed and tested, so we know which technologies are the most promising. In a previous publication, we analyzed the rise and fall of interest in graphene science and technology as a whole (Nguyen et al., 2020) before we investigated the shift in interest from one graphene topic to another in a follow-up study (Nguyen et al., 2022). After completing our two earlier studies, we realized that there are many exciting directions we can follow up on. In this paper, we will focus on the evolution of graphene technology from a regional competition perspective.

Concerning the regional competition in the graphene arena, the scientific questions that arise naturally are as follows. (i) In the technological competition between regions on graphene, who are the leaders at the aggregate level? (ii) How do the regions get to their leadership positions? Are the top regions always on top, or were there ranking changes? (iii) Are the leaders diversified across many graphene technology areas, or are they specialized? Guided by the above questions, we developed a working hypothesis that whoever commits the most funding in a given area will win the technological competition, and regions with limited budgets specialize in one or two areas to avoid excessive competition. Unfortunately, when we set out to directly answer these questions, we found that bits and pieces of data are available for a few regions, but in general comprehensive sets of the different types of data (public or private funding) needed are not readily available. We therefore proceeded to do the next best thing possible.

After a survey of relevant literature in Section 2, we partitioned graphene patents into different technology areas to track how interests shift from area to area, and whether there are regional differences in these histories. To do this, we describe in Section 3 methods that we used to identify different graphene technology areas, and allocate patent credits to assignees, assignee categories, and regions. After that, we quantified how diverse a given assignee/category/region is for the different graphene technology areas. After listing the graphene technology areas that we identified in Section 4.1, we examined how the importance of these areas change over time in Section 4.2. We then identified in Section 4.3 who the regional leaders were at the aggregate level and for different areas, and whether there had been ranking changes. In line with the scientific questions that we posed above, we would like to distinguish between public and private funding, and a simplistic way to do so would be to look at patents held by universities (mostly the results of public funding) and patents held by corporations (the outcomes of private funding). This led us to examine the proportions of university and corporate patents in different technology areas, and how these proportions evolved in Section 4.4. We then checked in Section 4.5 how diverse the patents in different categories are, for universities and corporations, using a heatmap visualization that we would describe later in the paper. We mapped this diversity at the aggregate level but also from year to year to see if the diversity changed with time.

Here we found that corporations and universities of different sizes evolve differently regarding the diversity of their patents. Corporations also do not grow as fast in diversity as compared to universities.

This led us to ask (iv) why university patents are more diverse than corporate patents. Our hypothesis on why this is true can be stated as follows.

H1. To develop products, *corporations* use bundles of *closely related patents* for each product.

H2. If possible, *corporations* want to avoid paying licensing fees to use someone else's patents. So instead, they *buy patents to complete product bundles* and *sell patents they are unlikely to use*.

H3. *Universities* do not develop products, so they *keep all their patents* and *focus on licensing*.

We tested this hypothesis in Section 5 and found evidence at the aggregate level supporting (H2) and (H3). These two parts of the hypothesis were further confirmed using SAMSUNG and RICE UNIVERSITY as case studies (the former for corporations, the latter for universities). Finally, we showed that (H1) is valid by examining the connections between Dyson's patents and its products. We then conclude in Section 6.

## 2. Literature Review

In this paper, we focus our survey on technological competition in general and in specific technologies. In this literature, Chakrabarti can be considered one of the pioneers (Chakrabarti, 1991). First, investigating technological positions between the US, United Kingdom, France, West Germany, Canada, and Japan between 1975 and 1988 in high-tech industries (which were divided by Lawrence into equipment, consumer durable, non-durable, and intermediate products (Lawrence, 2010)), he found Japan replacing the US as the leader in all four high-tech areas during this period. Following this, Amable and Boyer ranked European countries, Japan, and the US in terms of their competitiveness and growth in the 1980s (Amable & Boyer, 1995). They concluded that the European countries lagged behind Japan and the US.

Focusing on three critical emerging waste heat recovery technologies (thermoelectric generators, Rankine cycle, and organic Rankine cycle systems) of internal combustion engines, Karvonen et al. reviewed 15,575 PATSTAT patents from different regions between 1993 and 2012. They confirmed the leadership of Japan and the US in all three technologies (Karvonen et al., 2016). 7 of the top 10 patent asset holders were Japanese entities. In 2018, Sampaio et al. analyzed 22,682 records from the Derwent Innovation patent database to map out the development of photovoltaic technologies (Sampaio et al., 2018). They found that the US, China, Japan, Germany, and South Korea occupied the top regional positions regarding the number of patent applications between 2004 and 2013, while the top assignees were the US and Japanese institutions and companies.

More recently, Yang et al. studied the technological competition in graphene biomedical technology using the databases from Innography and Derwent Innovation (Yang et al., 2019), focusing on the identification of the leading regions, assignees, and hot areas in graphene technology. Their results also showed an increase in the number of patents since 2012 and the dominant positions of China, the US, South Korea, and Japan. Furthermore, the top assignees identified were also entities based in the US (LOCKHEED MARTIN CORPORATION), South Korea (SAMSUNG ELECTRONICS CO LTD), and China (universities). Finally, Huang et al. used patent indicators (numbers of patent applications over time and countries, patent citations, international patent classification (IPC), and life cycle) on 330 related patents up till 5 December 2019 to explore trends in the application of blockchain technology in power generation and distribution (Huang et al., 2020). The authors concluded that blockchain technology for power grid applications was in a 'growth' stage, with the US leading Germany, China, the United Kingdom, and Israel patent applications.

Next, we survey previous works investigating how the top regions became leaders in the technology ranking. In this area of the literature, Ames and Rosenberg were the first to write down a theory of how a region can become a technological and industrial leader (Ames & Rosenberg, 1963). In their 1963 paper, the authors formulated an economic theory of 'leadership' in industry, considering the output level (per capita) and state of technology (backwards, forward) of the country in question. They hypothesized three ways for latecomers to overtake early starters: (i) latecomers move through all stages of economic development faster than the early starters because they can avoid the mistakes that early starters have to make; (ii) the different stages of development are not equally demanding, and early starters may get stuck at one or more bottlenecks that late comers somehow navigated with ease; and (iii) early starters may stop developing their technologies further, because of bottlenecks or other reasons, allowing latecomers to overtake them. Testing these hypotheses is a non-trivial task, requiring economic development data from early starters and latecomers. One way to circumvent this limitation would be to use data on trades. This is what Wright did in 1990 when he analyzed the performance of the US against foreign competitors using US trading data on manufactured goods in the period 1879-1940. In a nutshell, Wright concluded that the US built its industrial leadership on the wealth of its non-renewable natural resources (Wright, 1990). Publishing in the same year, Nelson pointed out that the main reasons for US leadership in the technological competition during the 1950s and 1960s were its advanced mass-production techniques and industries (steel, electrical engineering) surpassing Europe since the start of the First World War, and its dominance in high-technology industries (computer, semiconductor) resulting from heavy investments into science and technology (jet engines, space systems, semiconductor) during the post-war era (Nelson, 1990). In this paper, Nelson also worried about the erosion of this technological edge from 1970 onwards, in reaction to the many predictions that the US would fall behind Japan and Western European countries due to the slower investment rate into new plants and equipment.

More recently, Pasierbiak studied the technological development of Japan in the international ranking and suggested that the leadership of Japan in the second half of the 1980s was the result of its successful international export of technologically advanced goods (Pasierbiak, 2013). However, from 2000 onwards, Japan's leading position in high technology goods (aerospace, computers, pharmaceuticals, scientific instruments, and electrical machinery) was seriously challenged by other regional competitors (Korea, China). Analyzing patents from 12 regions (Japan, US, Germany, France, United Kingdom, Taiwan, South Korea, Canada, Switzerland, Australia, Israel and China) that were filed with the USPTO between 1980 and 2011, Yamashita compared the citation performances of their patents before, during, and after the Japanese asset price bubble (1986-1991) (Yamashita, 2021). He concluded that the stagflation of the 1990s caused a dramatic erosion of the technological leadership position enjoyed by Japan before the crisis.

Finally, let us review the literature on the diversity of patent portfolios, be it at the single assignee level, group of assignees (e.g., universities) level, or regions. This review will also include previous works discussing how patent owners (e.g., corporations) manage their patent portfolios by making deliberate decisions to buy, sell, or license. During the 20$^{th}$ century, there were two significant perspectives on the relationship between patent portfolio management and business success: (1) technology portfolios should be managed strategically, or (2) technological investments should be diverse. From the first point of view, Porter and Millar (Porter & Millar, 1985), together with Kanter (Kanter, 1990), argued against the diversification of patent portfolios and reasoned based on the theory that specialization enables synergy. Others thought strategic technology management led to competitiveness in specific areas (Lubatkin & Chatterjee, 1994; Norton, 1994). Ultimately, both (Lang & Stulz, 1994) and (Barney, 1991) argued that it is evident that the focus of patent portfolios was for firms to build "their core competencies" in "fields that they can do best". These studies are the earliest papers that encouraged technology specialization. On the other hand, various other studies also supported portfolio diversification. Amit and Livnat analyzed data on business segments of individual firms from the COMPUSTAT business segment dataset. They suggested investment in different industrial strategies to trade part of their returns for a lower risk overall (Amit & Livnat, 1988). Later, Granstrand et al. (Granstrand et al., 1997) investigated how technology diversity catalyzed “corporate growth; increasing R&D investment; increasing external linkage for new technologies by various means (such as acquisitions, alliances, licensing); and opportunities to engage in technology-related new business”. They collected patenting activities data on 440 large, technologically active firms worldwide and classified the patenting activities into 34 different fields. They concluded that large corporations indeed invested broadly beyond their “core” competencies because, unlike the outsourcing of productions, there are no savings from outsourcing technologies and competencies. Moreover, from a risk assessment perspective, Sadowski and Roth (Sadowski & Roth, 1999) suggested a strong connection between firm activities and technology portfolio strategy, in which the holding of a variety of technologies, markets, and resources reduces corporate risks and increases business opportunities. In their research, technology leaders (characterized by high R&D investment in technology development) surpassed their opponents in profits and revenue growth by strategically allocating their technological resources and using future technology roadmaps.

More recently, Lin et al. studied the impact of composition and diversity in patent portfolios on the performance of firms (Lin et al., 2006), using data from 94 entities, each of them having more than 100 US patents in the period between 1985 and 1999. Based on the return on assets (ROAs) and Tobin’s Q of the firms, as well as a measure of patent diversity, the authors concluded that firms with high-technology stocks should aim for high patent diversity. In contrast, firms without high-technology stocks should seek to specialize their technology portfolios. In more recent research, Wuyts & Dutta investigated how diversity in patent portfolios strengthens superior product innovation (new products outperforming available products from the customer's perspective) (Wuyts & Dutta, 2014). In their study, patent data between 1985 and 1999 were collected from a variety of sources: Recombinant Capital (Recap), U.S. Patent and Trademark Office (USPTO), National Bureau of Economic Research (NBER), and FDA Drug Approvals. They unexpectedly found an inverted U-shaped effect, where superior technological products can result not only from very high diversity in the patent portfolio but also from patent portfolios with very low diversity as well. In 2019, Appio et al. collected data on patent portfolios of 391 international firms between 2000 and 2010 to analyze the correlation between portfolio diversity and firms' profitability (Appio et al., 2019). Their findings supported the inverted-U relationship at the IPC section level and the return on assets (ROA). In the same year, do Canto Cavalheiro et al. studied the patent portfolio of Universidade Federal Fluminense (UFF) as a case study to assess the technological capacity of universities in developing countries in terms of intellectual property assets (do Canto Cavalheiro et al., 2019). Their results indicated an increase in the number of patent applications in universities from developing countries, especially patents on emerging human technologies (hygiene, medical and veterinary science) in the case of UFF.

## 3. Data and Methods

### 3.1 Data

As explained in Section 1, we used patents as a proxy for the outcome of technological innovation to provide a lens on the competition between regions. Using “graphene” as the search topic, we downloaded 176,193 patent publication records between 1986 and 2017 from Derwent Innovation. Some of these records have the same application codes. For example, patents US20080193827A1 (2008) and US9318762B2 (2016) have the same application number US2007704873A “Conducting polymer-transition metal electro-catalyst compositions for fuel cells”. These refer to documents submitted at different stages of the patent approval process. Therefore, we combined records with the same application codes to finally obtain 139,899 distinct patent applications with application years from 1985 to 2017. Since our goal is to identify the graphene technology areas and how they evolved, we need to be able to discriminate between technology areas. The first two ways to do so that we tried were unsuccessful (as described in Section 3.2.1), and we finally settled on classifying technology areas using the International Patent Classification (IPC) subclasses. Hence, we exclude records without available IPC subclasses and refer to the remaining 139,264 graphene patent applications as the G-T collection. In the IPC classification scheme, there are a total of 642 subclasses. Out of these, 584 subclasses can be found in G-T records. In some G-T records, we found a minimum of 1 subclass listed, whereas for other G-T records, the maximum number of subclasses listed is 19.

## 3.2 Methods

### 3.2.1 Co-Clustering

One of the standard approaches for decomposing a corpus with explicit citations into topical areas is to perform clustering analysis on the citation network (Small, 1973), the co-citation network (Janssens et al., 2009; Liu et al., 2010), or the bibliographic coupling network (Kessler, 1963). Unfortunately, for our G-T dataset, we found that each G-T patent cites 8.28 patents and is cited by 2.36 patents on average (Nguyen et al., 2020). In fact, out of the 139,899 G-T applications, only 65% of them (91,027) have identified references – whether they were patents or journal papers. Furthermore, only 54,743 G-T applications were cited by other patents. Therefore, the citation network of G-T patents is too sparse for us to do citation-based clustering.

In the bibliometric literature, clustering based on words found in the titles and abstracts of papers and patents has been done (Callon et al., 1983). Many text-based clustering methods have been developed (Boyack et al., 2011; Boyack & Klavans, 2010), and we initially experimented with a method based on word co-occurrence network (Blondel et al., 2008), and a method based on block clustering algorithm (Ailem et al., 2015, 2016) in our previous publication (Nguyen et al., 2022). However, when applied to patents, the results were unsatisfactory. First, the modularity obtained is low (0.27-0.28). Second, when we extracted the community keywords, it was unclear what technology areas each community represented. This may be because of the styles that patents were written in. Eventually, we applied the block clustering algorithm CoClus proposed by (Ailem et al., 2015, 2016) to the document-code matrix, whose rows are G-T patents and whose columns are IPC subclasses listed on the patents. In this matrix, a matrix element is non-zero if code $j$ is found in patent $i$, but zero otherwise.

In the CoClus algorithm, our goal is to divide a set of objects, $O = (o_1, \dots o_i, \dots, o_n)$ and a set of attributes, $P = (p_1, \dots, p_j, \dots, p_d)$ into $g$ non-overlapping object-attribute clusters. To do this, an object is represented as a vector of attributes,

$$o_i = (o_{i1}, \dots, o_{ij}, \dots, o_{id}),$$

where $d$ is the total number of attributes and a non-zero $o_{ij}$ means that attribute $p_j$ can be found in the object $o_i$. The algorithm adapts the modularity measure defined by (Newman & Girvan, 2004), to make the clusters as distinct as possible. The modified modularity for CoClus is constructed a

$$Q(A, C) = \frac{1}{\sum_{i,j} a_{ij}} \sum_{i=1}^{n} \sum_{j=1}^{d} (a_{ij} - \frac{\sum_{i=1,\dots,n} a_{ij} \sum_{j=1,\dots,d} a_{ij}}{\sum_{i,j} a_{ij}})\, c_{ij},$$

where $A$ is the object-attribute matrix, $A = (o_1, \dots, o_n) = \{o_{ij}\}_{i=1,\dots,n; j=1,\dots,d}$, and $C$ is a block seriation such that $c_{ij} = 1$ if attribute $p_j$ and object $o_i$ are in the same cluster, and $c_{ij} = 0$ otherwise. The best partition, equivalent to a maximum value of the modularity $Q$, is then identified using the method of integer programming, for different values of $g$. In this paper, we applied CoClus to the patent-code matrix of 139,899 G-T patents and 584 IPC subclasses to obtain an optimal number of meaningful technological clusters.

### 3.2.2 Topic Identification

After partitioning our G-T collection into different groups $\{O^g\}_{g \in \{0,1,\dots\}}$ using the co-clustering method described in Section 3.2.1, we want to extract their meanings from the linguistic information in titles and abstracts of those patents. To do so, we assumed a null model where each word appears equally in any patent record. Considering a given word $w_k$ in a cluster $g$ or $O^g$, let $m_k$ and $M_k^g$ be the number of G-T patents containing this word in the whole G-T collection, and in $O^g$ respectively. In the null model, the probability that this word $w_k$ appears in a patent is $p_k = m_k/n,$ where $n$ is the size of G-T collection. As a result, the expected average number of patents containing $w_k$ in the cluster $g$ is then $\mu_k^g = |O^g| \cdot p_k$ and its standard deviation is $\sigma_k^g = [|O^g| \cdot p_k \cdot (1 - p_k)]^{\frac{1}{2}}$. If this word $w_k$ is highly important in group $O^g$, we expected it to appear more frequently in $O^g$ than in other groups, or $M_k^g \gg \mu_k^g$ in relation to $\sigma_k^g$. We used the z-score

$$z_k^g = \frac{M_k^g - \mu_k^g}{\sigma_k^g}$$

to quantify the importance of a word in a cluster, and words with the largest z-scores to represent its technological topic. The results will be shown in Section 4.1.

#### 3.2.3 Regional Credits for G-T Records

In addition to identifying the emergent technology areas in the G-T patents at the aggregate level, we also analyzed how these technology areas evolved in different regions and assignee categories. In particular, for analysing G-T patents from different regions, we first extract the regional codes for patent applications from the addresses of the first members in the lists of original assignees. For example, in patent CN201620685858U, the list of original assignees included Nanchang Oufei Biometric Technology Co. Ltd., Nanchang O-Film Technology Co. Ltd., Shenzhen O-Film Tech Co. Ltd., Suzhou O-film Tech Co. Ltd. The address of Nanchang Oufei Biometric Technology Co. Ltd. is Nanchang, Jiangxi, 330013, CN. Therefore, we say that this patent application is from CN. We do not want to assign regional credit based on the patent office the application is from because after the invention in KR, for example, SAMSUNG ELECTRONICS CO LTD will file patent applications in multiple patent offices.

#### 3.2.4 DWPI Assignee Codes for Assignee Names

We used the Derwent World Patents Index (DWPI) Assignee Codes to identify assignees uniquely. In our dataset, most records will contain the DWPI assignee names paired with the respective DWPI Assignee Codes. Sometimes, we will also find the original first assignee names. Notwithstanding the supposed uniqueness of the DWPI Assignee Codes, we found it necessary to do further cleaning because of two problems: (i) a single entity has multiple DWPI assignee names as well as DWPI Assignee Codes, and (ii) different entities (with their own DWPI assignee names) were assigned to the same DWPI Assignee Codes. For example, (i) Ocean's King Lighting Science & Technology Co., Ltd has many DWPI assignee names, including SHENZHEN OCEANS KING LIGHTING ENG CO LTD with SHEN|N as its DWPI Assignee Code, and OCEAN'S KING LIGHTING SCI & TECHNOLOGY CO with OCEA|N as its DWPI Assignee Code, and (ii) both UNIV HEIBEI NORMAL and UNIV HERIOT-WATT were assigned to the same DWPI Assignee Code, UYHE|N. Therefore, we adopted the cleaning process described as follows.

##### 3.2.4.1 Between DWPI Assignee Names with the Same DWPI Assignee Code

First, we examined whether DWPI assignee names with the same DWPI Assignee Code are the same entity by computing their similarity scores. For two assignee names, the similarity score between them is defined to be the maximum value between three different Levenshtein ratios: (i) the similarity ratio between two strings (normal ratio), (ii) the similarity ratio of the most similar substring (partial ratio), and (iii) the similarity ratio of the most similar substring after words are tokenized and sorted (partial token sort ratio).

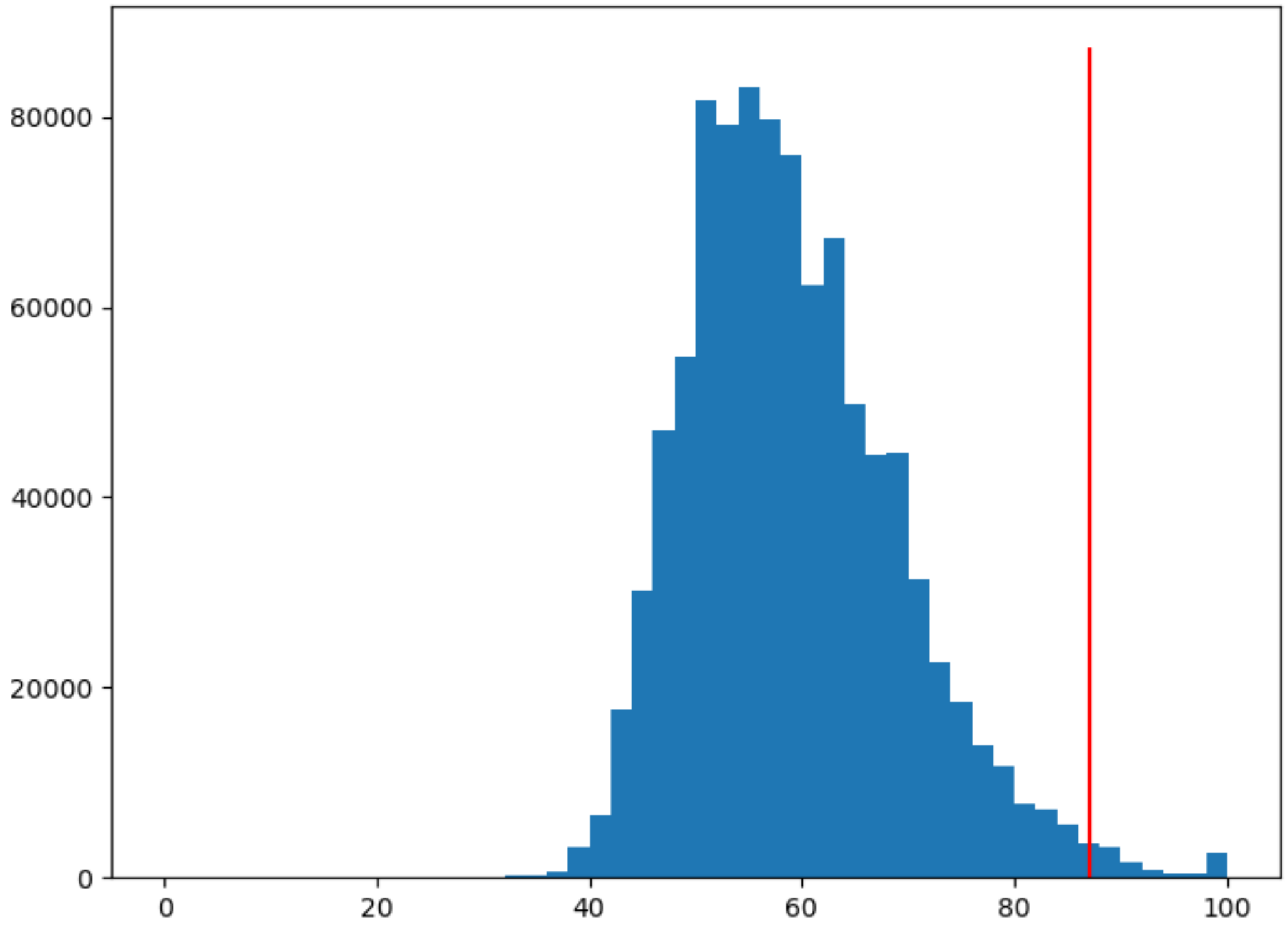

Figure 1: The histogram of similarity scores between DWPI assignee names under the same DWPI assignee codes.

From Figure 1, we see that the distribution of similarities among DWPI assignee names belonging to the same entities peaked sharply at a similarity score of around 50-60. For each DWPI Assignee Code, we constructed a network of assignee names weighted by their similarity scores. Suspecting that only highly similar assignee names refer to the same entity, we chose to retain links whose similarity scores are above a percentile value $85 \leq p_0 \leq 99$. With this strict criterion, each network of assignee names breaks into sub-networks of various sizes. After that, each sub-network for the given DWPI Assignee Code would be assigned an additional unique identifier. After our data processing at the highest percentile value $p_0 = 99$, different sub-networks of highly similar assignee names represent different unique entities, even though they share the same DWPI Assignee Code. This solves problem (ii).

Sometimes, a given DWPI assignee name can be matched to multiple DWPI Assignee Codes. We do not know the reason for this phenomenon but choose the most likely DWPI Assignee Code for the given DWPI assignee name. To do so, we first went over all sub-networks of all DWPI Assignee Codes, to extract the list of sub-networks that contain the given DWPI assignee name as one of their nodes. Next, we compared the similarity scores of this list of sub-networks and identified the one with the highest similarity score. Here, the similarity score of a sub-network containing the given DWPI assignee name is the maximum of the similarity scores between the given name and the rest of the names in the sub-network. If there is more than one sub-network that satisfies the two previous conditions (identical maximum similarity scores), we chose the one(s) with the highest average similarity scores. Finally, we prioritized standard forms (-C) over non-standard forms (-N) over other forms. After these have been done, every DWPI assignee name would be matched to one DWPI Assignee Code, and we believe this matching is highly accurate.

#### 3.2.4.2 Between Original Assignee Names and DWPI Assignee Codes

If a G-T application has no pre-assigned DWPI assignee names or Assignee Codes, we would examine the original name of its first assignee to find a matching DWPI Assignee Code. However, in the G-T patent records from Derwent Innovation, not all entities in the list of DWPI assignee names/ DWPI Assignee Codes are original assignees because some patents may have gone through a change of assignees. Therefore, we compared the first original assignee name with the list of DWPI assignee names. Due to the large number of records and the even more significant number of pairs of original assignee names and DWPI assignee names, we limited the comparison of pairs to within the same patent publication records only. The empirical distribution of their similarity scores is shown in Figure 2. After that, we computed Otsu's threshold to separate correct and incorrect matches of the DWPI assignee names to the original assignee names.

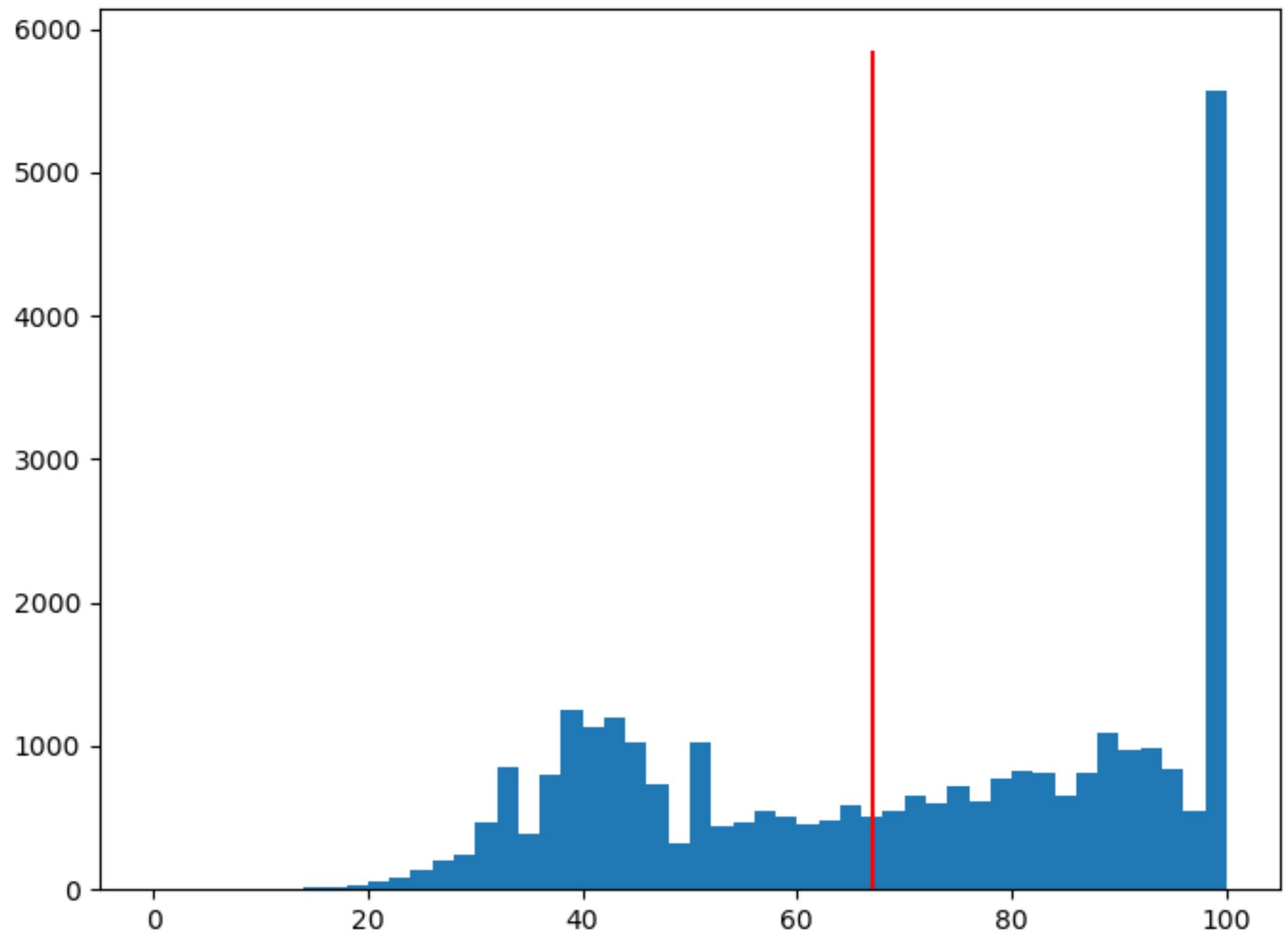

Figure 2: The histogram of similarity scores between the original and DWPI assignee names for the same patent publications.

If an original name was not matched to any DWPI assignee names from the same record, we calculated the similarities between it and all DWPI assignee names from other records. Using the same threshold value, we identified at most five DWPI assignee names with the highest similarity scores. We then went back to the selection process described in Section 3.2.4.1 to find the best match.

#### 3.2.4.3 Between US Assignee Names and DWPI Assignee Codes

To test the patent portfolio hypotheses described in the Introduction Section, we needed to examine the reassignment of patents concerning the categories of their assignees or the licensing of patents by their assignees. Unfortunately, in our dataset, only the 30,227/34,435 US patent publications contain reassignment and licensing information. In a field called "Reassignment (US)", we found data strings with a dictionary-like organization. A typical data string would look like "Assignee | Assignor | Assignee Date | Assignee Year | Document Number | Document Date | Document Year | Reason(s) | Legal Agent", and we extracted the relevant information from such a data string. We then examined 'Reason(s)' to determine whether the data string refers to a reassignment or a licensing. For example, for patent application US13048952A, we found that it has undergone one reassignment from DYNALLOY INC to GM GLOBAL TECHNOLOGY OPERATIONS LLC in 2011.

In these reassignment data strings, the names of the first assignees are given, but these are not matched to their DWPI Assignee Codes. Therefore, we repeated the procedure described in Section 3.2.4.2 to match these first assignee names to DWPI assignee names, and after that, to the DWPI Assignee Codes. We then used the same threshold for distinguishing between DWPI assignee names under the same DWPI Assignee Codes on the histogram of similarity scores shown in Figure 3, and the same selection process as in Section 3.2.4.1 to obtain the best DWPI Assignee Code for each first assignee name. For a given choice of percentile value $p_0$ (red vertical line in Figure 3), we obtained the matching between assignee names (DWPI assignee names, original assignee names, US assignee names) and their respective DWPI assignee codes.

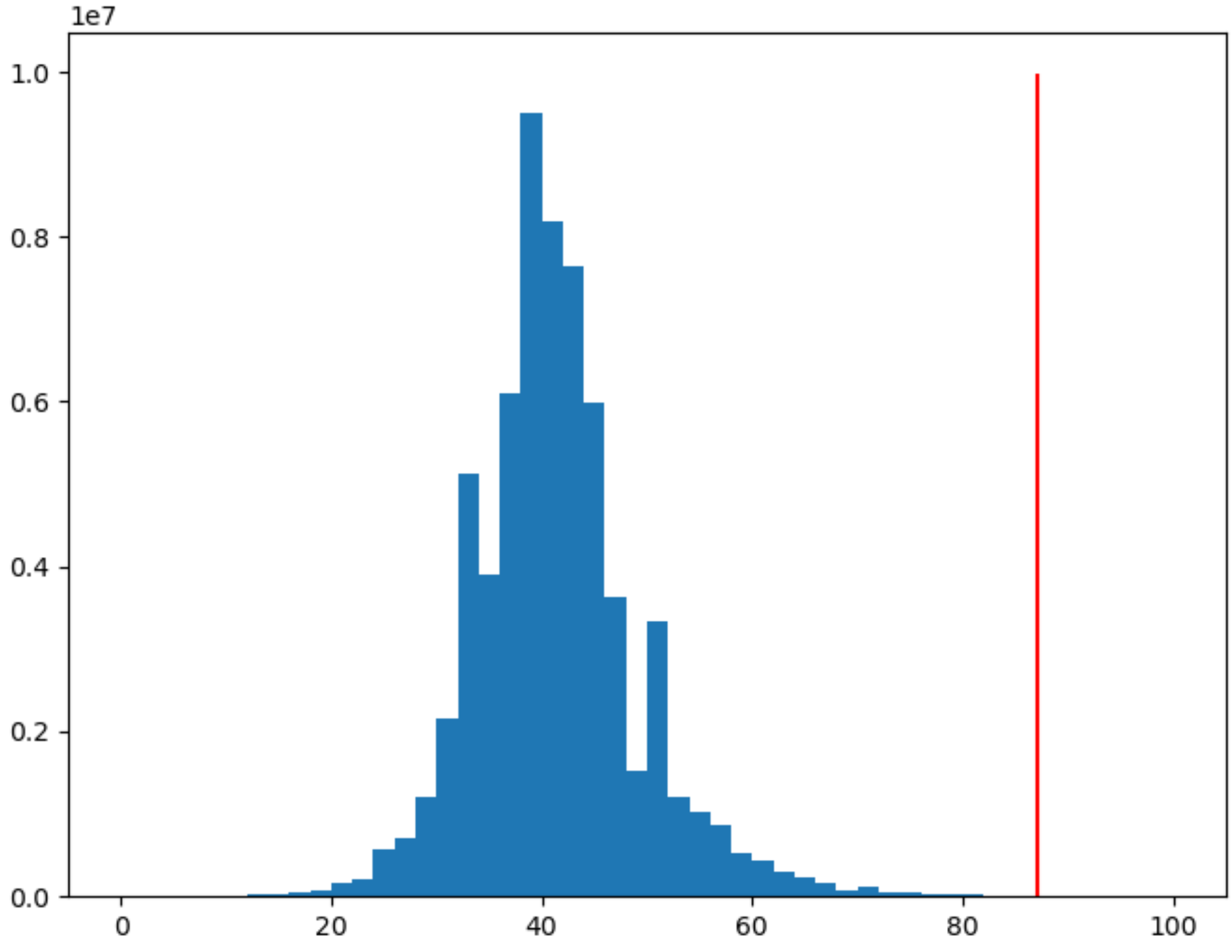

Figure 3: The histogram of similarity scores between the US and DWPI assignee names.

After the processing described in Sections 3.2.4.1, 3.2.4.2, and the current sub-section, we decided for each G-T patent its first assignee code(s) using the order of priority: US first DWPI Assignee Code(s), original first DWPI Assignee Code(s), and finally DWPI Assignee Code(s).

### 3.2.5 Categorical Credits

Finally, to test the hypotheses that corporations manage their patent portfolios differently from universities, we need to assign a unique category to each DWPI Assignee Code, which is now mapped to a bundle of assignee names. We did this by categorizing every name into a university, a corporation, or others based on words found within the name. For example, TOEP|C0 is associated with TOKYO ELECTRIC POWER CO, TOKYO ELECTRIC POWER CO INC, and CENTRAL RESEARCH INSTITUTE OF ELECTRIC POWER INDUSTRY. Next, we categorized those names into corporations and universities, before assigning the category of every DWPI Assignee Code based on the most common category credits of their DWPI assignee names. If we needed to decide whether a G-T patent belongs to a corporation, a university, or another, we would choose the most common category from its DWPI Assignee Codes.

## 4. Results

### 4.1 IPC Co-Clustering Results

As shown in Figure 4, the modularity increases monotonically up to seven clusters. After that, the modularity remains more or less the same up to nine clusters. In Figure 23 of Appendix A, we also show that the modularity remains more or less constant up to 20 clusters. This suggests that the clusters are highly heterogeneous, and any descriptions in terms of 7 or more clusters are meaningful in their own ways. For this study, we chose to work with the 7 clusters shown in Figure 5, since this represents the most concise, meaningful description of the patents.

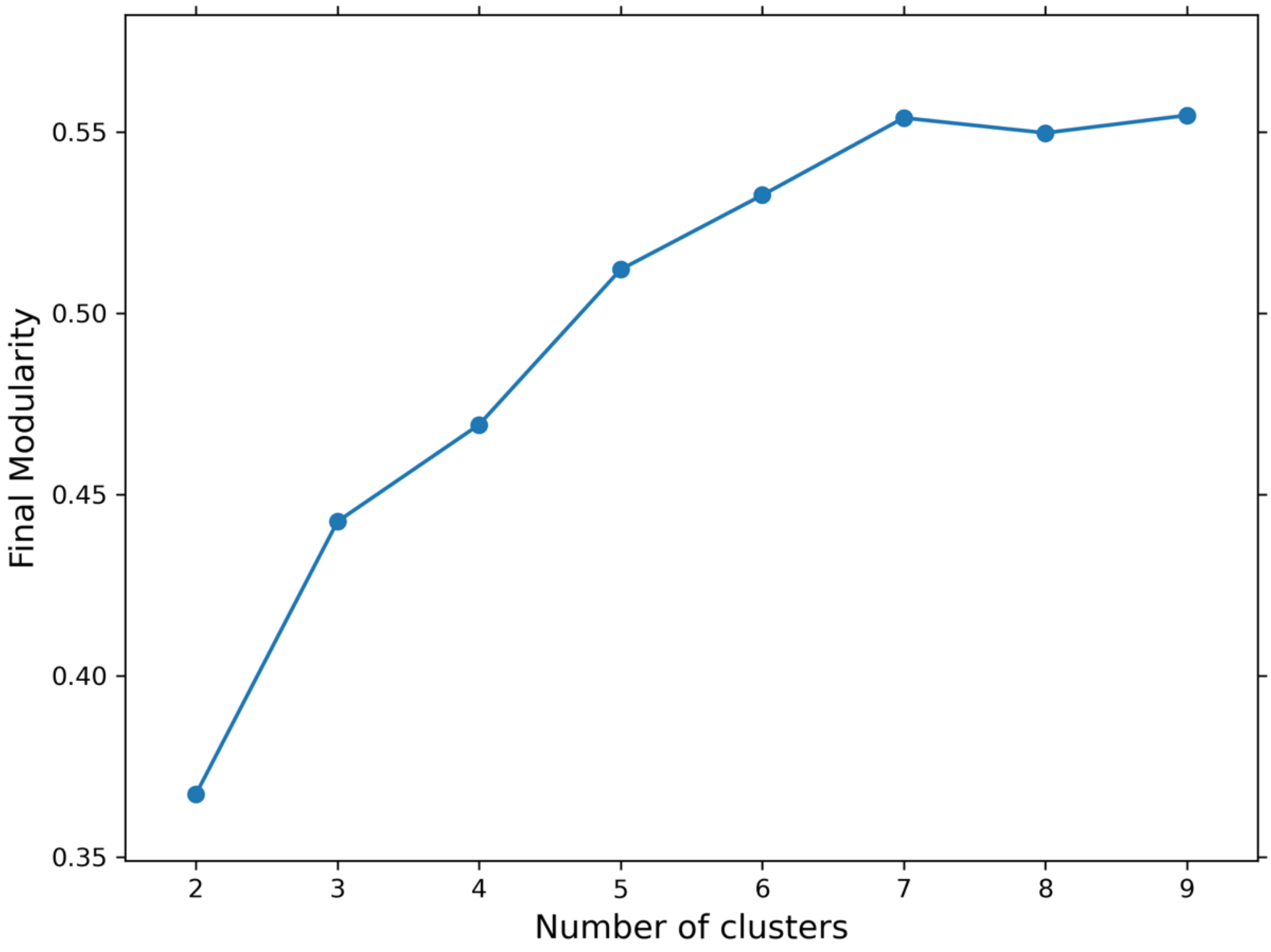

Figure 4: Plot of modularity values determined by CoClus algorithm for the organization of G-T application patents into different numbers of clusters based on their IPC subclasses.

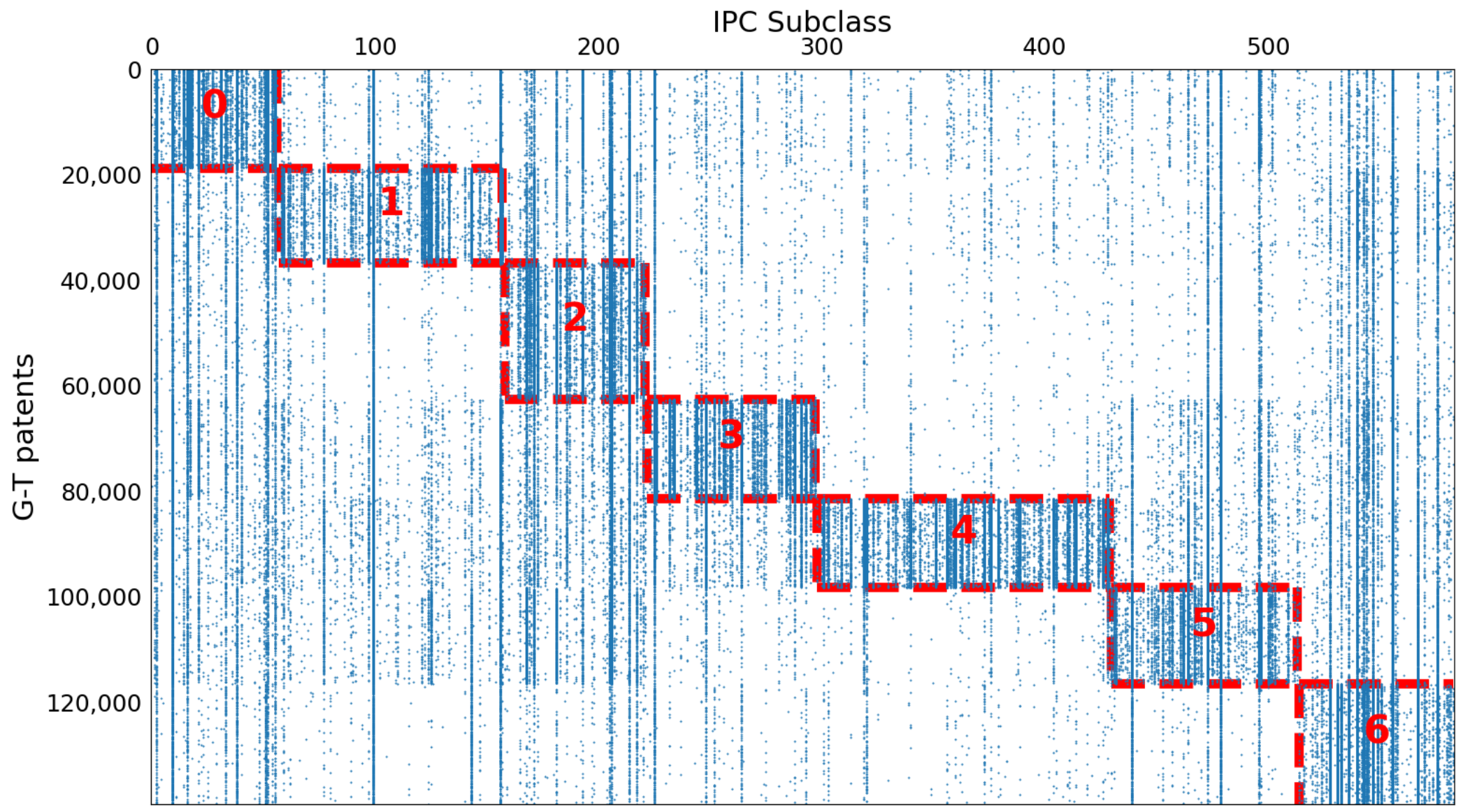

Figure 5: Sparsity plot of G-T patents and IPC subclass codes that appear in them after they have been clustered into $n = 7$ patent-IPC subclass communities by the CoClus algorithm. In this figure, the entry at $(i, j)$ is coloured blue if IPC subclass code $j$ appears in patent application $i$, or white otherwise.

To determine what technology areas these seven clusters represent, we tested words from titles and abstracts of the G-T patents, using the procedure described in Section 3.2.2. In this procedure, we first filtered out all stop-words. We then check which words are over-represented in each cluster by comparing their occurrence frequencies to those expected from a null model where all words are equally likely to appear in all clusters.

In Figure 6, we see that group (2) and group (6) are strongly characterized by the keywords 'battery', 'lithium', 'electrolyte', ..., and 'semiconductor', 'transistor', 'layer', 'gate', ... respectively. These keywords frequently occur in their respective groups but rarely in other groups. In contrast, the keywords of the other groups are weaker, but we were able to identify the technology areas of these groups based on the top few keywords in them. All in all, these technology groups are (0) *composites*, (1) *synthesis*, (2) *batteries*, (3) *sensors*, (4) *water treatment*, (5) *catalyst*, and (6) *devices*. Again, this is a testament to the effectiveness of using IPC subclasses for co-clustering.

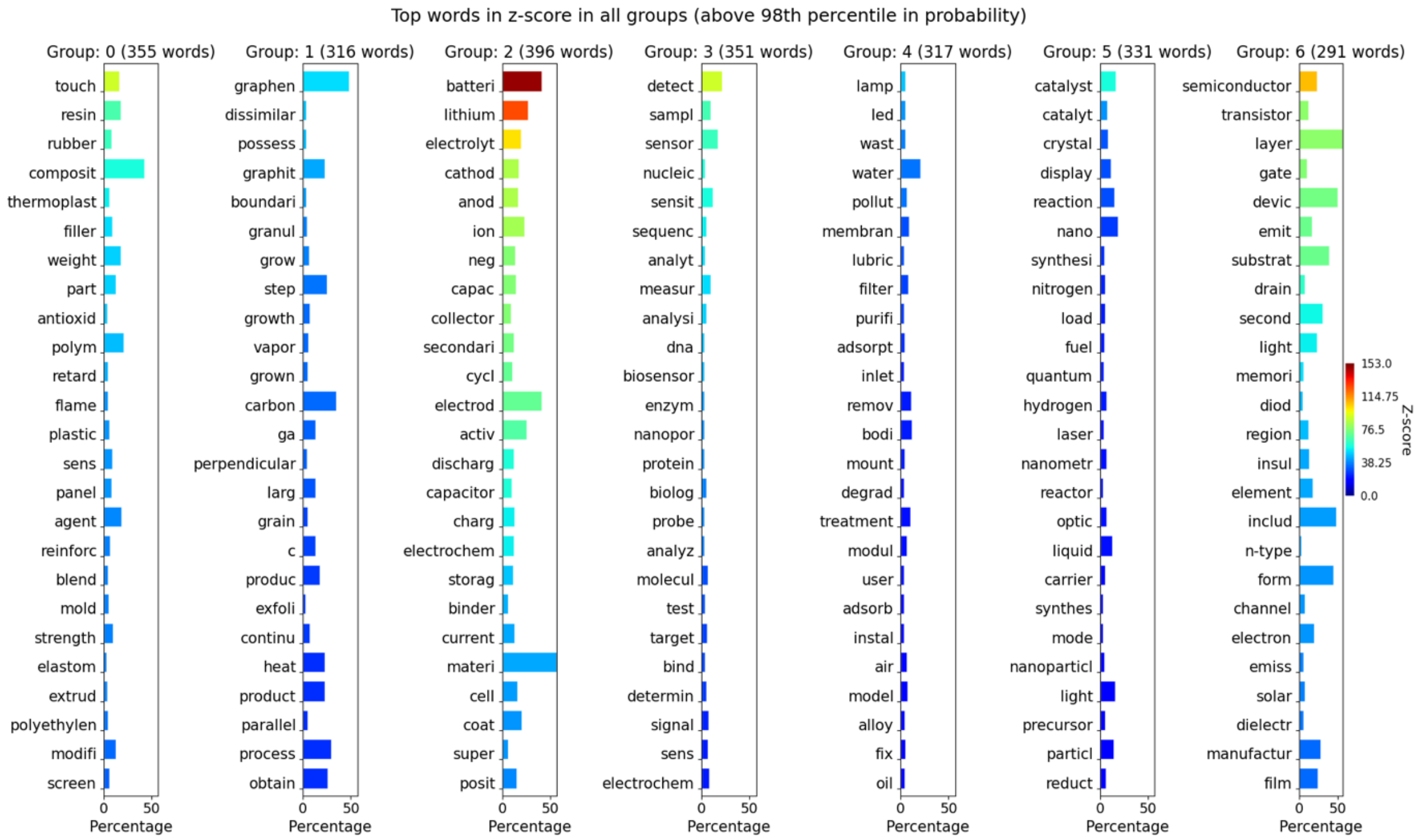

Figure 6: The list of top 25 keywords for the seven G-T clusters, sorted according to their z-scores. In this figure, the z-score of a word is shown in colour (blue is close to 0, red is high, and green is intermediate), while the bar length tells us how frequent the word is in the cluster considered.

## 4.2 Temporal Analysis

In Section 4.1, we broke the G-T patents down into seven clusters according to their technology areas. Here, we will organize patents in the seven clusters according to their application years to see how their numbers and proportions change over time. When we look at the numbers of patents in these groups over time in Figure 7(A), they are very similar to each other except for (2) *batteries*, which dominates after 2012. In general, it is challenging to compare growing trends. Therefore, we look at the proportions in Figure 7(B) instead. Most of the proportions look noisy, but we found three clear trends: (i) group (1) *synthesis* was the first dominant group, but it has been on the decline since 2004; (ii) group (6) *devices* was the second dominant group; it grew and fell, peaking in 2009; and finally (iii) the third dominant group was (2) *batteries*, which started growing in 2008, and seemed to be still rising in importance. Another noteworthy group is group (4) *water treatment*. It was never the dominant group but seemed to have a clear rising trend since 2008. Compared to the evolution sequence of scientific topics that we found in our previous paper (Nguyen et al., 2022), where the first topic that emerged is *theory and experimental tests*, followed by *synthesis and functionalization*, *sensors*, and finally, *supercapacitors and electrocatalysts*, we see that the technology area evolution (1) – (0) – (3) – (6) – (5) – (2) – (4) agreed roughly with what was happening in graphene science. This agreement is rough because there is no one-to-one mapping from scientific to technological areas. For example, in graphene technology, *devices* form their own group, whereas in graphene science, we found their keywords within the *synthesis and functionalization* cluster. In graphene technology, *catalysts* form their own cluster, but they are merged with *sensors* in graphene science. In graphene science and technology, *batteries* and *supercapacitors and electrocatalysts* were among the last to emerge. The main surprise here is the emergence of *water treatment* as a graphene technology area. We have yet to detect signs of such a topic emerging in graphene science.

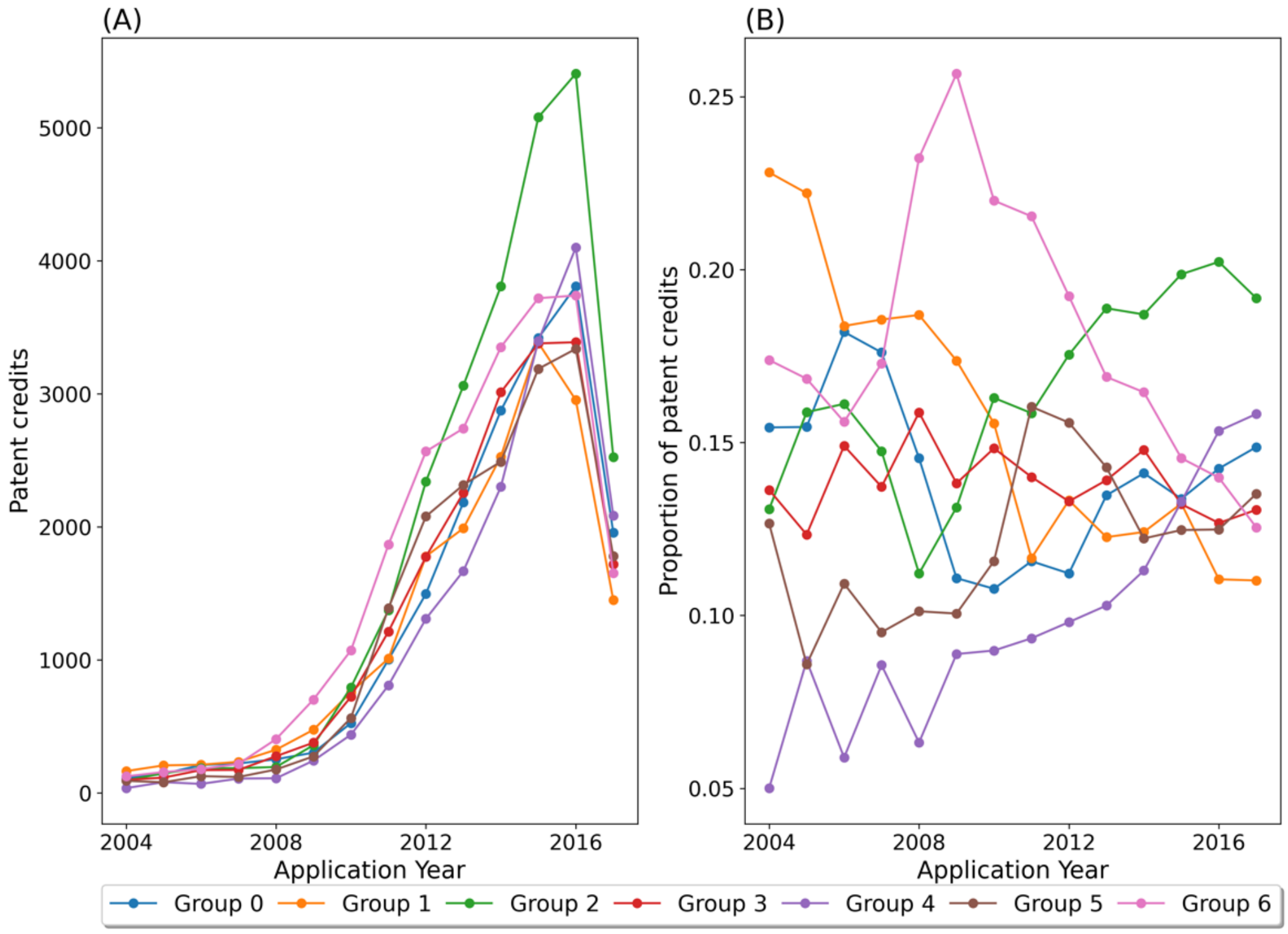

Figure 7: (A) the number of G-T applications every year from seven groups between 2004 and 2017, and (B) the proportions of G-T applications every year from seven groups between 2004 and 2017. We can see that the first most crucial technology area is group (1) *synthesis*, the next technology area that took over is group (6) *devices*, and the last one is group (2) *batteries*. We believe the next important technology area would be group (4) *water treatments* based on apparent trends.

### 4.3 Regional Analysis

After knowing the temporal sequence for the emergence of different technology areas, we next want to investigate how the technology evolution varies over space (geographical regions). The key question we would like to ask here is whether all regions followed the same global technological sequence: (1) *synthesis* – (6) *devices* – (2) *batteries* – (4) *water treatments*, or they became specialized at different times. For this part of our study, we focused on the top six regions based on the number of patents they are credited with (using the credit allocation method described in Section 3.2.3). As shown in Figure 8, these regions are, in descending order based on patent credits: China (CN: 50,384), United States (US: 26,059), South Korea (KR: 15,824), Japan (JP: 12,097), Germany (DE: 3,871), and United Kingdom (GB: 2,030).

To see whether all six regions follow the same sequence of technological evolution, we plot the proportions of patents in the seven technology areas for each region in Figure 9, and contrast these dynamics against that shown in Figure 7. As we can see, different regions seem to have different preferences for technological groups to invest in. We see that US, JP, and KR are similar in that they first invested in (1) *synthesis*, followed by (6) *devices*, and followed (2) *batteries*. On the other hand, CN followed a trajectory that is most different from the global sequence: CN does not have a (6) *devices* peak, and after a short dominance by (0) *composites*, CN was the earliest region to invest in (2) *batteries*. However, this early interest seemed to fizzle out quickly, with (1) *synthesis* becoming hot before (2) *batteries* came back into dominance later. In the top 6 regions, KR also had a brief early interest in (2) *batteries* to be replaced by (6) *devices* for an extended period. After 2008, (2) *batteries* returned with their steadily increasing trend. From Figure 7, (4) *water treatment* seems to be the next big technology area, but from Figure 9 up till 2017 only CN and US seemed to be investing in it.

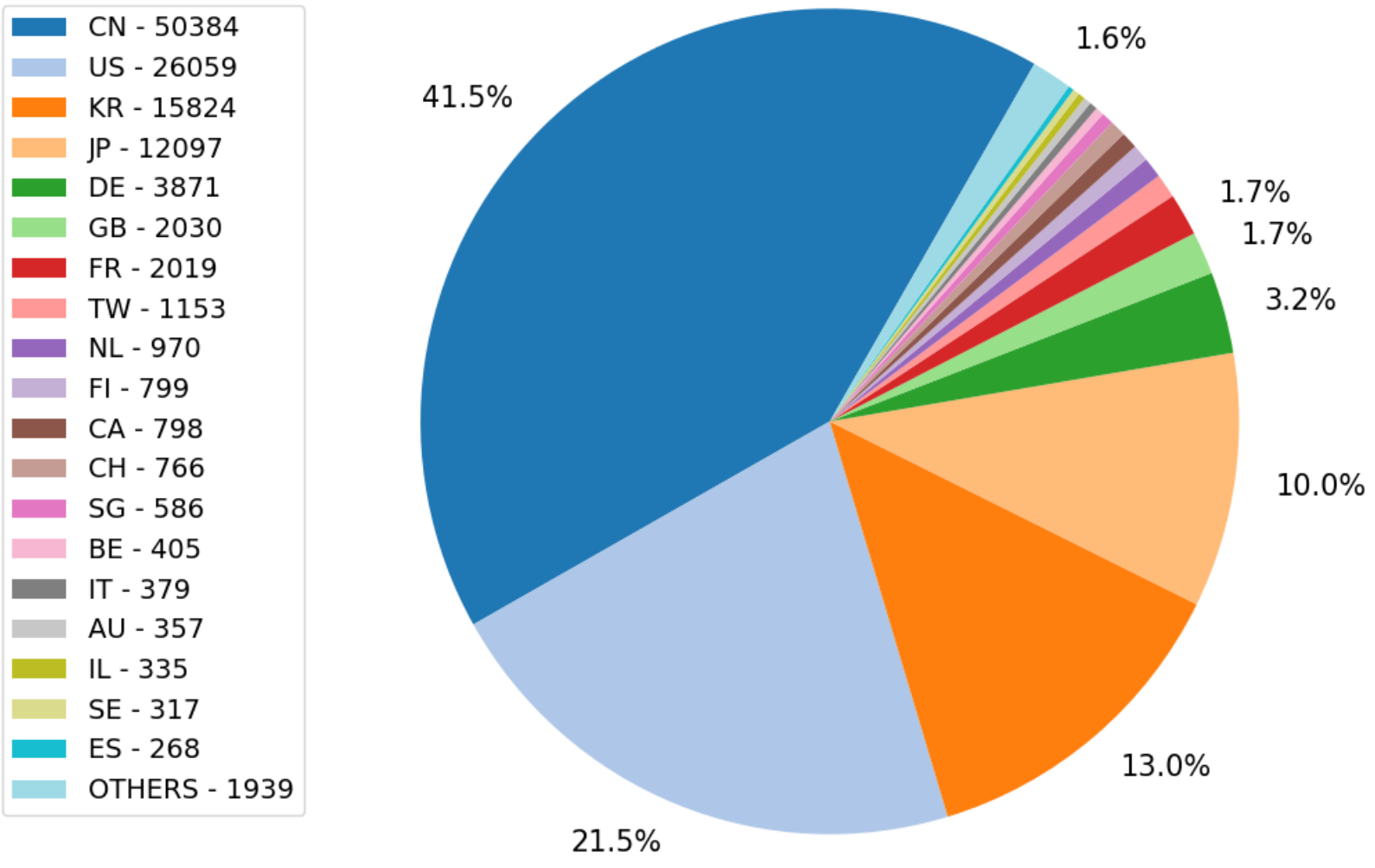

Figure 8: The pie chart of application proportions of the top 19 regions. This pie chart only shows regions with at least 1% of the total number of applications.

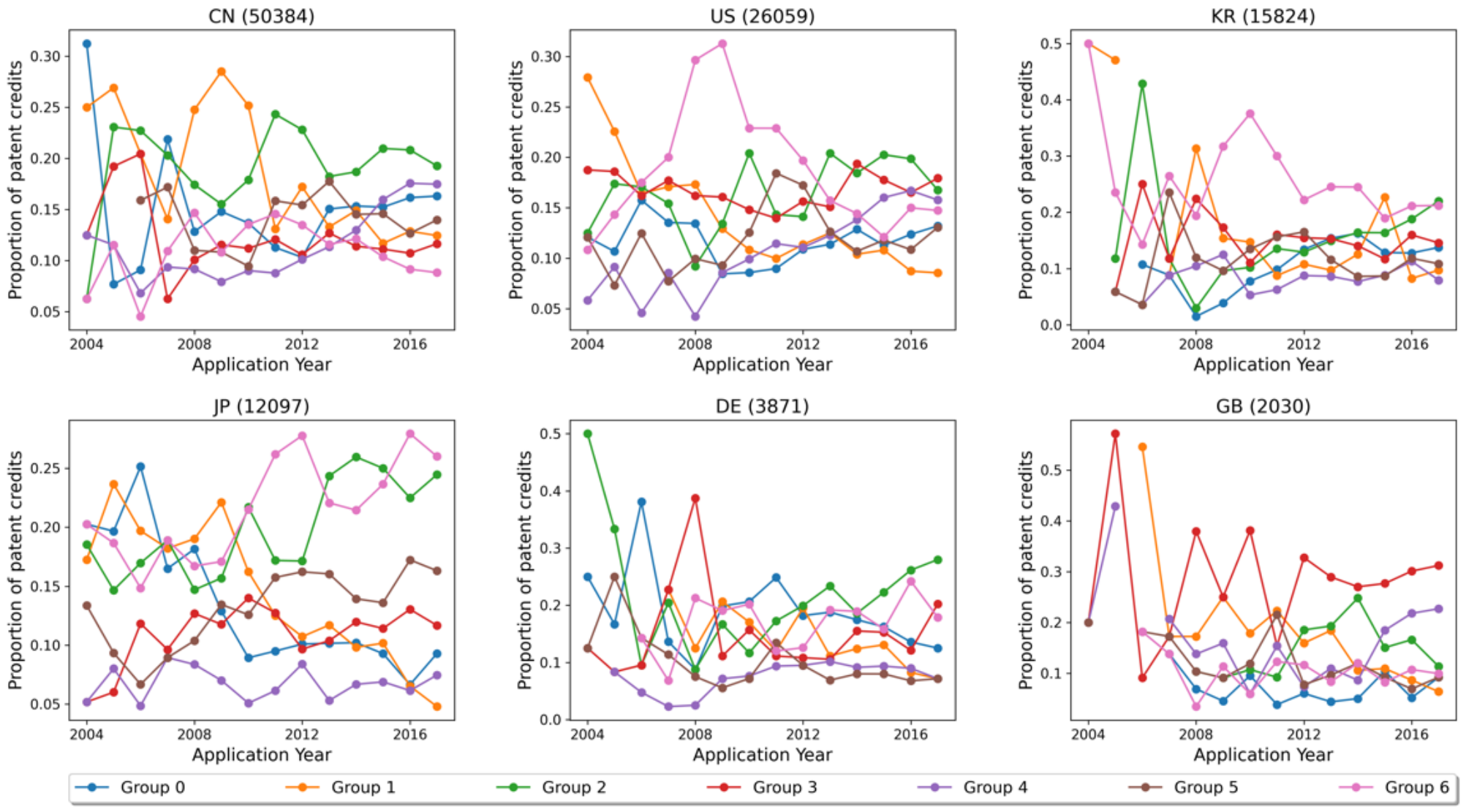

Figure 9: The proportions of patent applications in the seven technology areas of the top six regions.

One of the goals of our study is to examine the technological competition between regions in graphene. Therefore, it is not enough to know how the technology areas evolve in each region — we must also assess how effective a region is in promoting a technology area. We show this by plotting the ranks of different regions in different years, first aggregating over all technology areas, and after that focusing on individual technology areas.

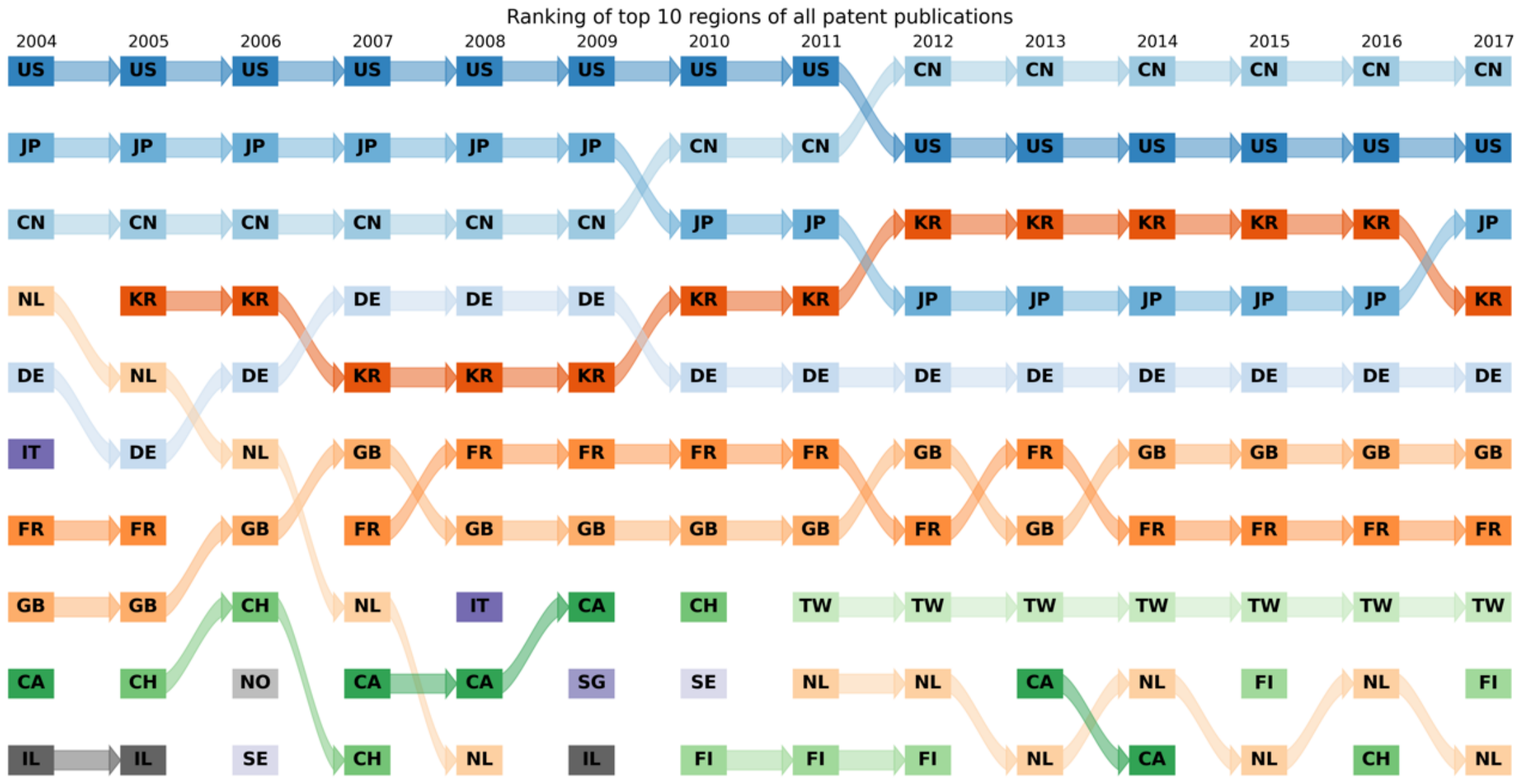

Figure 10: Ranking changes of the top 10 regions regarding the number of applications.

As we can see from Figure 10, when we compare across regions for all patents, the US was the overall leader between 2004 and 2011. CN took this overall leadership position from 2012 onwards. From 2004 to 2009, JP was ranked second for graphene patents before CN and KR displaced it in 2010 and 2012 respectively. In 2017, JP regained the third spot in the ranking. The top four regions we identified agree with those identified by (Yang et al., 2019) using databases from Innography and Derwent Innovation. Following these top four regions, we found DE followed by GB and FR.

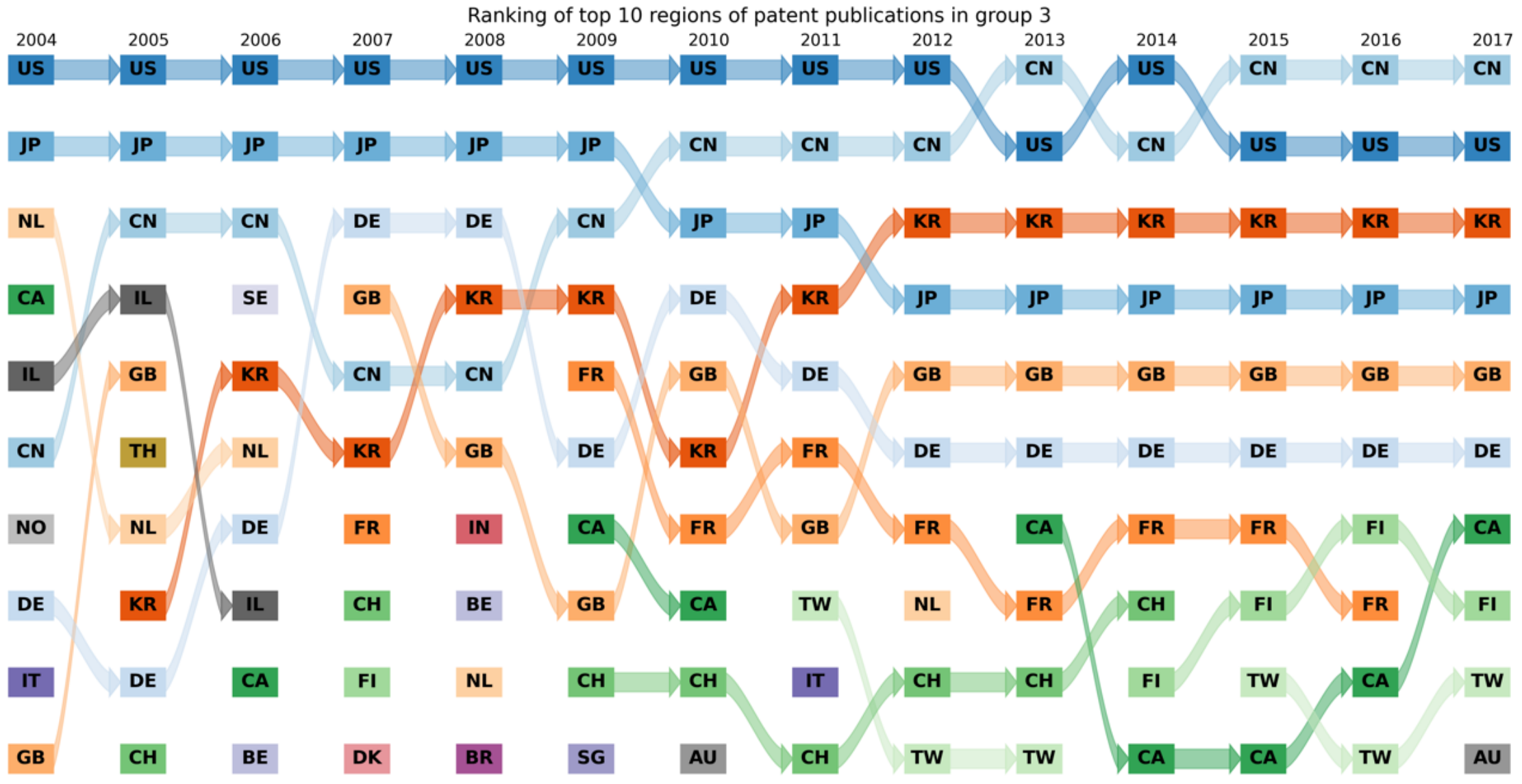

Figure 11: Ranking changes of the top 10 regions regarding the number of applications in (3) *sensors*.

When we examined the rankings at the individual technology area level, we found from Figure 11 that in (3) *sensors*, the US was also the leader for a long time, until CN briefly displaced it in 2013 and then permanently from 2015 onwards. Between 2004 and 2009, JP was number 2 in this technology area, but from 2012 onwards, JP fell below KR, which became the clear number 3 after CN and US. Apart from these differences when the regions displaced each other, the top four regions remained the same. Similar ranking changes also appeared in (1) *synthesis*, as we can see from Figure 31 in the Appendix.

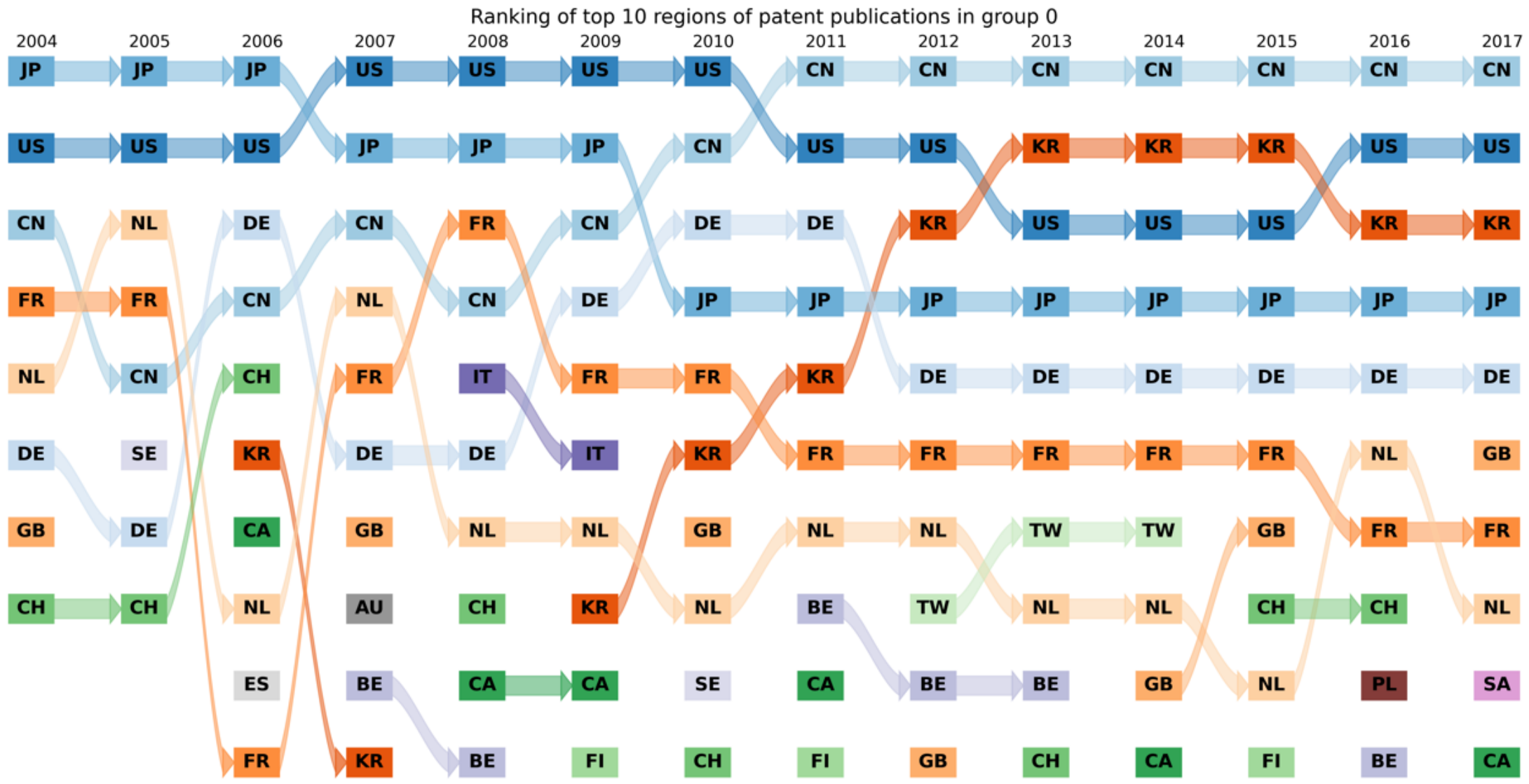

Figure 12: Ranking changes of the top 10 regions regarding the number of applications in (0) *composites*.

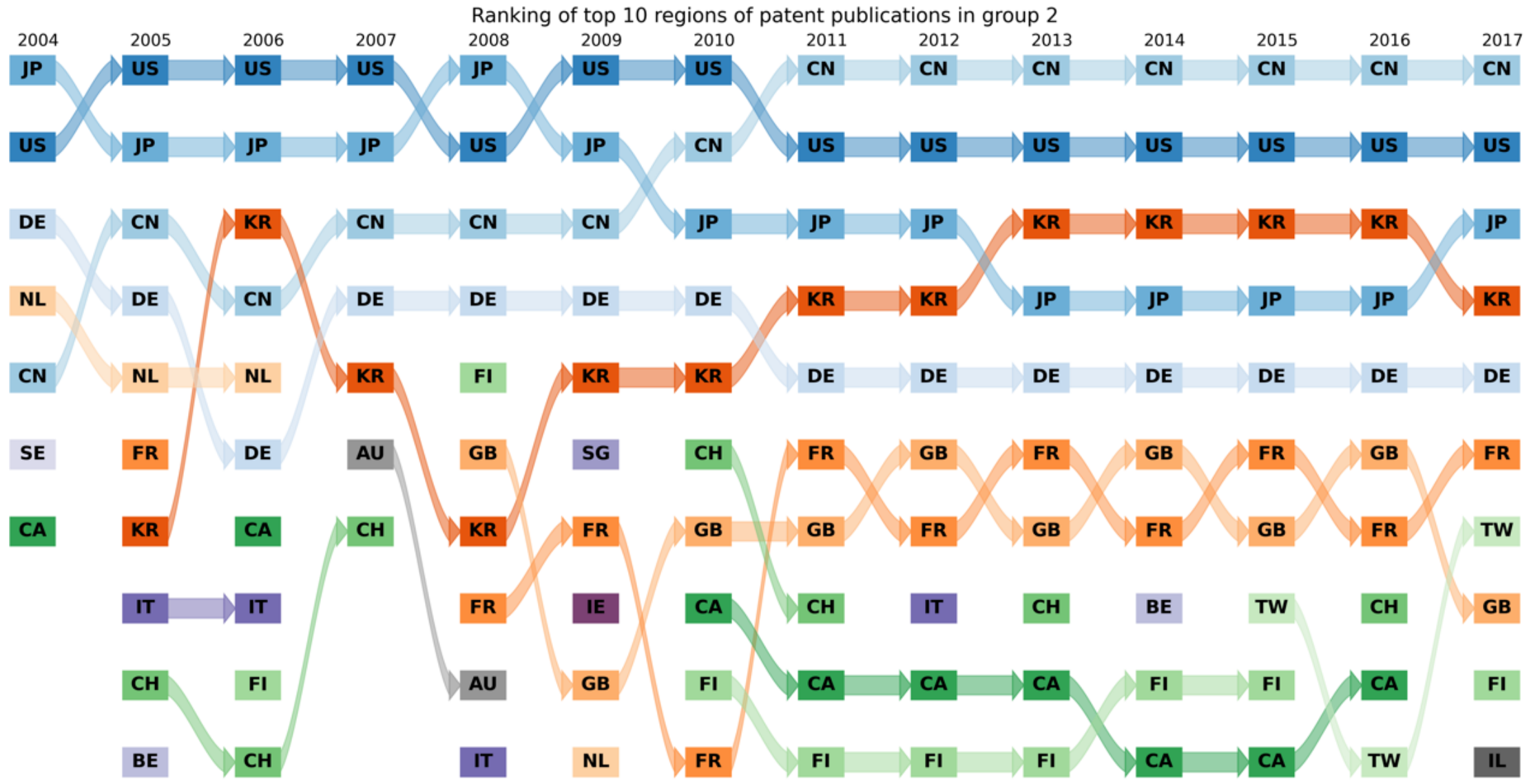

Figure 13: Ranking changes of the top 10 regions regarding the number of applications in (2) *batteries*.

For (0) *composites*, we see from Figure 12 an overall sequence of leadership changes JP – US – CN: JP was initially the leader between 2004 and 2006 before it was overtaken by the US, who was overtaken by the eventual leader CN in 2010. We see also from Figure 12 that after some initial uncertainties, KR rose steadily in ranking from 2009 to become the region with the third most patents in (0) *composites*. The same story is valid for (5) *catalysts*, as shown in Figure 33 in the Appendix. For (2) *batteries* and (4) *water treatment*, there was fierce competition between JP and US to be the leader from 2004 to 2010, before CN obtained the eventual leading position (Figure 13 and Figure 32 in the Appendix).

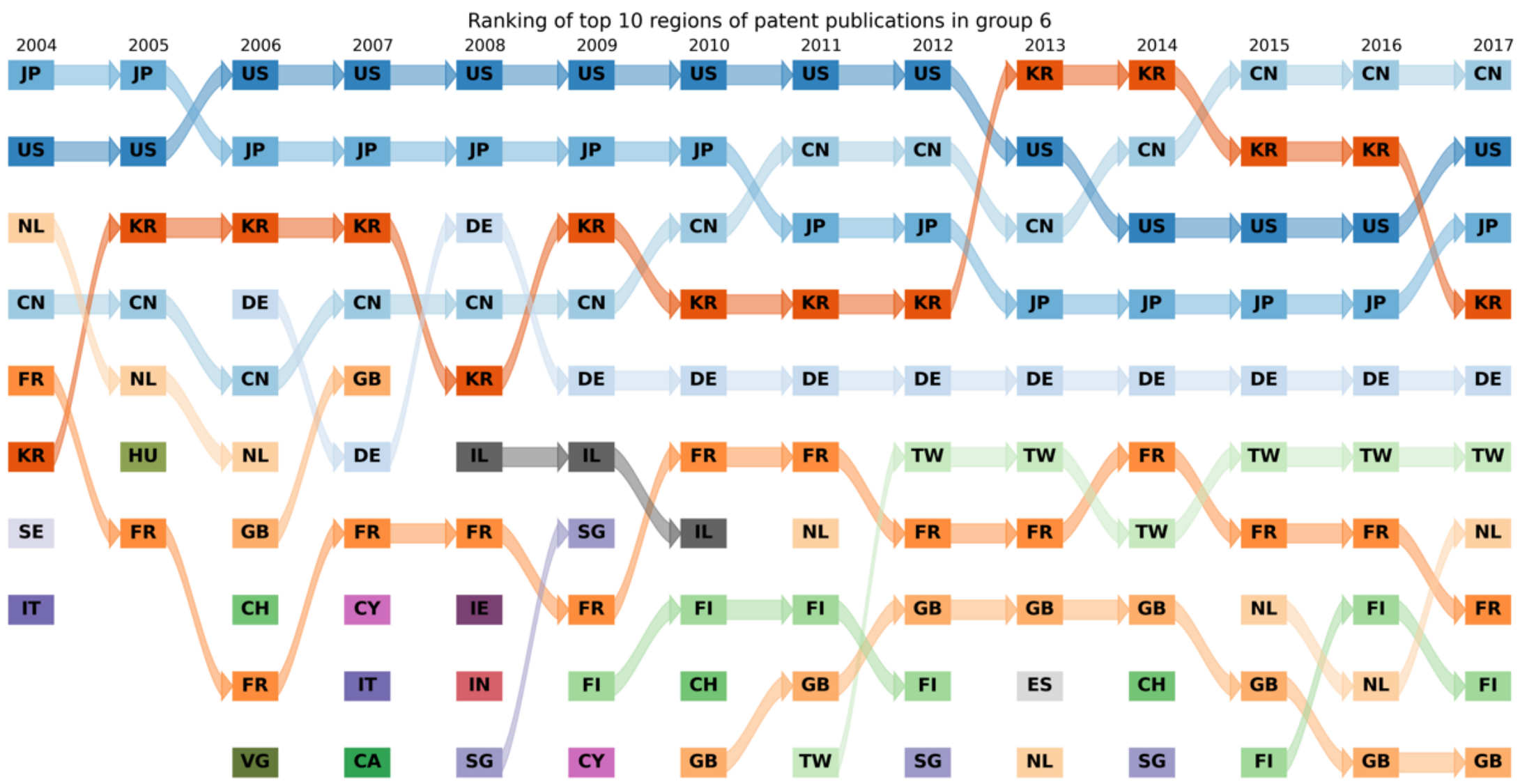

Figure 14: Ranking changes of the top 10 regions regarding the number of applications in (6) *devices*.

For (6) *devices*, we see from Figure 14 that the top four regions remained as CN, US, JP, and KR. However, there were two surprises. First, after lagging US and JP, KR briefly led the ranking in 2013 and 2014 before CN displaced it. Second, TW first made it into the top 10 in 2011 before ascending to sixth place by 2017. This is probably related to TW's strength in the semiconductor field.

By visualizing the rankings of different regions for the various technology areas, we see that the overall picture is consistently CN rising to the top after several years. This ascent is likely the result of substantial government funding in graphene research. However, obtaining data on the financing level in CN to test this hypothesis is challenging. More importantly, we can see crossings clearly from the ranking plots, but we cannot see these from the plots of proportions. We can also see from the ranking plots whether the ranking changes were sudden or gradual. For example, KR started its ranking later than CN but rose quickly and occupied the number 3 spot ahead of JP in (0) *composites* and (3) *sensors*. Here, we see that the top regions are consistently the top economic leaders worldwide as of 2017. CN was the second largest economy in the world, but it beat the US (the world's largest economy) in all graphene technology areas. If data becomes available, we would like to test whether this is the outcome of a more significant investment in graphene research. KR was another outstanding example. Ranked twelfth in the world in terms of its GDP, KR was consistently in the top 3 or 4 of all patent rankings. Going further down the rankings, we see that NL was also an overperformer because it was within the top 10 in some graphene technology areas, despite being $18^{th}$ in the world in terms of its 2017 GDP (*GDP by Country - Worldometer*, n.d.).

### 4.4 Categorical Analysis

By dis-aggregating the patent data into different years and regions, we find the exciting story of a technology area displacing the previous technology area and then being superseded by the following technology area. We believe this is a general feature in the evolution of technology, whereby innovators switch their attentions from one area with decreasing returns to another showing greater promise. However, we also noticed regional differences in the technological evolution sequence. Since university researchers form a globalized community, the sequences reflecting their interest in successive technology areas should be the same. We believe this is not the case for corporations, which have to compete with each other within the same regions, or on the global stage. Therefore, they are more likely to invest in technology areas that help them avoid competition at the regional or international levels.

This means that it is potentially possible for the technology evolution sequences to be different from region to region if we consider only the corporations.

To check whether there are systematic differences between universities and corporations, we further dis-aggregate the patents of different technology areas, based on the following three types of assignees: (0) corporations, (1) universities, and (2) others. After that, we plot the proportions of patent applications within the three assignee types in Figure 15. As we can see, for corporations the technology evolution sequence consists first of (1) *synthesis*, then (0) *composites*, then (6) *devices*, and after that (2) *batteries*. Of the remaining technology areas, we see that (4) *water treatment* has not yet peaked by 2017, but its proportion of patents was already approaching that of (6) *devices*. When we look at the universities, the technology evolution sequence started first with (1) *synthesis*. After that, we see a close competition between (1) *synthesis* and (6) *devices*. For (5) *catalyst*, we see that the proportion of corporate patents peaked in 2011, ahead of the proportion of university patents, which peaked in 2012. The university technology evolution sequence also ended with (2) *batteries*, just as for the corporate technology evolution sequence. We found (4) *water treatment* gaining interest at the end of the studied period.

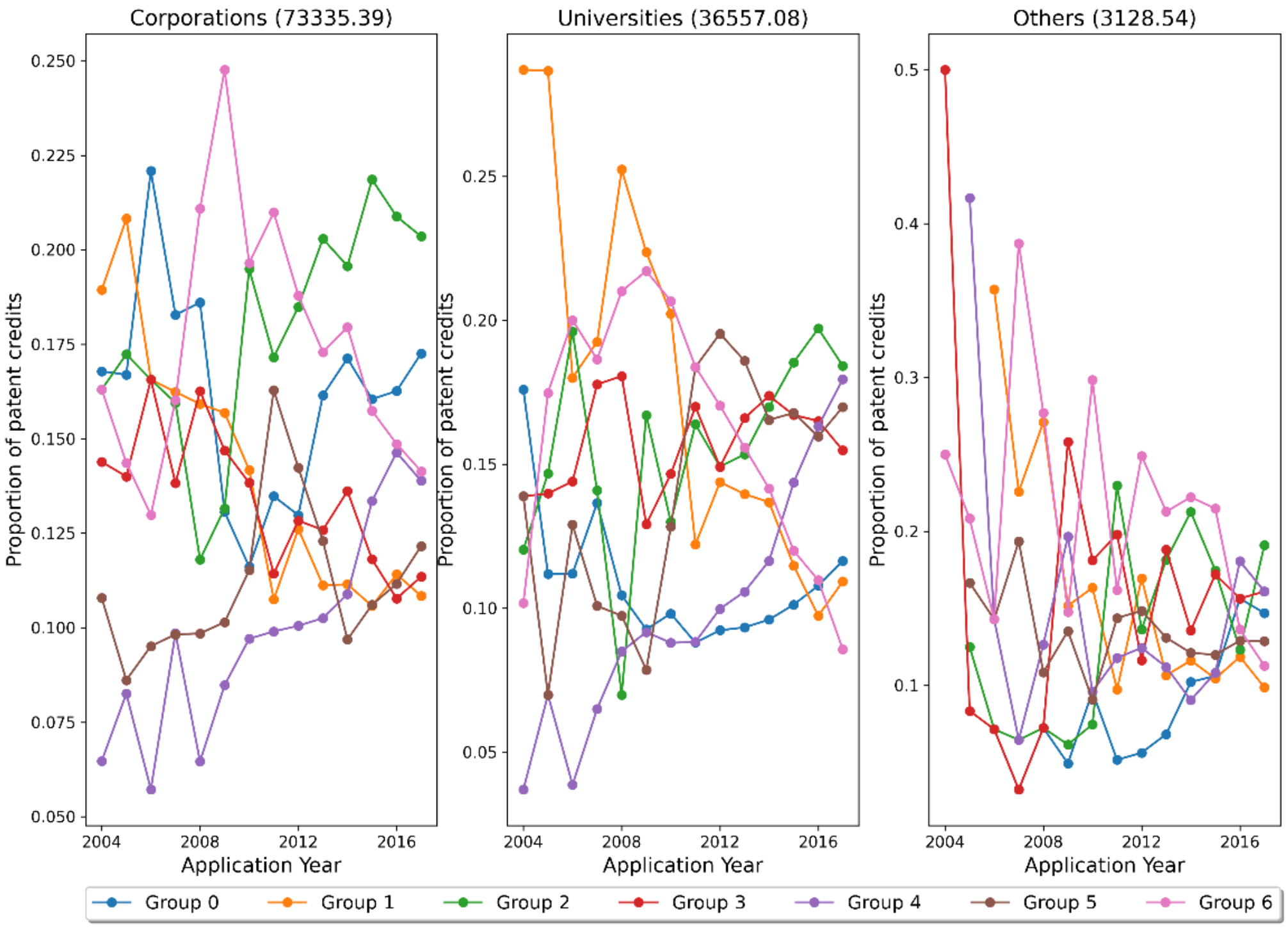

Figure 15: The proportions of G-T applications in seven groups of three categorical assignee groups.

The next natural step would be to further dis-aggregate the patents for different assignee categories into individual assignees. If we can do this, we would be able to show which assignees are diversified (expected for universities and large multi-national corporations, owning patents across multiple technology areas) and which are specialised (expected for small regional corporations, holding patents in one or two technology areas). Unfortunately, even for the top six regions that we are interested in, the number of assignees is significant for each assignee category. We show such plots in Figure 34 in the Appendix, but most of the graphs are messy. The same problem occurs for universities (Figure 35) and Others (Figure 36). We also worry that focusing on the top companies or top universities with the most patents may introduce bias into our understanding.

While it is possible to see how diverse the eco-systems were by looking at the proportions of corporate and university patents in Figure 15, this graph tells us more about changes in time (which technology areas dominated what periods and were replaced by which other technology areas), but not whether an assignee category was diverse or specialized. The main challenge here is that an assignee can be specialized if all its patents are in the same technology area. It can also be diversified if its patents are all in different technology areas. This distinction between specialized or diversified can apply to assignees with few patents, but it can also apply to assignees with many patents. In this sense, an assignee's degree of diversification (or specialization) is a quantity independent of its total number of patents. Let us write the patent portfolio of an assignee as $\vec{p} = (p_0, p_1, \dots, p_k, \dots, p_6)$, where $p_k$ is the proportion of its patents in the technology area $k$. The most commonly used measure of the diversity of the patent portfolio is the entropy function

$$S = -\sum_{k=0}^{6} p_k \log(p_k) \,.$$

To investigate the distribution of diversities over all corporations and universities and over all sizes of patent portfolios, we go to the entropy plots shown in Section 4.5.

## 4.5 Entropy

In Figure 16, we plotted the entropies of all assignees of a given category against their accumulated patent credits up till 2017, in the form of heat maps. As explained in Section F of the Appendix, red regions have a high density of assignees on these heat maps, green regions have an intermediate density of assignees, and blue regions have a low density of assignees. Focusing on the heap maps of corporations (left) and universities (middle) in Figure 16, we see that the maximum log patent counts for the two categories are about the same, but a more significant fraction of university patent portfolios are diversified. On the other hand, corporations see a wide range of entropy (from very diversified to very specialized) in their patent portfolios.

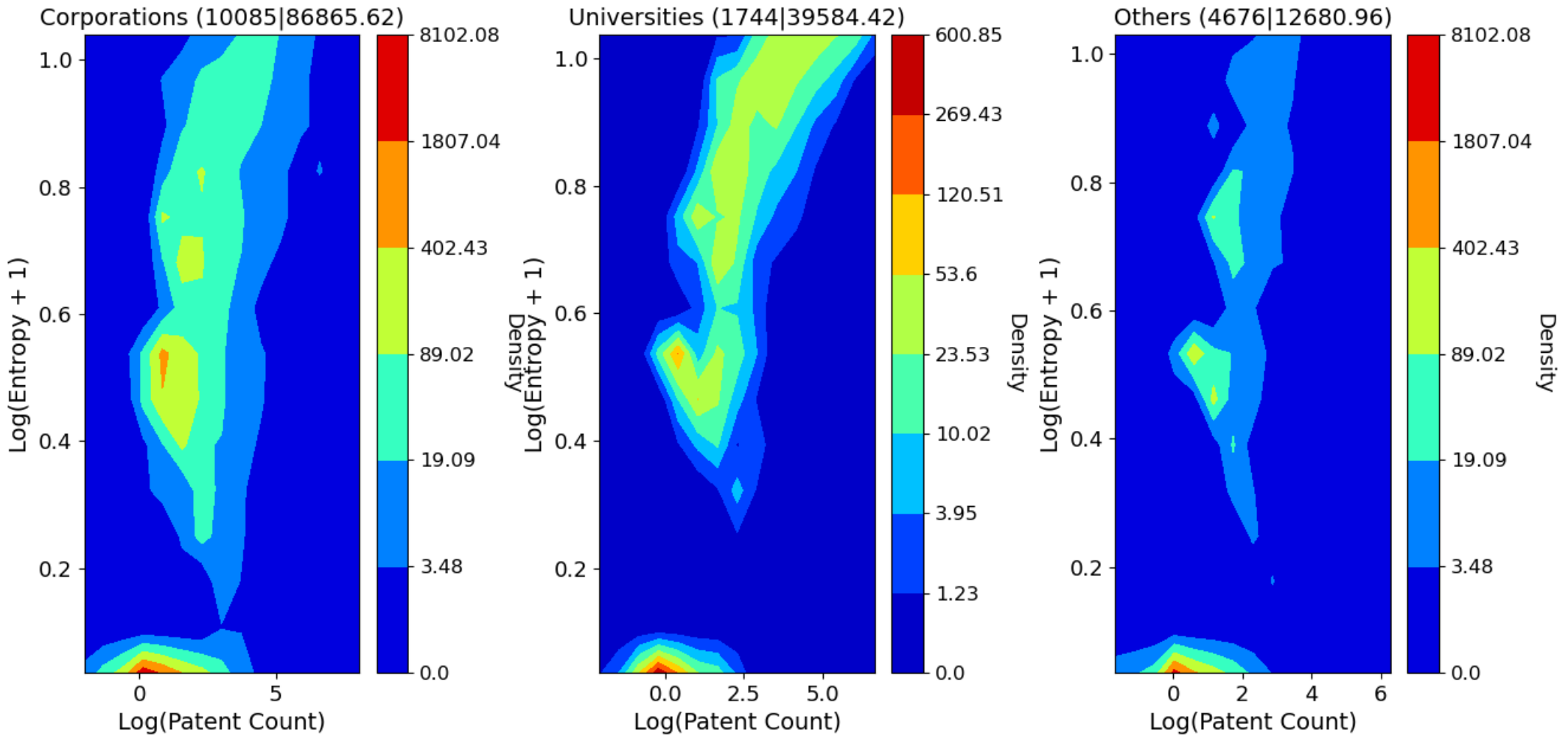

Figure 16: The technological entropy versus the number of patents in different entities from three categories.

In creating heat maps like these, the price we pay is the loss of temporal information (i.e., how the patent portfolios change over time). It is possible to put such dynamics back in two ways. First, plot the heat maps for every year or every two years. Second, track how the assignees manage their patent portfolios every year. We settled for the second way to avoid having an excessive number of figures.

To follow the temporal change in assignee $i$, let us use $(n_i^t, S_i^t)$ to denote its accumulated number of patents and its entropy of patents in year $t$. For example, in year $t+1$, this 'coordinate' of assignee $i$ would be $(n_i^{t+1}, S_i^{t+1})$. A vector from $(n_i^t, S_i^t)$ to $(n_i^{t+1}, S_i^{t+1})$ shows how the entropy of assignee $i$ changes when its patent portfolio increases from $n_i^t$ to $n_i^{t+1}$. After that, the time evolution for all patent portfolios of a particular category would be shown as a vector field, where for each assignee $i$ we have one or more arrows pointing from $(n_i^t, S_i^t)$ to $(n_i^{t+1}, S_i^{t+1})$ for different $t$. However, this is very noisy, so we created a two-dimensional histogram. We then averaged all vectors within each histogram bin and plotted the average vector at the centre of the bin. Because the assignees are not uniformly distributed across the phase space, we used different colours to indicate densities.

### 4.5.1. Entropy Changes for Different Combinations of Entropy and Accumulated Number of Patents

In Figure 17, we used the entropy versus accumulated patents graphs to track the entropy movements of entities from different assignee categories over 14 years in the 2004 - 2017 period. However, this plot is still not informative because an assignee that is moving from the bottom left corner (few patents on the same topic) to the upper right corner (many patents, many topics) cannot be distinguished from an assignee that is stuck in the middle (number of patents growing, but the number of topics stagnant) of the phase space. Therefore, to tease more information from the data, we classified the assignees into three groups based on their accumulated numbers of patents at the end of 2017. The first group is the lower quartile of all assignees, the second group is the interquartile group, and the third group is the upper quartile.

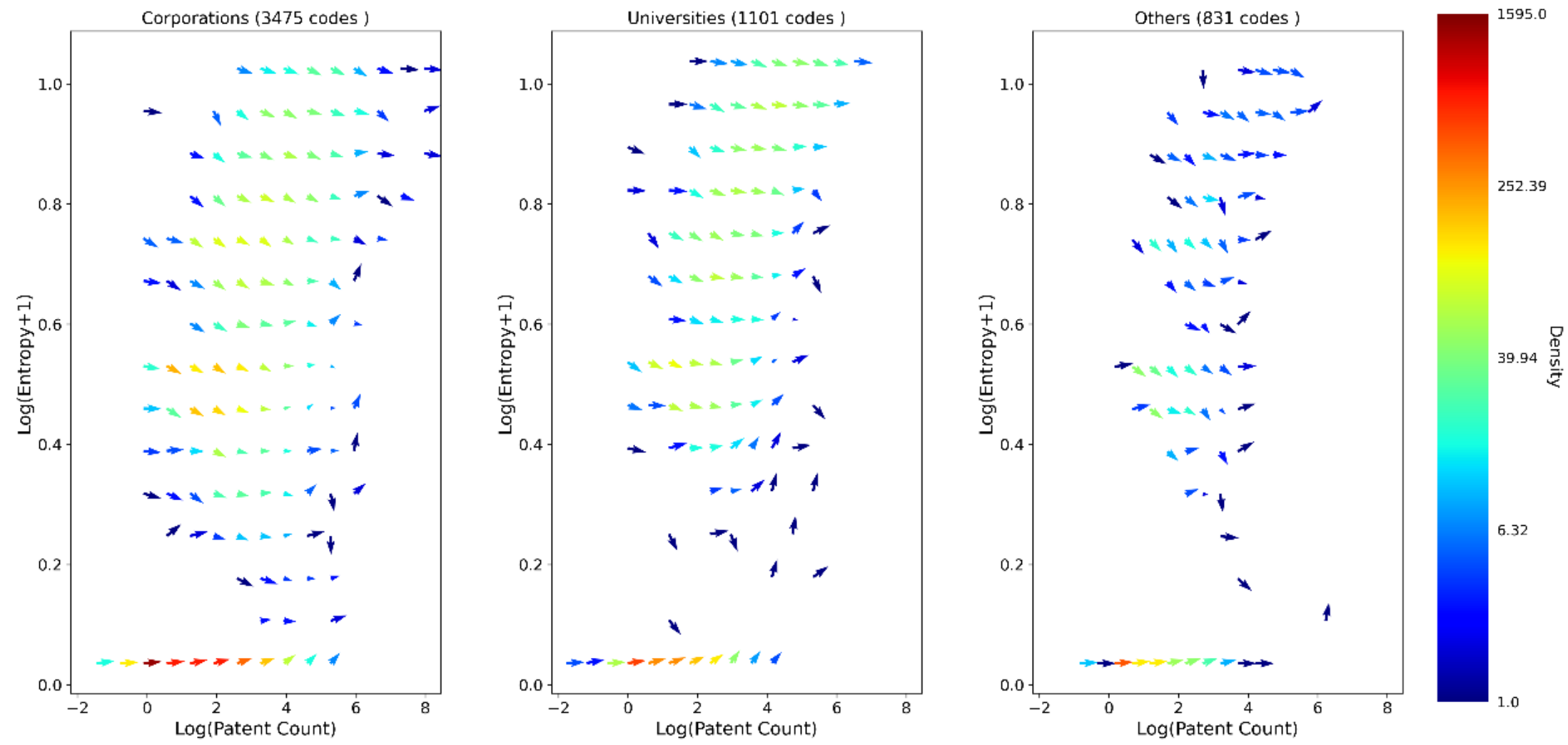

Figure 17: Plots of entropy versus the accumulated number of patents showing entropy changes across 14 years from 2004 to 2017 for entities in three categories. In this figure, a blue vector results from averaging over a low density of vectors, a green vector comes from averaging over an intermediate density of vectors, and a red vector comes from averaging over a high density of vectors.

In Figure 18, for both corporations and universities, the lower quartile groups showed clear decreasing trends (i.e., over time individual entities became less diverse technologically as they accumulated more patents). For the corporate interquartile group, the individual entities also became less diverse technologically over time. For the university interquartile group, we found that the most diverse entities became slightly less diverse over time, whereas the less diverse entities became somewhat more diverse over time. Unfortunately, the dataset is not large enough to ascertain the existence of a fixed point of intermediate diversity that all entities evolve towards. Finally, the vector fields of the upper quartile groups tell us that entities ending up with large patent portfolios eventually will also become highly diverse technologically.

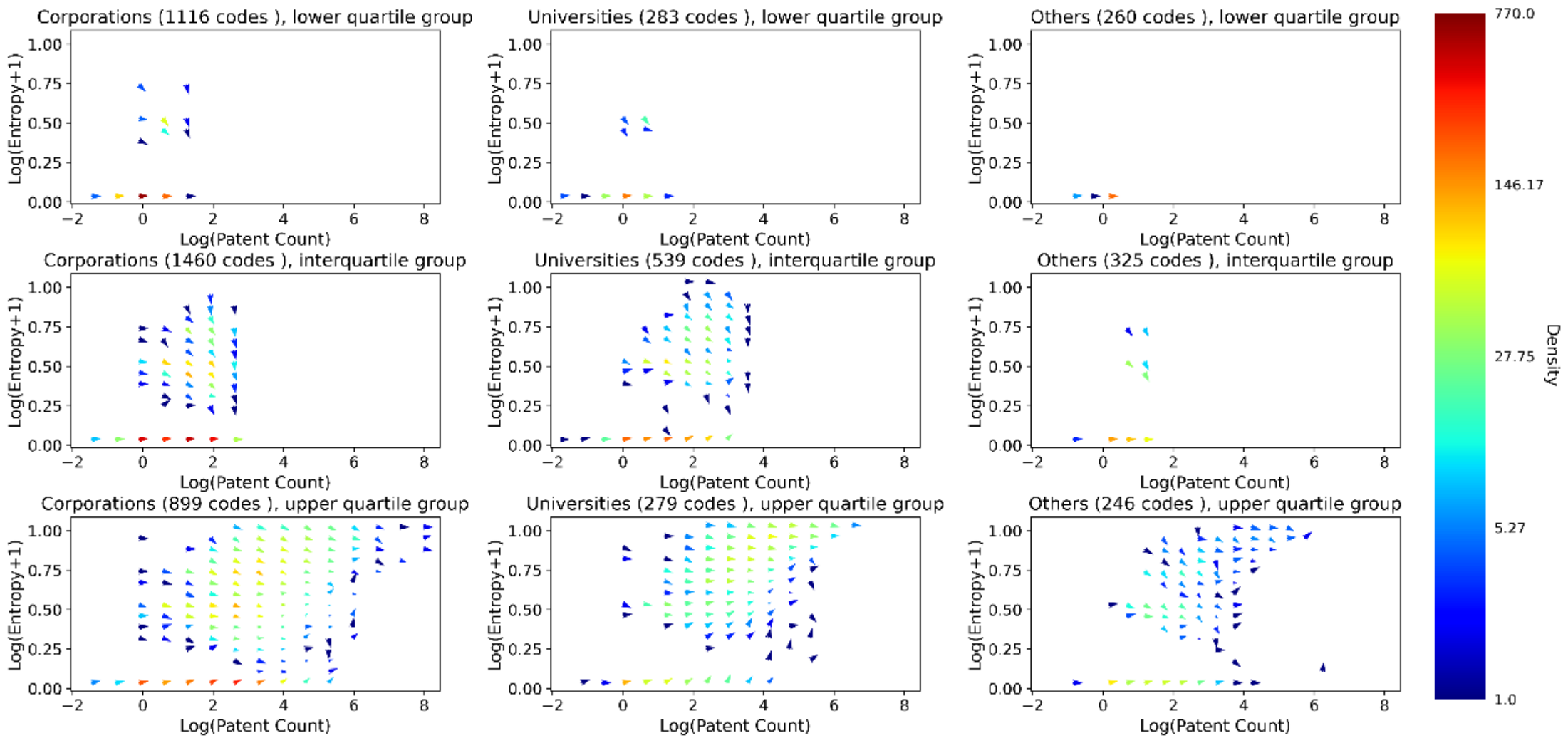

Figure 18: Plots of entropy versus the accumulated number of patents showing entropy changes across 14 years from 2004 to 2017 for entities in three quartile groups and three assignee categories. In this figure, a blue vector results from averaging over a low density of vectors, a green vector comes from averaging over an intermediate density of vectors, and a red vector comes from averaging over a high density of vectors.

Here, let us note the finite size effect due to the discrete portfolio used to compute entropy (~ 0.25/4 or 0.25/5) is roughly 0.05. This means the gaps seen in Figure 18 cannot be explained as being due to finite size effect. In the case of the corporate interquartile group, the diversity of an entity hardly changes as its patent portfolio increases in size. On the other hand, the patent portfolio diversity of a successful entity in the upper-quartile group increases with time, suggesting that this entity tends to venture into new technology areas.

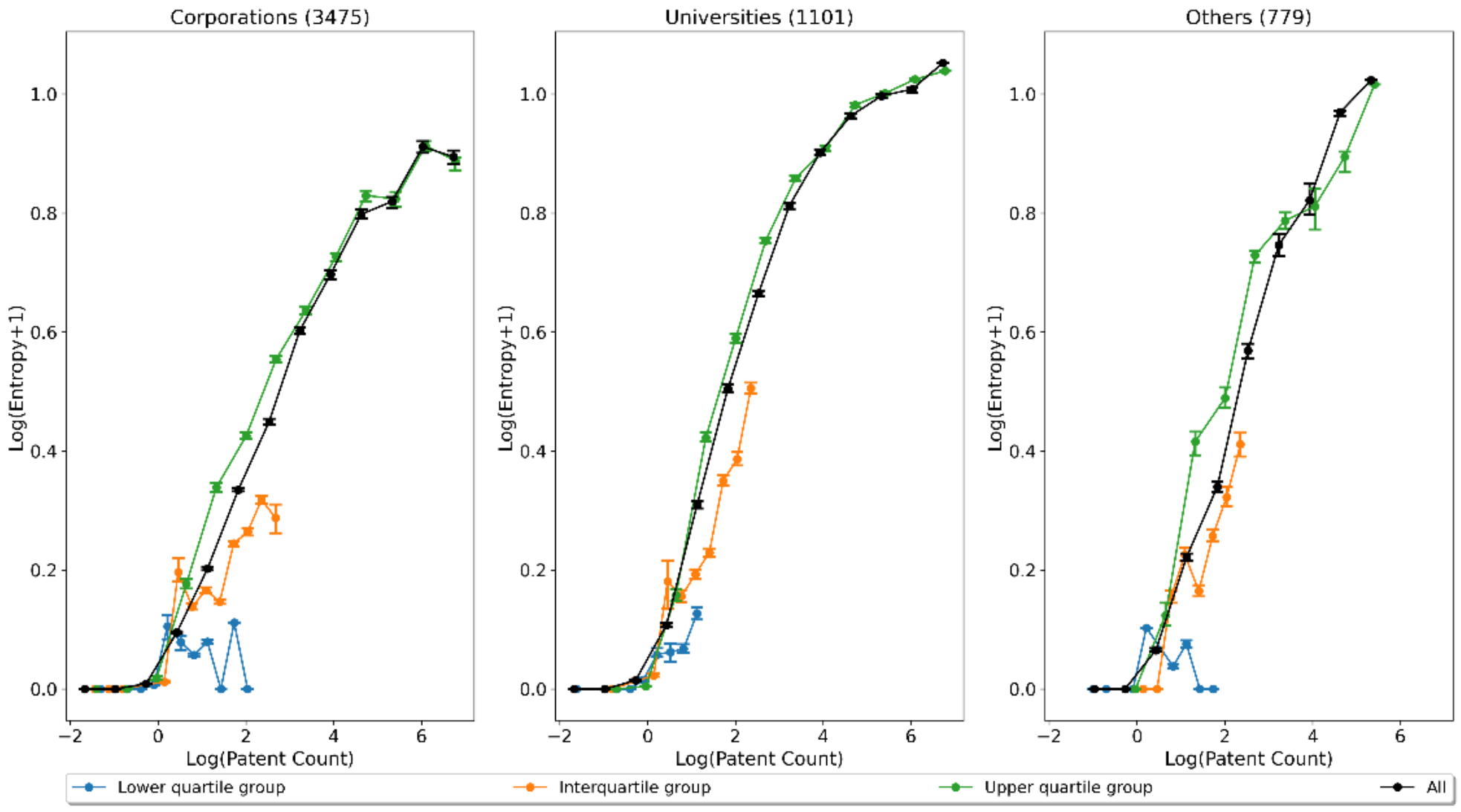

Figure 19: The plots of the average entropy of all entities (black) and the average entropies of entities from three assignee percentile groups in different assignee categories for 2014 – 2017.

Here, let us note that this method of visualization using vector fields has a natural bias, in the sense that an entity with many patents has its trajectory stretched along the patent count axis. In contrast, an entity with fewer patents would have its trajectory compressed along the patent count axis. This introduces a distortion of the vector field that makes it difficult for us to compare the time evolution of these two entities. Although by plotting the average log entropy over the three quartile groups we see a clearer time evolution of the patent diversity in Figure 19, the problem of difference in scales of patent counts remained.

### 4.5.2. Entropy versus Relative Years

Instead of plotting entropy values versus accumulated patent credits over 14 years in Section 4.5.1, we replaced the accumulated patent axis by time in years relative to 2004, which we considered to be the starting year of graphene technology. We then visualized the changes in entropy across different relative years for all entities in the three assignee categories, as shown in Figure 20. As with Figure 17, it is complicated to visualize patent diversity changes from these accumulated log-scale plots.

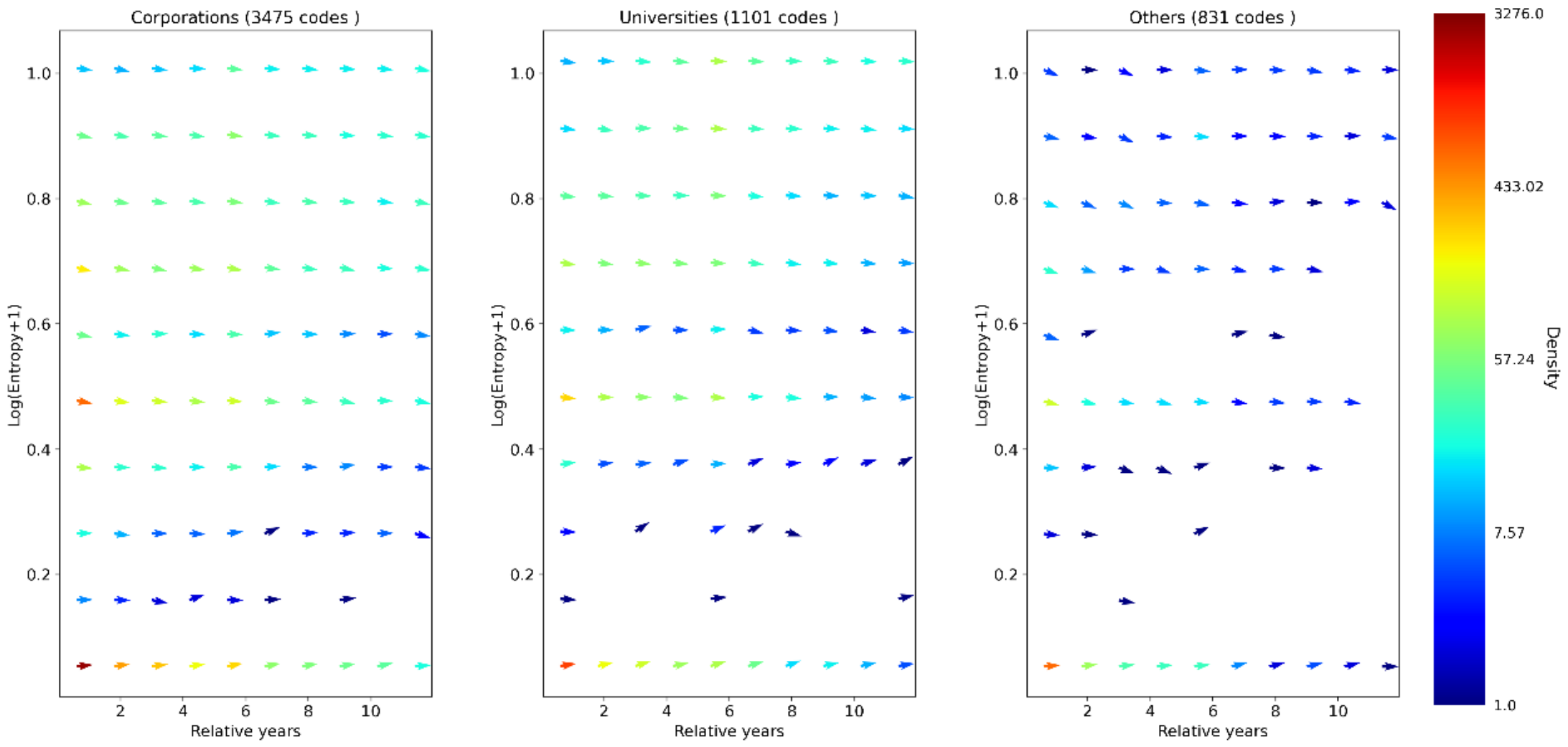

Figure 20: Plots of entropy versus relative years to 2004 showing entropy changes across 14 years for entities in three categories. In this figure, a blue vector results from averaging over a low density of vectors, a green vector comes from averaging over an intermediate density of vectors, and a red vector comes from averaging over a high density of vectors.

Therefore, we again classified entities from every assignee category into three quartile groups based on their total numbers of G-T patent applications. Then, we showed their entropy movements in Figure 21. The lower quartile groups of corporations and universities evolved to lower entropy values. For the interquartile groups, the corporate entities moved towards a narrow range of intermediate entropy values, whereas the university entities evolved towards a broader range of entropies. Finally, for the upper quartile groups of corporations and universities, the distribution of entropies remains broad after increasing slowly for ten years. Based on the colour of the arrows, upper-quartile corporate entities started with a wide distribution of entropy values (the arrows were associated with intermediate (green) to high (red) densities), and this distribution remained broad after ten years (the arrows were associated with intermediate (green) densities). In contrast, the distribution of entropy values for upper-quartile university entities has significantly narrowed (the arrows for log entropy larger than 0.7 were associated with intermediate (green) densities, while the arrows for log entropy smaller than 0.7 were associated with low (blue) densities) after ten years, even though they started with the same broad distribution of entropy values. This suggests that the most successful corporations managed their patent portfolios differently from the most successful universities.

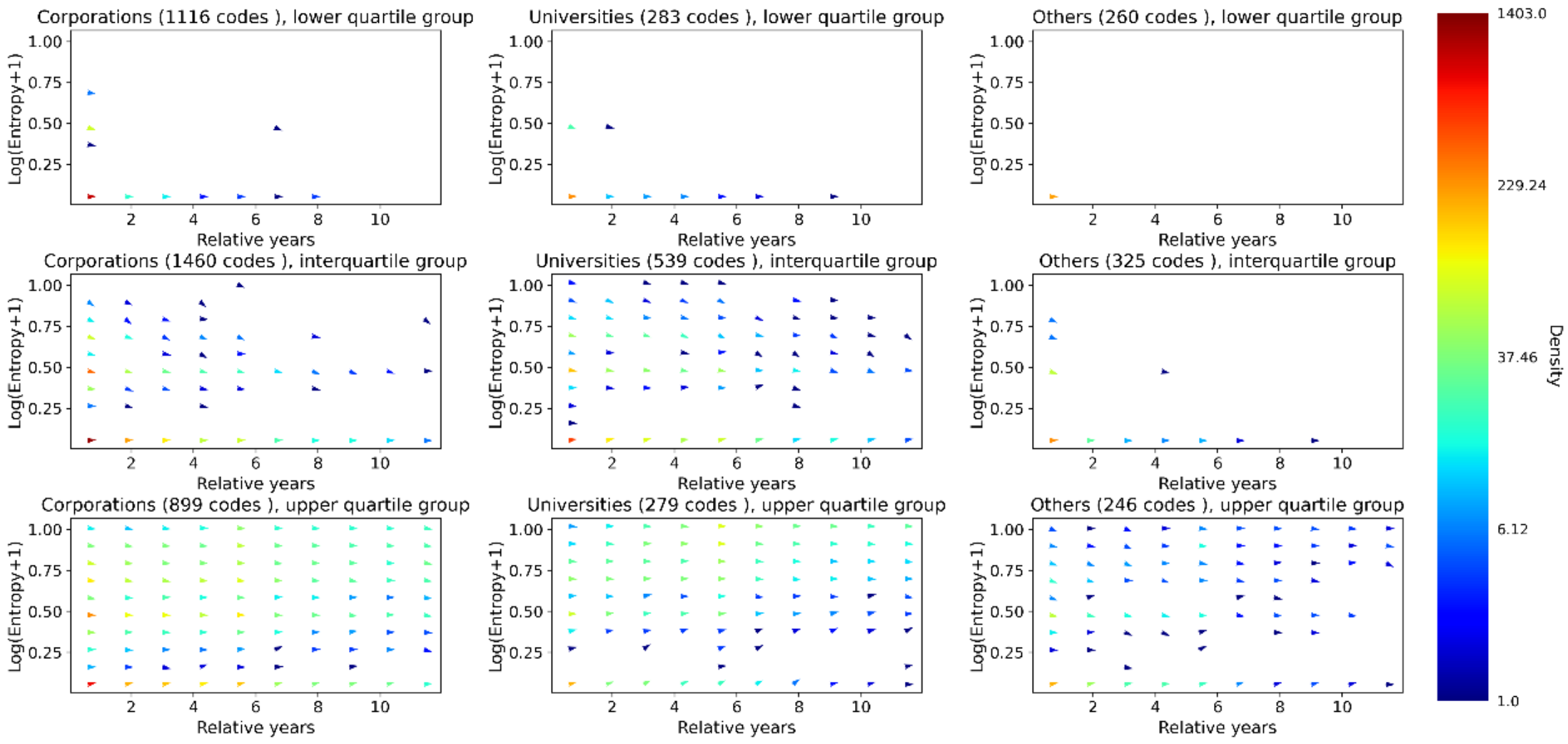

Figure 21: Plots of entropy versus relative years to 2004 showing entropy changes across 14 years for entities in three quartile groups and three assignee categories. In this figure, a blue vector results from averaging over a low density of vectors, a green vector comes from averaging over an intermediate density of vectors, and a red vector comes from averaging over a high density of vectors.

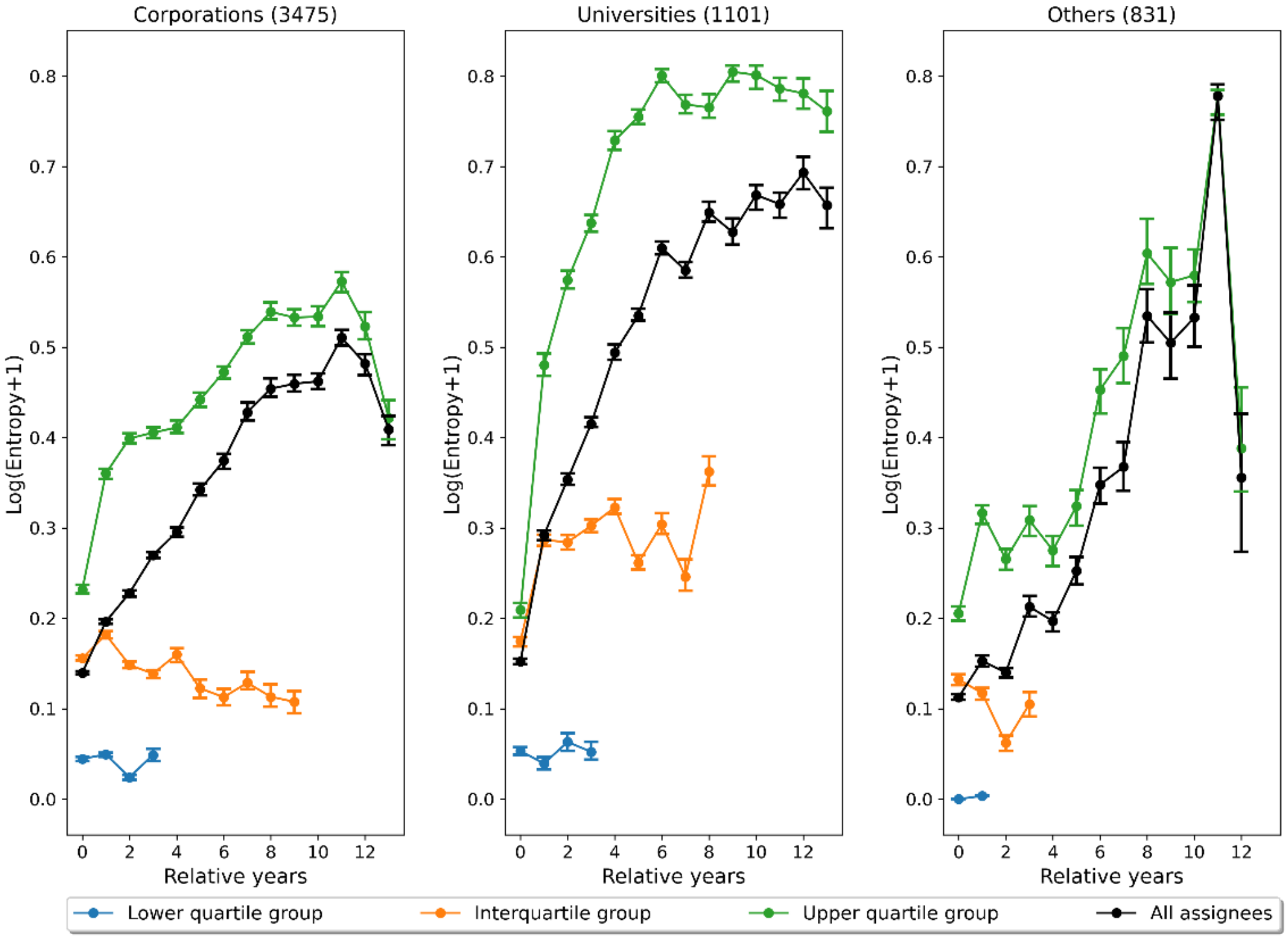

Figure 22: The average entropy across 14 relative years to 2004 and in three quartile groups.

We can show these differences in time evolution between corporations and universities more clearly by plotting the average log entropy as a function of year for the three quartile groups (instead of over the number of patents), as shown in Figure 22. For the lower quartile group, the average log entropy of universities was comparable to that of corporations. This behaviour should be contrasted to the average log entropy of universities in the interquartile group, which is significantly higher than that of corporations. Finally, for the upper quartile group, the average log entropy of corporations and universities started around 0.2, but that of corporations rose to just below 0.6, whereas that of universities rose close to 0.8. This is a big difference in the diversity of their patent portfolios, suggesting that these are managed differently.

#### 4.5.3. Portfolio Management Hypothesis

To explain the results in Section 4.5.2, let us ask what differences in the management of patent portfolios can give rise to these observed differences between diversities. Our hypothesis is (i) to develop a product, a group of patents is needed, (iia) *corporations* use bundles of closely related patents to develop products. If possible, they want to (iib) avoid paying to license patents. Therefore, they (iic) buy patents to complete product bundles and sell patents they have no chance of using for products. In contrast, *universities* do not develop any products, so they (iiia) keep all their patents and (iiib) focus on licensing them. We will test this hypothesis by analyzing reassignment and licensing records in Sections 5.1 and 5.2 respectively, and examine three representative case studies in Section 5.3. Moreover, we also investigate whether there may be different regional realizations of this hypothesis in Section 5.4.

## 5. Discussion

To test our patent management hypothesis, we were forced to use only US-filed patents because only these have the necessary reassignment and licensing information. Out of 20,843 G-T patents filed in the US, 15,650 (75.1 %) patents were initially filed by corporations, while 3,807 (18.3 %) were from universities. Since only a tiny proportion of US G-T patents (0.18%) has a mix of university, corporation, and/or others as their initial assignees, we can discuss corporate patents and university patents in a well-defined manner. For G-T patents with mixed categories of corporations and universities, we consider the patent a university patent if there are more university assignees, a corporate patent if there are more corporate assignees, and others otherwise. Assuming that the hypothesis is true, we expect to find few or no reassignments of university patents but more licensing of such patents. On the contrary, we expect to see few or no licensing of corporate patents both ways (i.e., a corporation is unlikely to license out its patents or to license patents from another corporation) but buys many corporate/university patents and sells some of its corporate patents.

### 5.1 Reassignment

We first looked at the reassignment from one entity to another in all US-filed G-T patents from both corporations and universities to find evidence supporting hypotheses (iic) and (iiia), respectively. In 15,650 patents initially assigned to corporations, 12,785 did not go through assignee change, while the remaining 18.3% accounted for 3,808 transactions. In the list of reassignments (some patents may be reassigned more than once), 81.9% were reassigned from one corporation to another, and 13.6% were reassigned from corporations to universities. For example, 189 patents from HON HAI PRECISION INDUSTRY CO LTD were reassigned to TSINGHUA UNIVERSITY, 14 patents were transferred from UT-BATTELLE LLC to OAK RIDGE ASSOCIATED UNIVERSITIES, and six patents were reassigned from SAMSUNG ELECTRONICS CO LTD to RESEARCH & BUSINESS FOUNDATION SUNGKYUNKWAN UNIVERSITY.

On the other hand, out of 3,807 patents initially assigned to universities, 82.6% of them (3,144 patents) did not go through any assignee changes, while the remaining 663 patents (17.4 %) accounted for 793 transactions. In the list of reassignments, 45.5% of them were reassigned from universities to corporations (361), and 36.3% were reassigned from universities to universities (288). For example, one patent belonging to the UNIVERSITY OF CALIFORNIA was transferred to LAWRENCE LIVERMORE NATIONAL SECURITY LLC. In contrast, RESEARCH & BUSINESS FOUNDATION SUNGKYUNKWAN UNIVERSITY moved seven patents to SAMSUNG (5 to SAMSUNG ELECTRONICS CO LTD and 2 to SAMSUNG TECHWIN CO LTD) and 3 to GRAPHENE SQUARE INC. At face value, the number of changes of assignees for university patents seems to be many. Still, a significant number are between different entities within the same universities, and a small number are between various universities. It is also possible for a university to transfer its patents to a corporation that the university wholly or partially owns. This will become clearer after our discussion in Section 5.3.

Based on the aggregate statistics, the proportions of reassigned patents are more or less similar for corporations and universities. This is contrary to what we expect from our patent management hypothesis. Still, we will see from our case studies later that there is evidence to support the hypotheses (iic) and (iiia) at the individual corporate and university levels.

## 5.2 Licensing

Next, we investigated the licensing of corporate and university patents to validate hypotheses (iib) and (iiib), respectively. Out of 15,650 corporate patents, 430 patents were licensed once, 20 patents were licensed twice, and three patents were licensed three times. Only 2.9% of US corporate patents were licensed at least once. For comparison, we see that out of 3,807 university patents, 512 patents were licensed once, 78 patents were licensed twice, 21 patents were licensed three times, seven patents were licensed four times, and one patent was licensed six times. These total numbers make up 16.3% of US university patents. These results also confirm our hypothesis (iib) and (iiib) that universities are more likely to license patents while corporations avoid doing so.

The top 10 corporations in licensing out patents are MCALISTER TECHNOLOGIES LLC (93.0), UCHICAGO ARGONNE LLC (63.0), UT-BATTELLE LLC (52.0), LAWRENCE LIVERMORE NAT LAB (30.0), INT BUSINESS MACHINES CORP (27.0), BROOKHAVEN SCI ASSOC (18.0), NANOTEK INSTR INC (15.0), SANDIA CORP (15.0), GEORGIA TECH RES CORP (14.0), LOS ALAMOS NAT SECURITY LLC (13.0). Here, the number in the parenthesis following the name of an entity is its patent credit (see Section 3.2.5 on how this is computed). As we can see, most of these are US national labs. On the other hand, the top 10 universities in licensing out patents are MASSACHUSETTS INST TECHNOLOGY (60.0), UNIV RICE WILLIAM MARSH (59.0), UNIV NORTHWESTERN (55.5), UNIV HENGYANG NORMAL (43.0), BATTELLE MEMORIAL INST (37.0), CALIFORNIA INST OF TECHNOLOGY (27.0), HARVARD COLLEGE (26.0), UNIV PRINCETON (25.0), UNIV FLORIDA RES FOUND INC (20.0), and UNIV PENNSYLVANIA (20.0). Apart from BATTELLE MEMORIAL INST and UNIV FLORIDA RES FOUND INC, the remaining US-located universities from this top 10 list were also in the list of the top 50 US universities on the 2017 college ranking by Wall Street Journal/Times Higher Education (*Wall Street Journal/Times Higher Education College Rankings 2017*, 2016).

It is also interesting to see which entities were paying license fees to use G-T patents. Even though there were 1,244 licensing instances, there were only 22 licensees. These are US DEPARTMENT OF ENERGY (281+130), NATIONAL SCIENCE FOUNDATION (24+367), NATIONAL INSTITUTES OF HEALTH (8+91), ADVANCED GREEN TECHNOLOGIES LLC (94+0), NAVY (10+72), NASA (3+64), AIR FORCE (25+15), BLACK DIAMOND STRUCTURES LLC (23+0), ARMY (4+15), DARPA (2+6), MANOMECH INC (2+1), UNITED STATES PATENT AND TRADEMARK OFFICE (1+1), UNIV PENNSYLVANIA (0+2), LOCKHEED MARTIN CORPORATION (1+0), SUNEDISON SEMICONDUCTOR TECHNOLOGY PTE LTD (1+0), INTELLECTUAL DISCOVERY CO LTD (0+1). As we can see, most of these are US government entities. In the above list, the numbers in parentheses following the licensee name are the number of corporate and university patents licensed. Besides being the largest licensee, the US DEPARTMENT OF ENERGY also licensed roughly the same numbers of corporate and university patents. In contrast, other US government entities such as the NATIONAL SCIENCE FOUNDATION, NATIONAL INSTITUTES OF HEALTH, NAVY, NASA, and others licensed predominantly from universities.

## 5.3 Case Studies

To provide evidence for hypotheses (iic) and (iiia), we chose to examine the details of two case studies in graphene. The first case study is SAMSUNG, a multi-national corporation heavily invested in graphene technology. We chose this as our case study for corporations because it has the highest number of G-T patents worldwide when we aggregate all SAMSUNG subsidiaries. In terms of the number of patents acquired, SAMSUNG ranked third, after GLOBALFOUNDRIES (with mainly internal transactions and external transactions from IBM ("GLOBALFOUNDRIES Acquires IBM's Microelectronics," 2014)), and SANDISK (with mostly internal transactions from and external transactions from FUSION-IO INC (Hesseldahl, 2014)). After SAMSUNG on the ranking based on the total number of acquired G-T patents, we have TSINGHUA UNIVERSITY (with mostly external transactions from HON HAI PRECISION INDUSTRY CO LTD due to the set-up of the Tsinghua-Foxconn Nanotechnology Research Center with funding from the company (*Tsinghua-Foxconn Nanotechnology Research Center*, n.d.)), and NOKIA TECHNOLOGIES OY (with mostly internal transactions between NOKIA TECHNOLOGIES OY and NOKIA CORPORATION).

The second case study is RICE UNIVERSITY, a medium size private university in Houston, Texas in the US. RICE UNIVERSITY has the third highest number of G-T patents in the US, after UNIVERSITY OF CALIFORNIA and MASSACHUSETTS INSTITUTE OF TECHNOLOGY. We chose this as our case study for universities because it ranks highly in reassignment and licensing. Regarding the average number of licenses per patent, RICE UNIVERSITY ranked higher than the other two universities, at 52.6%, compared to 49.0% for the UNIVERSITY OF CALIFORNIA and 36.1% for the MASSACHUSETTS INSTITUTE OF TECHNOLOGY.

### 5.3.1 SAMSUNG

In the Derwent Innovation database, there are nine entities with 'SAMSUNG' in their names. These are SAMSUNG ELECTRONICS CO LTD, SAMSUNG DISPLAY CO LTD, SAMSUNG SDI CO LTD, SAMSUNG ELECTRO-MECHANICS CO LTD, SAMSUNG MEDISON CO LTD, SAMSANG ELECTRONICS CO LTD, SAMSUNG TECHWIN CO LTD, TOSHIBA SAMSUNG STORAGE TECHNOLOGY KOREA CORPORATION, and SAMSUNG UNIVERSITY. Therefore, we treated them as subsidiaries of the same multinational corporation and referred to their aggregate as SAMSUNG. As of December 2017, SAMSUNG has filed 1,276 US G-T patents and owns a more significant number of G-T patents (2,924) filed in other regions.

In our data, SAMSUNG acquired 242 patents from 58 entities. However, 79 patents were from 6 internal entities. For example, SAMSUNG ELECTRONICS CO LTD acquired 26 patents from SAMSUNG SDI CO LTD. For the remaining 163 patents, 53 were obtained from 15 corporations, while 81 were acquired from 22 universities. The remaining 29 patents were acquired from 15 individuals. The top five corporations that transferred patents to SAMSUNG were: UNIDYM (10), NANOTEK INSTRUMENTS INC (8), INTERNATIONAL BUSINESS MACHINES CORPORATION (7), XEROX CORPORATION (6), UNIST ACADEMY - INDUSTRY RESEARCH CORPORATION (5). The top five universities that transferred patents to SAMSUNG were: RESEARCH & BUSINESS FOUNDATION SUNGKYUNKWAN UNIVERSITY (18), SEOUL NATIONAL UNIVERSITY R & DB FOUNDATION (14), INDUSTRY-ACADEMIC COOPERATION FOUNDATION YONSEI UNIVERSITY (9), MIE UNIVERSITY (5), KOREA ADVANCED INSTITUTE OF SCIENCE AND TECHNOLOGY (5). Here, the numbers in parentheses following the names of corporations/universities tell us how many G-T patents they transferred to SAMSUNG. We checked that UNIDYM is a US company that makes carbon nanotubes. In contrast, NANOTEK INSTRUMENTS INC originated from Australia and has investments from SAMSUNG CORNINGS, but it does not seem to make any graphene products. Out of 81 patents SAMSUNG acquired from universities, 63 are from Korean universities/institutes, 11 are from US universities, one from a Chinese university (FUDAN UNIVERSITY), and five from a Japanese university (MIE UNIVERSITY). The top three Korean universities that SAMSUNG acquired the most patents from are: RESEARCH & BUSINESS FOUNDATION SUNGKYUNKWAN UNIVERSITY (SKKU) (18), followed by SEOUL NATIONAL UNIVERSITY R & DB FOUNDATION (14), and INDUSTRY-ACADEMIC COOPERATION FOUNDATION YONSEI UNIVERSITY (9). SAMSUNG has always benefitted from collaborations with Korean universities. Its first partnership was with SKKU from 1965 to 1977. After a 20-year lapse, SAMSUNG started re-engaging the Korean Universities in 1996 ("THE STRATEGIC ALLIANCE BETWEEN SUNGKYUNKWAN UNIVERSITY AND THE SAMSUNG GROUP," n.d.), starting with SKKU but expanding soon to other universities. More recently, SAMSUNG began several global initiatives, such as the Global Scholar Program, the University Partnership Program, and the Global Research Outreach Program to tap into the talents and ingenuity in universities worldwide.

Next, we found that SAMSUNG transferred 123 patents to 22 entities. Out of these, 79 patents were assigned to 5 internal entities. The remaining 44 patents were transferred to 17 entities, including 21 patents to 8 corporations and 23 patents to 9 universities. SAMSUNG transferred the most patents to RESEARCH & BUSINESS FOUNDATION SUNGKYUNKWAN UNIVERSITY (11), followed by S-PRINTING SOLUTION CO LTD (10), ULSAN NATIONAL INSTITUTE OF SCIENCE AND TECHNOLOGY (5), LOTTE ADVANCED MATERIALS CO LTD (4), and ROBERT BOSCH GMBH (2). The transfer of patents to universities seemed surprising initially but can be explained in terms of the partnership between SAMSUNG and the universities. In agreement with the hypothesis (iic), SAMSUNG does indeed buy and sell patents, presumably to manage their patent portfolio. However, the number of patents purchased and sold is small compared to the number of patents it holds, and we see that SAMSUNG bought more patents than it sold. Unfortunately, we cannot find the lists of patents used in individual SAMSUNG products. We thus cannot confirm whether there are products that were developed using patents acquired or the patents sold are in unrelated technology areas. Later in Section 5.3.3, we will use the products and patents of Dyson to support hypotheses (i) and (iia). Finally, we checked that there are no records in the 1,276 G-T US patents owned by SAMSUNG that the company has licensed its technology to other entities. There are also no records of SAMSUNG licensing technologies from other entities, as far as US patents are concerned. This tells us that SAMSUNG is not paying licensing fees to other corporations to develop and sell its products. On the other hand, other companies are also not paying to use SAMSUNG's technologies. This supports the hypothesis (iib).

### 5.3.2 RICE UNIVERSITY

In the Derwent Innovation database, Rice University is referred to as RICE UNIVERSITY or WILLIAM MARSH RICE UNIVERSITY. In our dataset, 157 patents were initially filed by RICE UNIVERSITY. The university then acquired 22 patents from 14 entities and transferred the ownership of 11 patents to 9 entities. Of the 22 patents acquired, 14 were from 7 corporations (6 from LOCKHEED MARTIN CORPORATION, three from M-I LLC, etc.). Seven were from 6 universities (UNIVERSITY OF TEXAS, NORTHEASTERN UNIVERSITY, SOUTHWEST RESEARCH INSTITUTE, THE METHODIST HOSPITAL RESEARCH INSTITUTE, THE TRUSTEES OF THE UNIVERSITY OF PENNSYLVANIA, UNIVERSIDADE FEDERAL DE MINAS GERAIS). It also acquired one patent from the US Air Force.

Presumably, the transfer of 6 patents from LOCKHEED MARTIN CORPORATION is part of the nanotechnology strategic alliance between LOCKHEED MARTIN CORPORATION and RICE UNIVERSITY (*Lockheed Martin and Rice University to Announce Strategic Nanotechnology Partnership*, n.d.). Of the 12 patents that RICE UNIVERSITY gave up, seven went to 5 corporations (2 to MONOLITHIC 3D, 2 to NANOHOLDINGS LLC, etc.), and four went to 3 universities (2 to BAYLOR COLLEGE OF MEDICINE, 1 to CENTRE NATIONAL DE LA RECHERCHE SCIENTIFIQUE, and 1 to THE METHODIST HOSPITAL RESEARCH INSTITUTE). The last patent went to the US Air Force.

Besides patent reassignment, RICE UNIVERSITY also licensed its patents 96 times to the US Government, including 53 times to the NATIONAL SCIENCE FOUNDATION, 19 to NASA, 17 to the US NAVY, and 3 to the US DEPARTMENT OF ENERGY. RICE UNIVERSITY did not license any patents from other universities and corporations. These observations support hypotheses (iiia) and (iiib).

### 5.3.3 Dyson

So far, we have evidence from the aggregate statistics or the case studies to support hypotheses from (iib) to (iiib). Unfortunately, we could not get data on the relationships between SAMSUNG's patents and the products they enable for hypotheses (i) and (iia). Generally, we expect each technologically sophisticated product to depend on several patents. Still, we could not find any previous studies confirming this, except for a paper claiming that 3.5 patents were required on average to develop a drug in 2005 (Ouellette, 2010). This observation that multiple patents are needed to produce one product agrees with what we found on Dyson's patent website (*Patents | Dyson*, n.d.). On this website, Dyson openly listed 344 patents associated with 88 of its products. Using this information, we found that on average, every product depends on 33.14 patents. Furthermore, we found that a patent can be used in one or up to 53 Dyson products, and every pair of Dyson products share between 0 and 57 patents in common. These support our hypotheses (i) and (iia).

## 5.4 Comparison Between CN and US

Finally, we attempted to gain further insights into the global competition over graphene technology by comparing the management of patents in CN and the US. These are the top two regions according to the number of G-T patents. Based on the overall dataset, at a very high threshold ($p_0 = 99$) to disambiguate assignee names, we found that CN companies owned around 4.12 patents only while US companies owned on average 10.68 patents. On the other hand, US universities owned 27.64 patents on average, while CN universities owned 25.04 patents. Since there is no licensing information on the G-T patents filed in CN, and most patents held by CN companies and universities were filed in CN (96.0 %), we do not know how many such patents were licensed for use or used by their assignees to develop products. This makes it difficult to compare the licensing rates between CN and US. In general, we know from the literature that the patent utilization rate for US universities was around 30% (Morgan et al., 2001), whereas the utilization rate for US corporations was between 48.9% - 74% (Griliches, 1990; Morgan et al., 2001; Svensson, 2007). We do not know what specific rates apply in CN, but the utilization rate for universities is likely lower than that for corporations.

In our dataset, the top 5 US companies are IBM (790.5, founded 1911), ELWHA LLC (445, founded 1931), BAKER HUGHES (392, founded 1907), LOCKHEED CORP (306, founded 1926), and SANDISK 3D LLC (301, founded 1988). We see that they are all established companies. In contrast, the top 5 CN companies are OCEAN'S KING LIGHTING SCI & TECHNOLOGY (567, founded 1995), BOE TECHNOLOGY GROUP CO LTD (542, founded 1993), TCL China Star Optoelectronics Technology Co., Ltd (261, founded 2009), O-FILM TECHNOLOGY CO LTD (256, founded 2002), and CHENGDU NEW KELI CHEM SCI CO LTD (241, established 1992). Thus, the top CN corporations holding G-T patents tend to be younger, even though they may be equal to or even more than the US corporations in terms of capitalisation.

TCL China Star Optoelectronics Technology Co., Ltd was started after the experimental discovery of graphene in 2004. Therefore, we can classify it as a start-up whose sole or primary business is semiconductor display (*About Us - TCL华星光电技术有限公司 - TCL China Star Optoelectronics Technology Co., Ltd*, n.d.). Other CN corporations include BOE TECHNOLOGY GROUP CO LTD, OCEAN'S KING LIGHTING SCI & TECHNOLOGY, O-FILM TECHNOLOGY CO LTD, CHENGDU NEW KELI CHEM SCI CO LTD were started before 2004. BOE TECHNOLOGY GROUP CO LTD also has its core business around semiconductor display (*About Us_BOE Official Website*, n.d.), but the remaining three corporations are in different business areas: OCEAN'S KING LIGHTING SCI & TECHNOLOGY started as a traditional lightning company (*Company Profile_Ocean's King Lighting Science & Technology Co., Ltd.*, n.d.), O-FILM TECHNOLOGY CO LTD focuses on multiple technologies from touch displays to sensors to smart cars to optics (*Nanchang O-Film Tech Co Ltd - Company Profile and News*, n.d.), and CHENGDU NEW KELI CHEM SCI CO LTD specializes in starch-based bioplastics with a patenting culture since 1993 (*公司简介-Company Abouts*, n.d.).

Let us contrast these corporate profiles against the top 5 US corporations in different business areas. IBM is an established high-technology company considering the application of graphene in transistors and photodetectors (*IBM | Graphene-Info*, n.d.), in particular as a candidate for replacing silicon in computer chips (*IBM to Invest $3 Billion Trying to Seek the Next-Gen Chip Technology, Graphene Is a Candidate | Graphene-Info*, n.d.). As an energy company, BAKER HUGHES has invested in graphene composites and functionalization, mainly for the 3D printing industry (*Baker Hughes and Würth Industry North America Announce Joint 3D Printing Service Offering*, 2020; *UW SER's Collaboration Proposal With Baker Hughes Selected for DOE Funding | News | University of Wyoming*, n.d.). As the technology areas (2) *batteries* and (4) *water treatment* grow in importance in the US (see Figure 9), we find LOCKHEED CORP investing in graphene membrane technology (*Perforene Graphene Membrane*, 2018) and power storage (*Lockheed Martin Partners with Elcora to Deliver Graphene-Enhanced Li-Ion Batteries | Graphene-Info*, n.d.). Lastly, SANDISK 3D LLC has a long business history in digital storage/ memory devices for nearly three decades, is in cooperation with RICE UNIVERSITY and TSINGHUA UNIVERSITY (*Investing In Graphene: The Good, The Bad And The Ugly | Seeking Alpha*, n.d.) on developing graphene technologies to consolidate their leadership position in their core business areas.

Ideally, we should do case studies of a CN corporation and a CN university, to compare against SAMSUNG (or one of the top US corporations) and RICE UNIVERSITY. However, to do the latter using the top 50 universities filing patents in the US, we would only choose between TSINGHUA UNIVERSITY (63 patents) and the CHINESE ACADEMY OF SCIENCES (54 patents). Their patent portfolios are comparable to that of RICE UNIVERSITY (156 patents). However, from our dataset we found no records they had licensed their US-filed patents. This should not be considered evidence against our hypothesis, since it is understandable that the US Government, who is the largest licensee of graphene technology, would not use technology from foreign universities. It is possible that the two CN universities license their technologies in other regions (in particular, CN), but such activities are not captured in the dataset we downloaded from Derwent Innovation. In terms of the number of US-filed patents, we could perhaps use BOE TECHNOLOGY GROUP LTD with 111 patents (compared to SAMSUNG's 1,276 patents) as a CN corporate case study. However, most of their transactions were between BOE internal entities. The only exceptions were four transfers from GENERAL ELECTRIC to BOE, five transfers from ORDOS YUANSHENG OPTOELECTRONICS CO LTD to BOE, which supports our hypothesis, and one licensing out to NIH, which contradicts our hypothesis.

Ultimately, given that 96% of the patents held by CN entities were filed in CN, we cannot conclude with any statistical confidence without reassignment and licensing information from these patents. We can only confirm that the top CN and US corporations have different core businesses that can benefit from graphene technologies. Still, most of them maintain highly diverse patent portfolios (entropies between 1.26 and 1.77) spanning all seven graphene technology areas. The exception is SANDISK 3D

LLC, whose entropy value of 0.42 is low (with no patents in (2) *batteries* and (3) *sensors*), making it the most specialized corporation with significant investment into graphene technology.

## 6. Conclusions

This paper uses patents as a proxy to investigate the technology evolution in graphene at the aggregate, temporal, geographical, and assignee category levels. We identified seven graphene technology areas by applying a block-clustering method to the IPC subclasses. We used the z-score method described in Section 3.2.2 to extract the most meaningful words from the titles and abstracts of the G-T patents. The technology evolution sequences discovered, namely *synthesis – composites – sensors – devices – catalyst – batteries – water treatment*, agrees roughly with the sequence of scientific evolution in graphene. We then performed a geographical analysis on the top six regions regarding the accumulated number of patents and found that different regions had different preferred technology areas to invest in. For example, CN seemed to be interested mainly in *synthesis* and *batteries*. The US, KR, and JP heavily invested in *devices* before shifting to *batteries*. Moreover, CN, US, KR, and JP were consistently ranked in the top 4 over all graphene technology areas. CN displaced the US as the leader in all technology areas starting around 2010.

When we attempted to understand how different assignees chose to invest in different preferred technology areas, we found that there are simply too many assignees, many of which do not have large patent portfolios. Therefore, we cannot analyze individual assignees' diversity in the same way that we analyzed individual regions. Instead, we grouped assignees into three categories (universities, corporations, and others) and plotted heat maps of their entropy distributions as a function of patent portfolio size. A patent portfolio is diverse if its entropy value is high; otherwise, it is considered specialized. We found a significant difference between the most successful universities and corporations regarding their distributions of diversities and how these diversities evolved. To explain these differences, we presented a hypothesis whereby corporations must build up lists of closely related patents for product development. Thus, they prefer filing or buying over patents to paying licensing fees to use patents they do not own. Corporations will sell unused patents to other entities to recover the costs of developing these as a corollary. In contrast, universities rarely get involved in commercialization and product development. Therefore, they keep all of their patents and focus on licensing instead.

To validate this hypothesis, we analysed the reassignment and licensing activities among US patents. We found evidence supporting some parts of the hypothesis. We confirmed the remaining analyses on three representative case studies: SAMSUNG for corporations, RICE UNIVERSITY for universities, and DYSON for product-patent relationships. Moreover, an additional comparison of the distribution of diversities between CN and US (the two leading regions) at the level of their top five corporations suggested that there is little or no regional difference in this hypothesis. Despite the top five companies in both regions having different investing preferences in the seven technology areas, most of them are technologically diverse with high entropy values (1.26 to 1.77). The only exception is SANDISK 3D LLC in the US. While it had a large number of G-T patents and invested in five out of seven technology areas, it is the most specialized corporation with relatively low diversity (0.42).

## Appendix

### A. Comparison of Technological Topics Between $n = 7$ and $n = 9$

We showed in Figure 23 the modularity values determined by CoClus algorithm for the partitioning of G-T patents into different number of topics, $n$ ($2 \leq n \leq 20$). From Figure 4 and Figure 23, we chose $n = 7$ as the value of modularity remains more or less the same from $n = 7$ to $n = 20$. In Figure 24, we visualized the overlaps $f(k_l, k) = \frac{|O_l^{k_l} \cap O^k|}{\max(|O_l^{k_l}|, |O^k|)}$ between document clusters $O_l^{k_l}, O^k$ for these two cases.

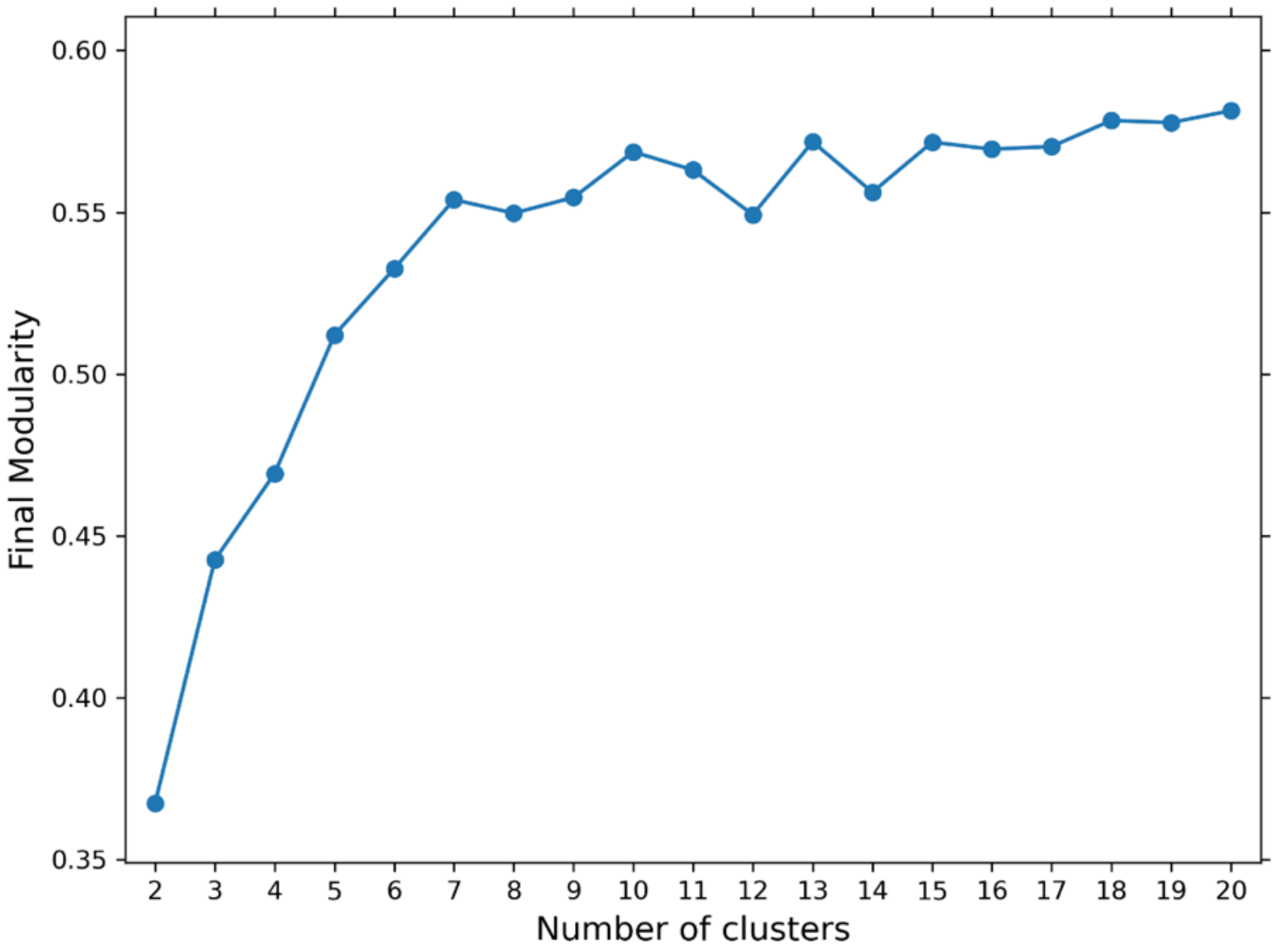

Figure 23: Plot of modularity values determined by CoClus algorithm for the organization of G-T application patents into different number of clusters from 2 to 20 based on their IPC subclasses.

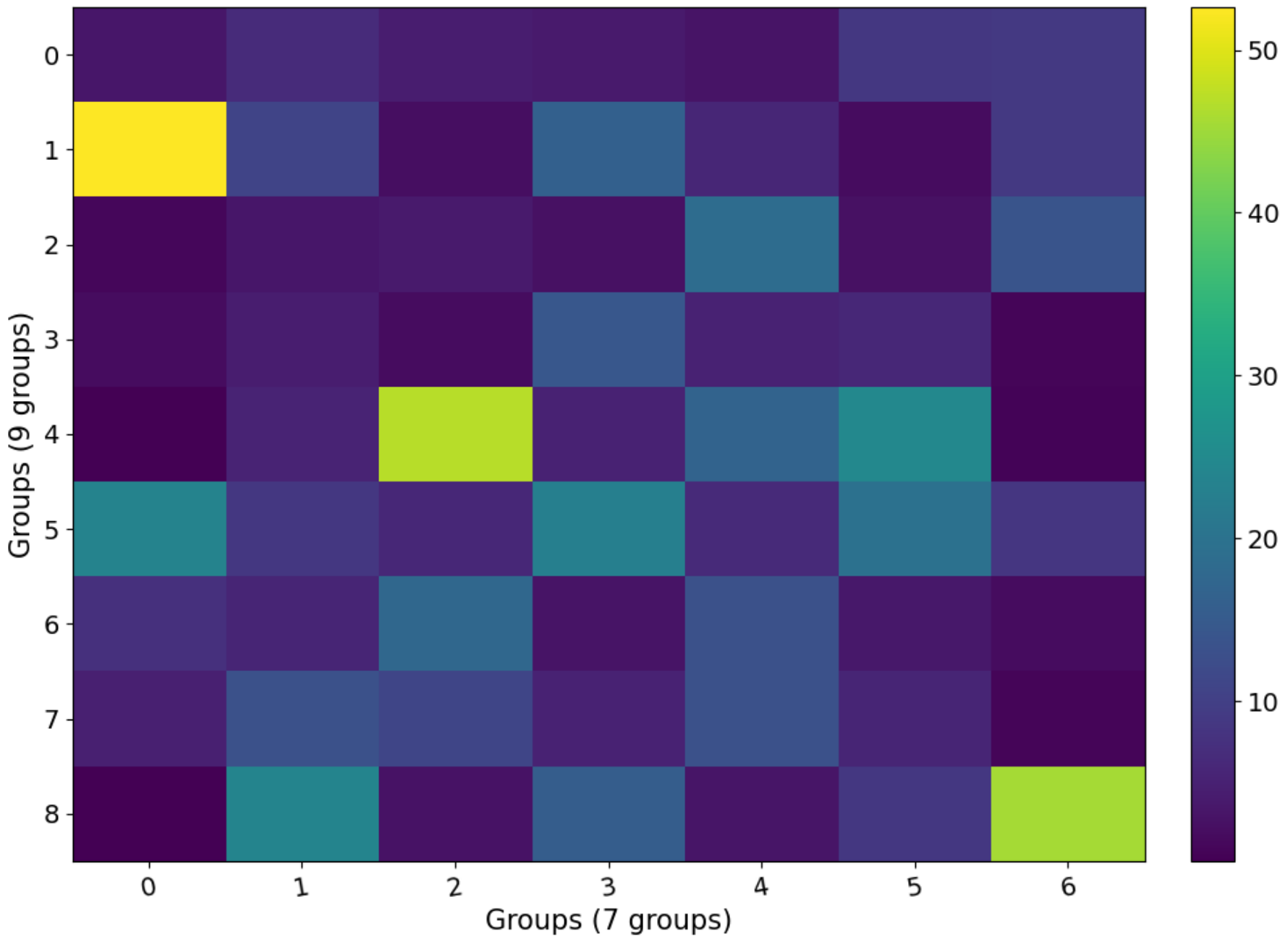

Figure 24: The overlaps $f(k_l, k) = \frac{|O_l^{k_l} \cap O^k|}{\max(|O_l^{k_l}|, |O^k|)}$ between cluster $O_l^{k_l}$ identified for $n = 7$ communities, and $O^k$ identified for $n = 9$ in the full dataset.

From Figure 24, we see that only group (0), group (2), and group (6) remain at least 50% the same from $n = 7$ to $n = 9$. In Figure 25, we used the flow visualization to show to overlaps in patent clusters and confirmed that when $n = 7$ group (0), group (2), and group (6) are highly overlapped with group (1), group (4), and group (8) in $n = 9$.

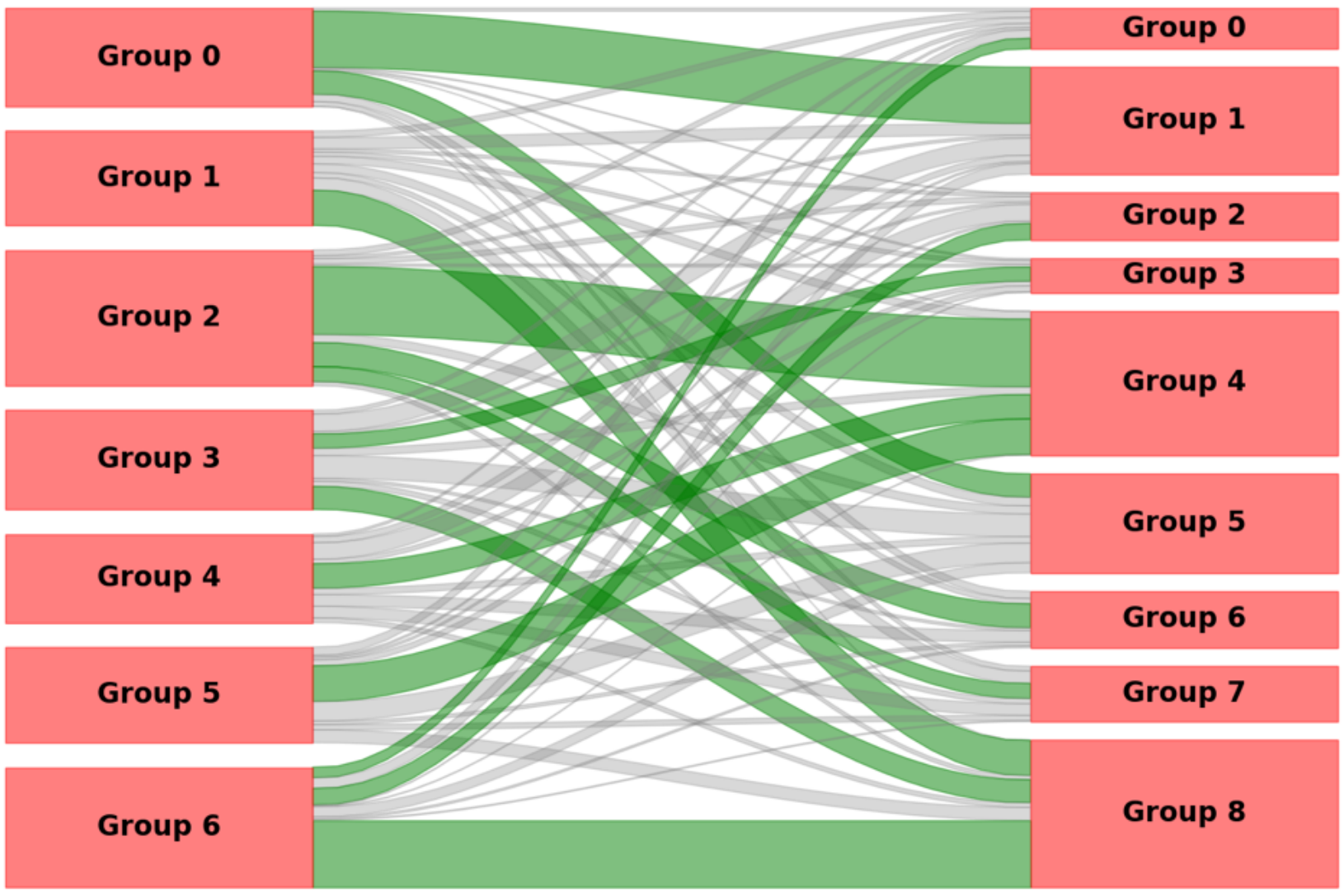

Figure 25: The overlapping documents between clusters for $n = 7$ and $n = 9$ communities. The flow between group $k_l$ and group $k_m$ represents the proportion of common documents between them.

## B. Sensitivity Analysis

### B.1. 90% of G-T Patents

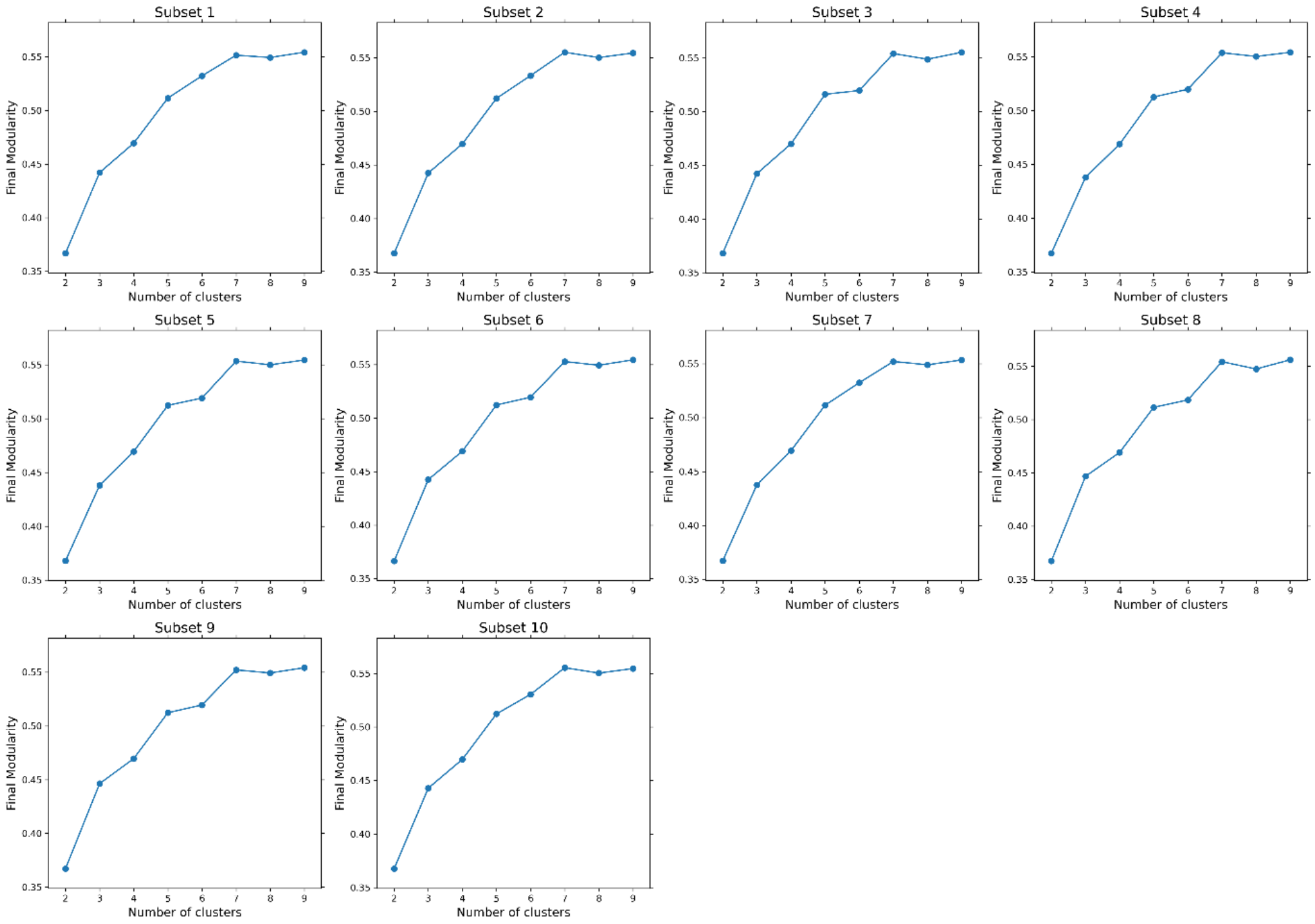

Figure 26: The plots of modularity at different number of clusters for ten subsets obtained by randomly selecting 90% of the G-T patents.

## B.2. 90% of IPC Subclasses

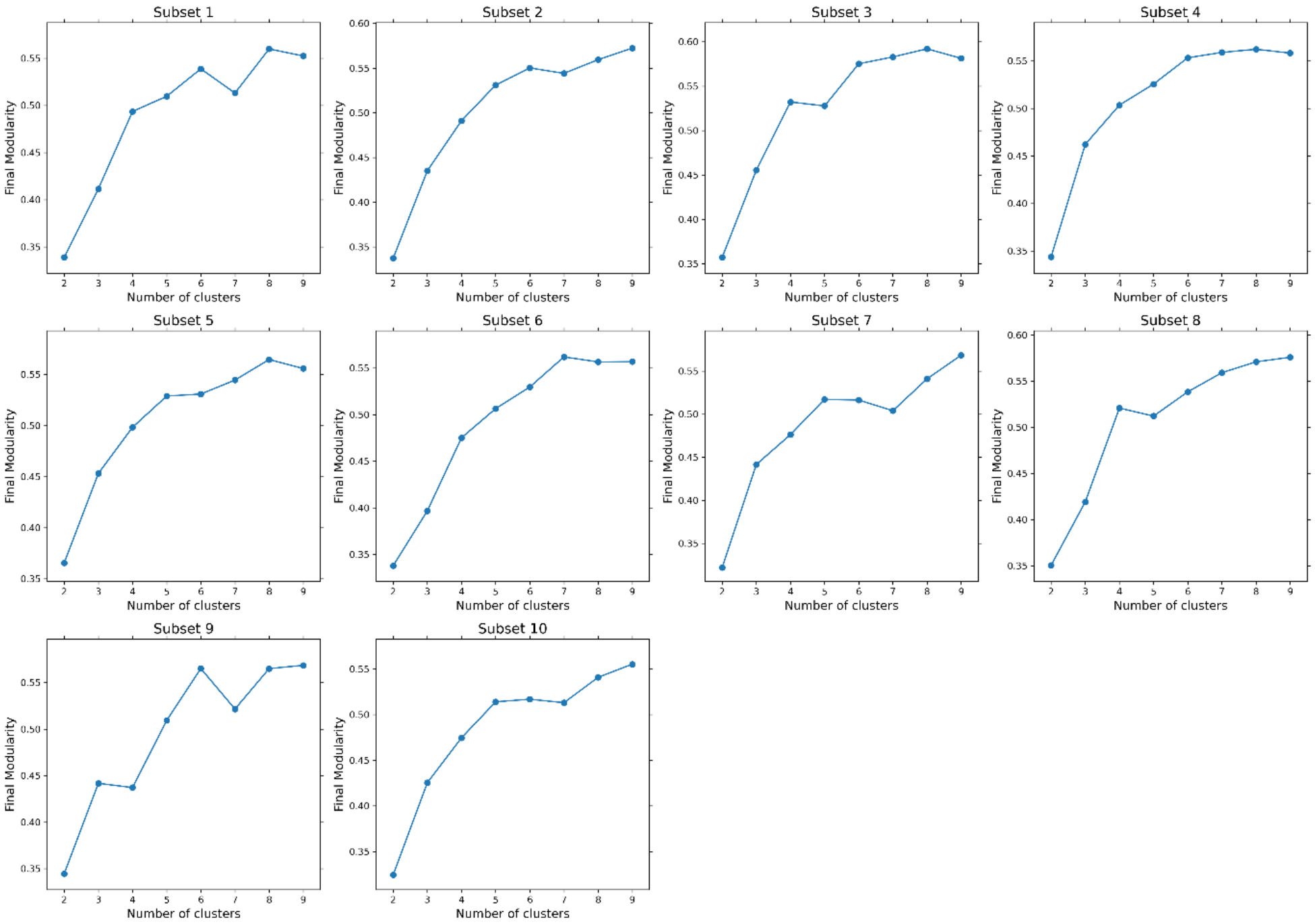

Figure 27: The plots of modularity at different number of clusters for ten subsets obtained by randomly selecting 90% of IPC Subclasses.

# C. Sensitivity Analysis

## C.1. 90% of G-T Patents

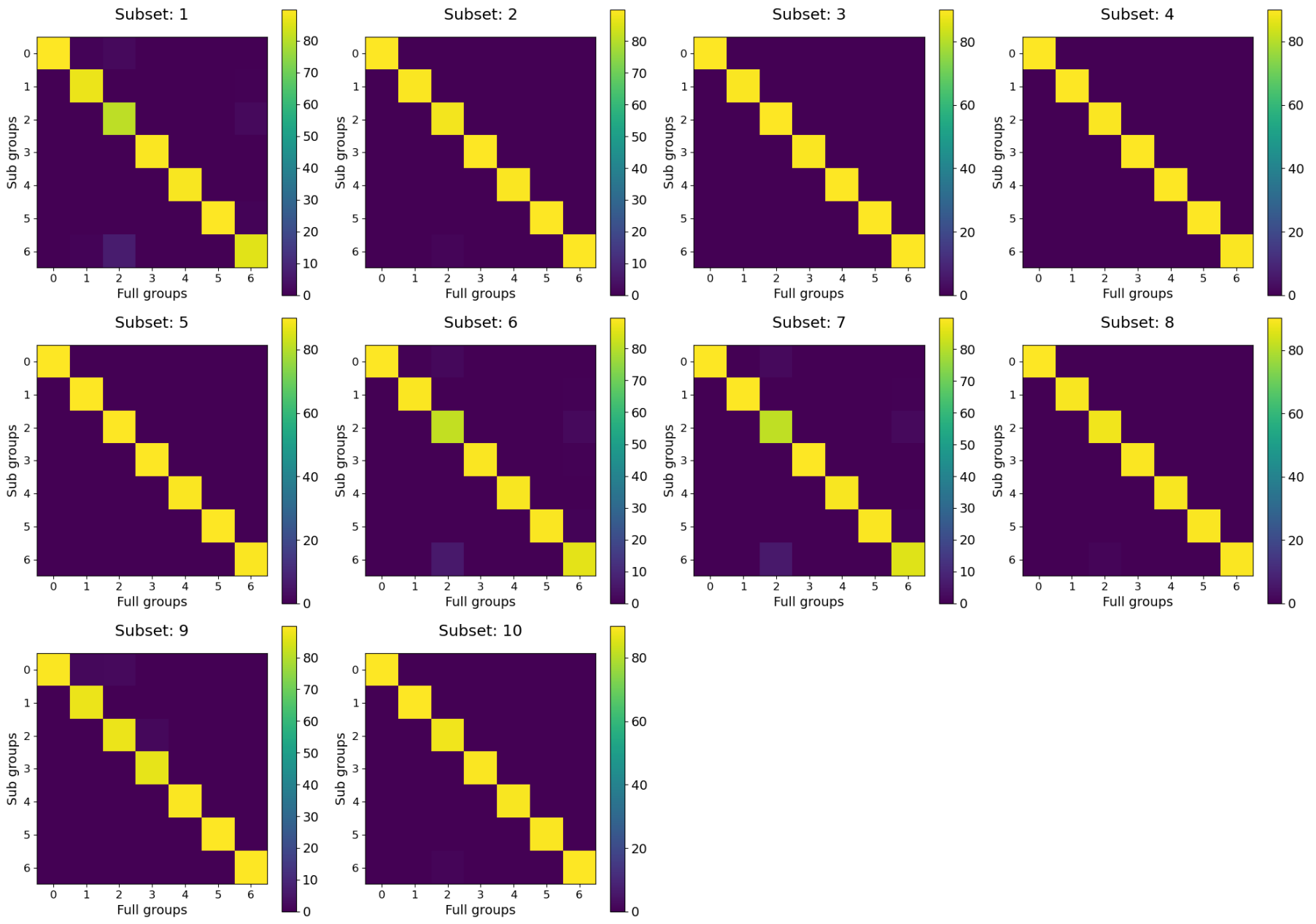

Figure 28: The overlaps $f(k_l, k) = \frac{|O_l^{k_l} \cap O^k|}{\max(|O_l^{k_l}|,|O^k|)}$ between cluster $O_l^{k_l}$ identified in the document subset, and $O^k$ identified in the full dataset.

## C.2. 90% of IPC Subclasses

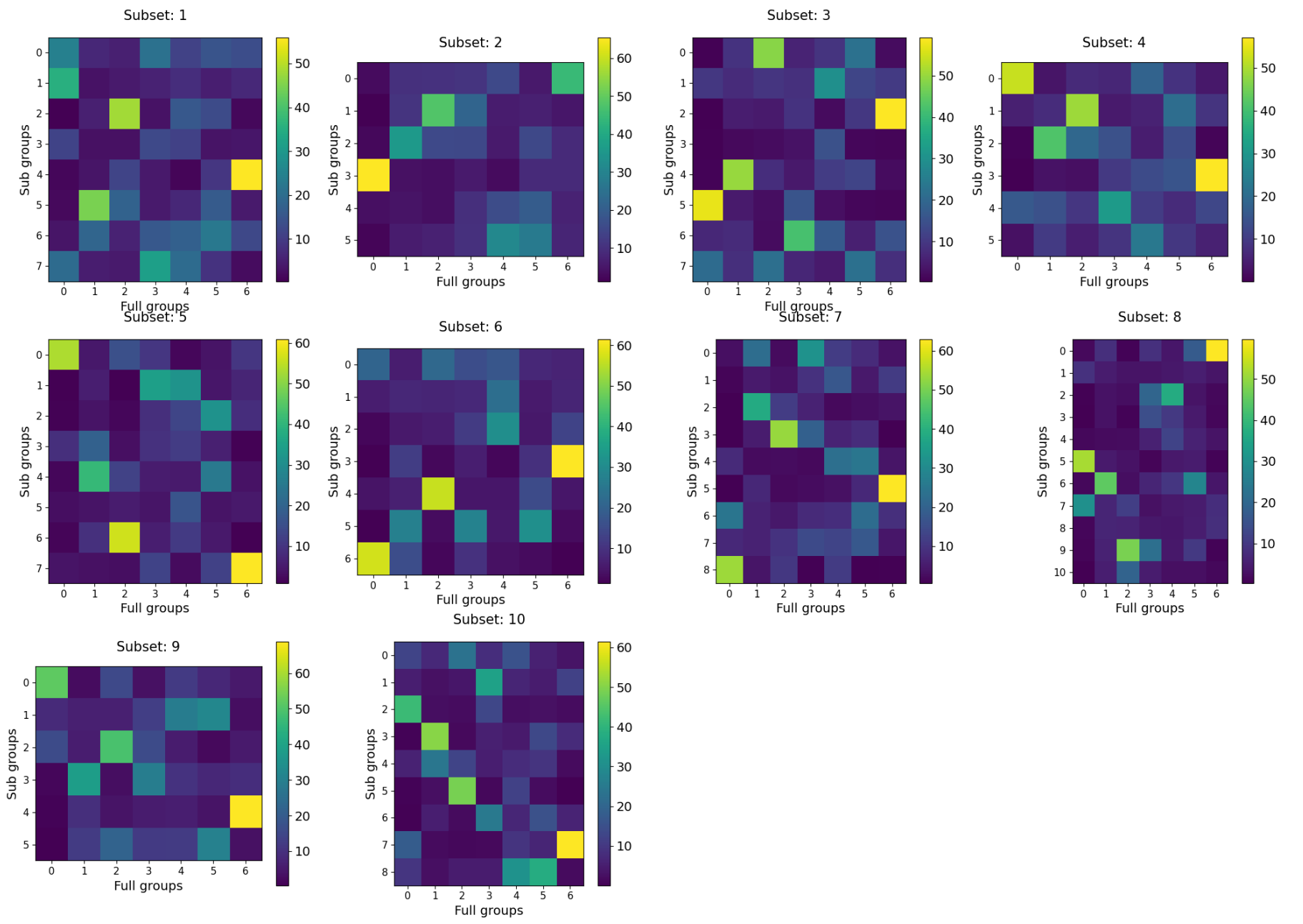

Figure 29: The overlaps $f(k_l, k) = \frac{|O_l^{k_l} \cap O^k|}{\max(|O_l^{k_l}|, |O^k|)}$ between cluster $O_l^{k_l}$ identified in the term subset, and $O^k$ identified in the full dataset.

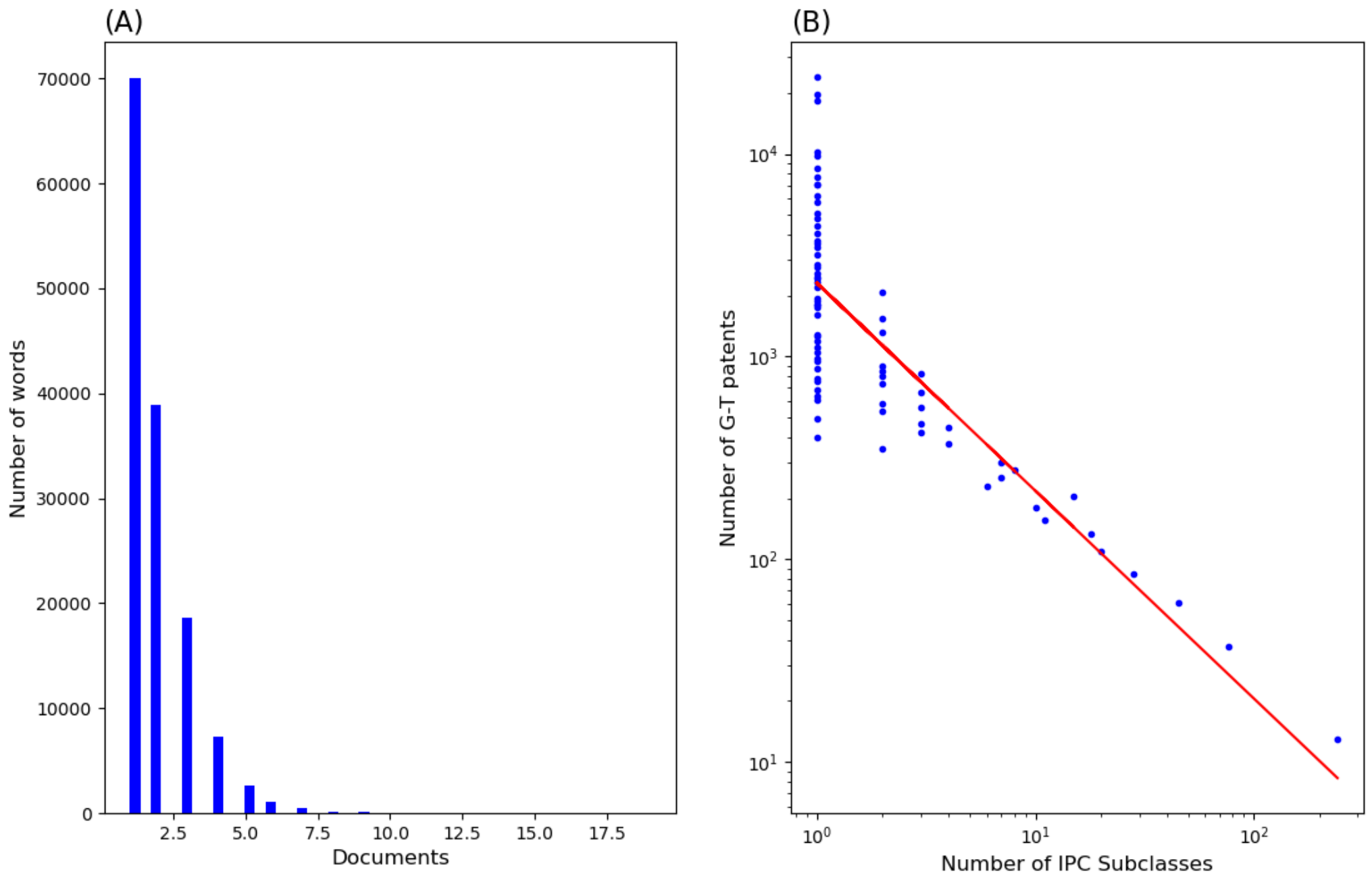

Figure 30: The degree distributions of patents (left) and IPC subclasses (right). The coefficient of the linear regression (right) is $-1.024$.

## D. Rankings in Graphene Technology Area

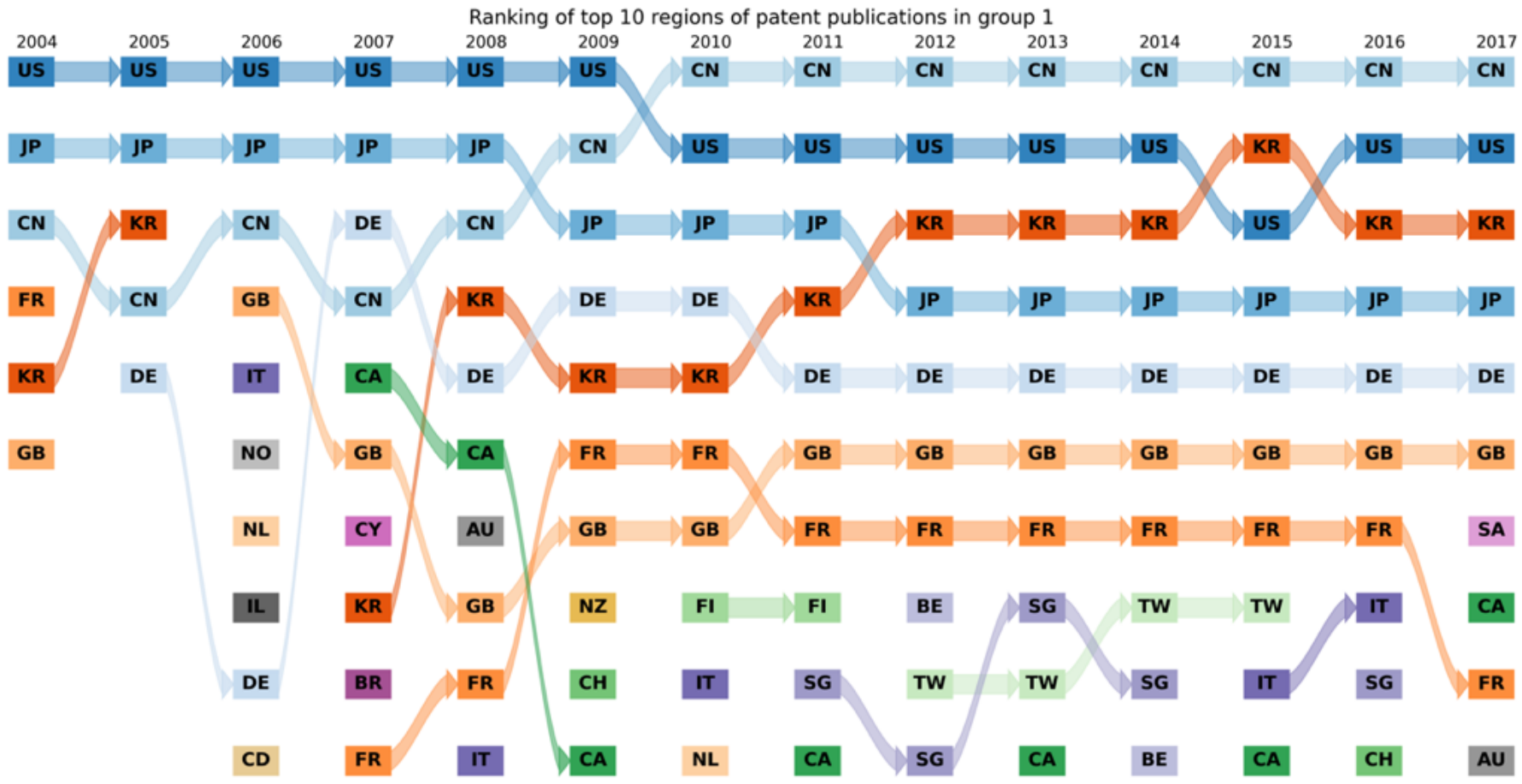

Figure 31: Ranking changes of the top 10 regions in terms of the number of applications in (1) *synthesis*.

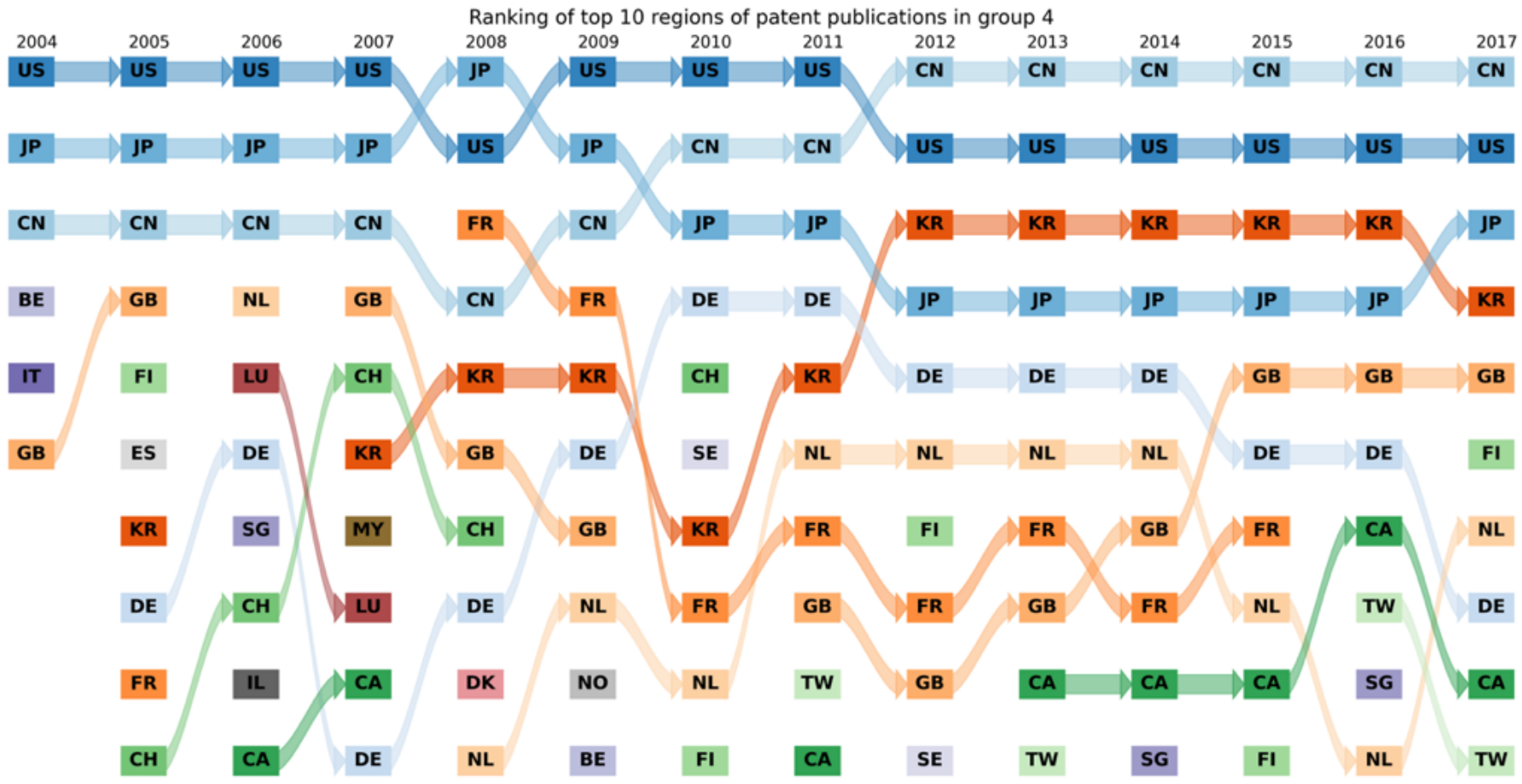

Figure 32: Ranking changes of the top 10 regions in terms of the number of applications in (4) *water treatment*.

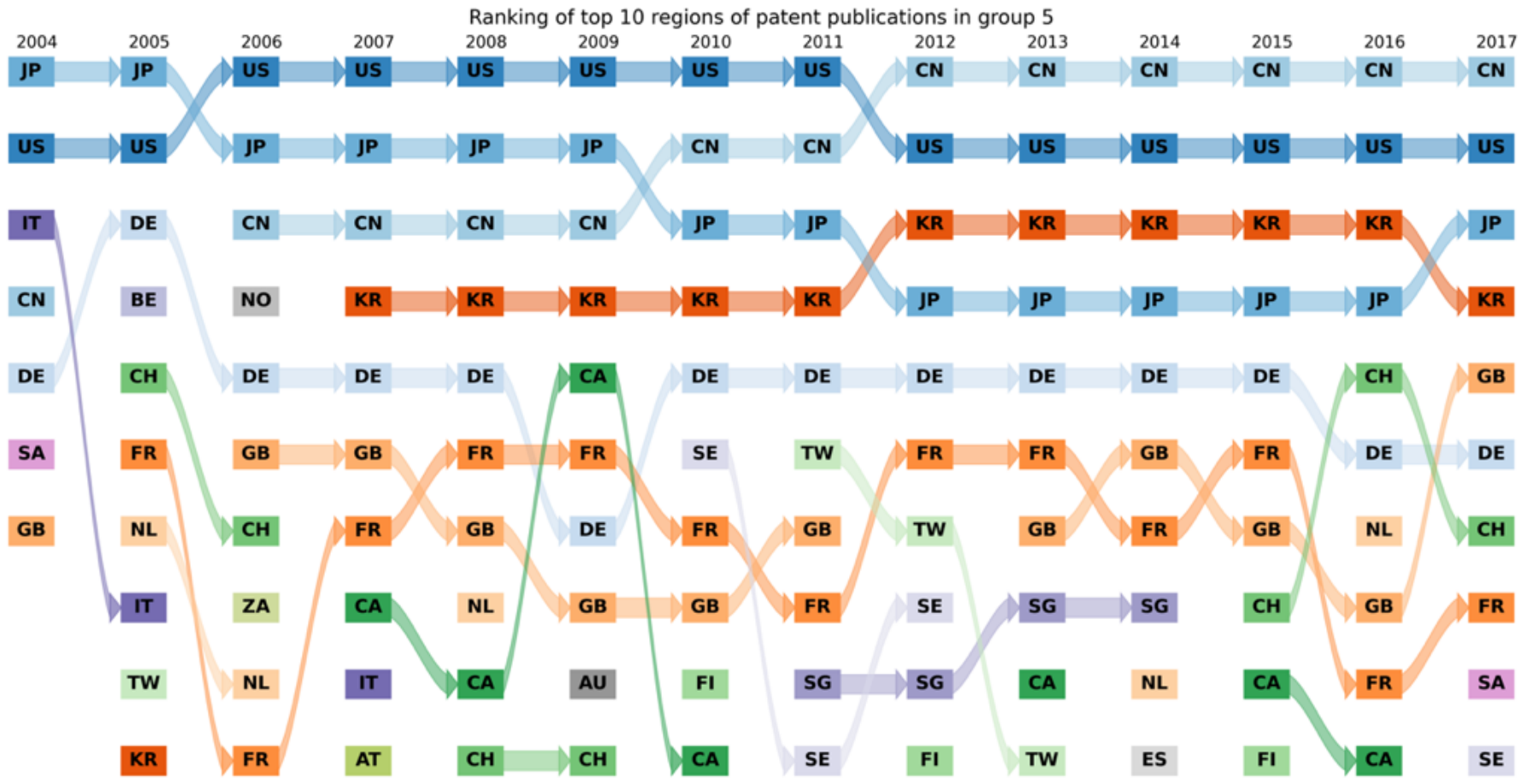

Figure 33: Ranking changes of the top 10 regions in terms of the number of applications in (5) *catalyst*.

## E. Technological Proportions of G-T Patents for Top Entities

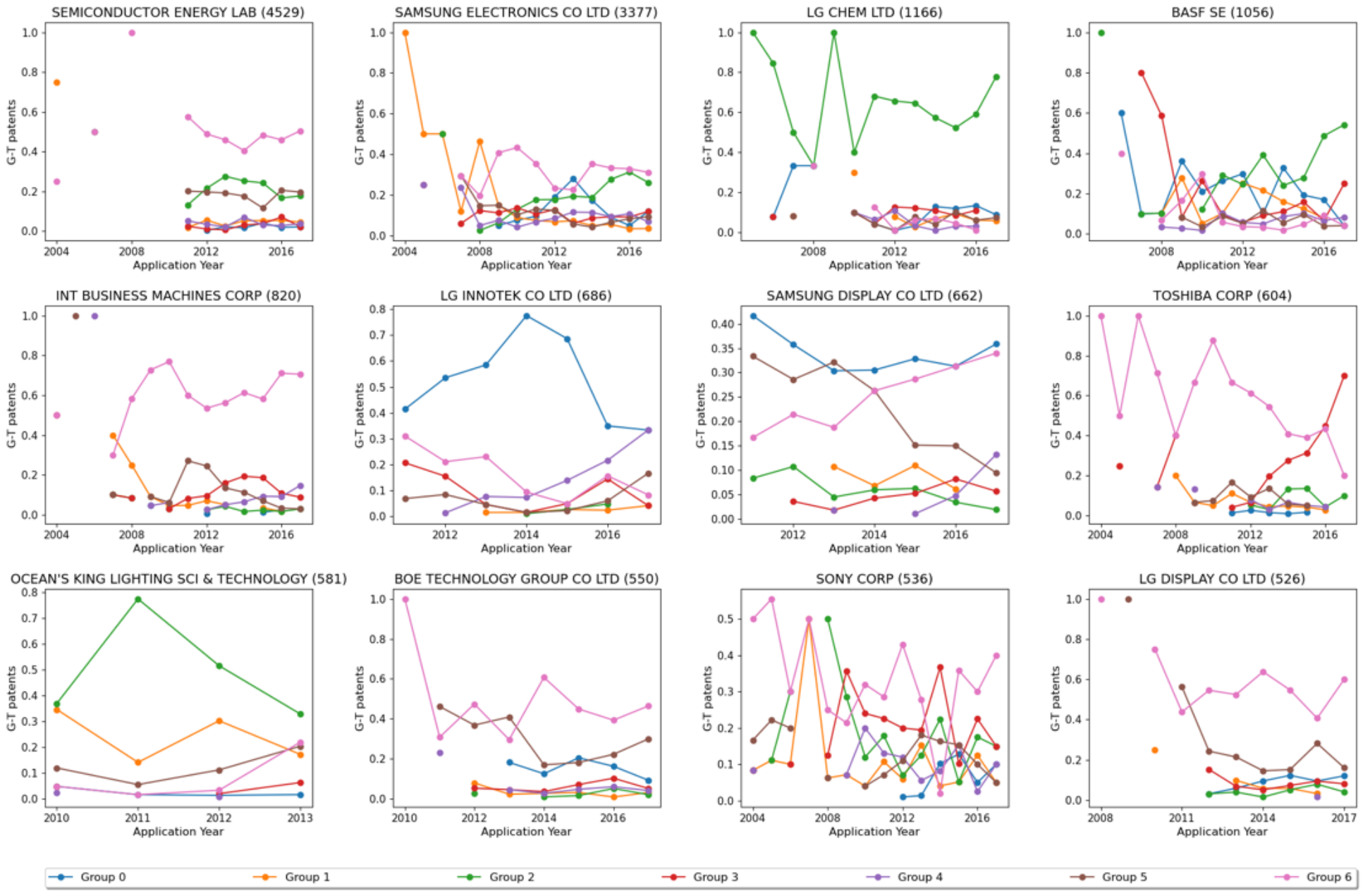

Figure 34: The technological proportions of G-T applications for the top 12 companies.

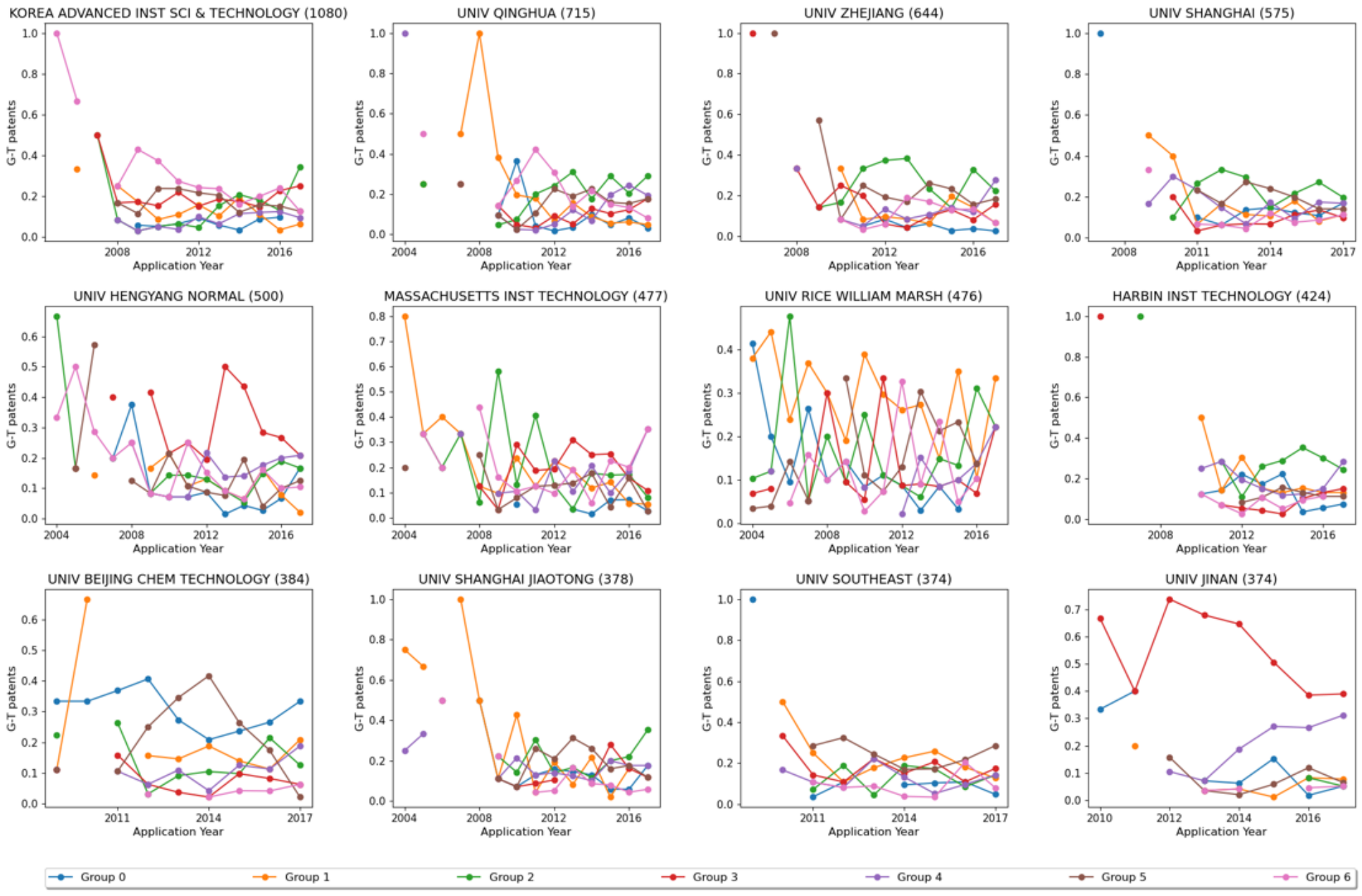

Figure 35: The technological proportions of G-T applications for the top 12 universities.

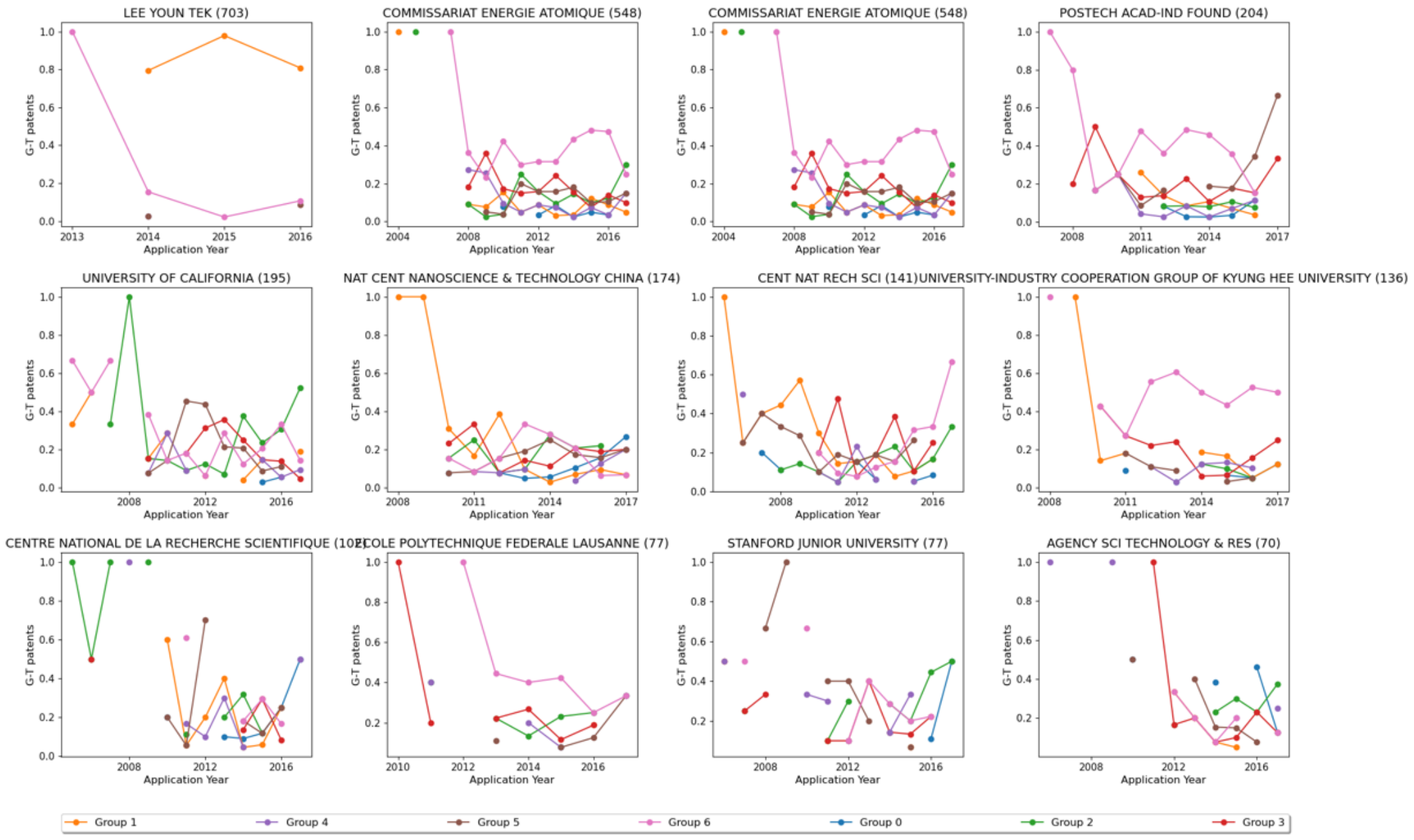

Figure 36: The technological proportions of G-T applications for the top 12 entities in Others.

## F. Entropy Heat Map

In this Section, we describe how to make the heat map from a scatter plot $(x, y)$. From Figure 37, we see that from the scatter plot (left), we use a pre-determined number of bins, $n$ to make each axis into a histogram. As a result, the plot turns into the $n \times n$ matrix in which each *entry* represents the number of points sharing the same entry coordinates. We use a heat map to visualize this 2D histogram in the log-log scale (middle). Finally, we obtain the heat map of all points (right) from the original plot (left), where the color indicates the density (red color means high density, and blue color means low density). The reason that the size of the heat map was smaller than the original map is due to the bin-location in the histogram process.

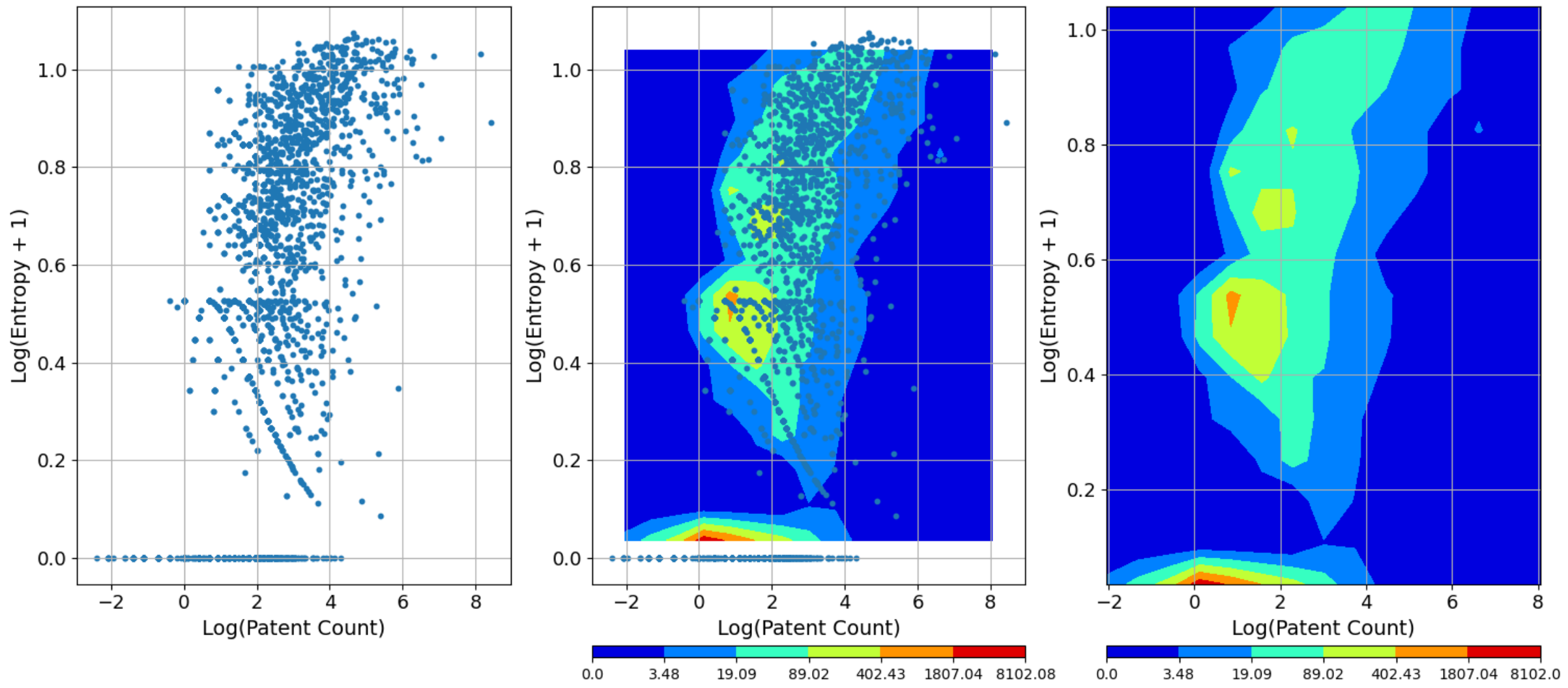

Figure 37: The plots of entropy versus patent credits in log-log scale as a scatter plot of all points (left), a scatter plot with heat map visualization (middle), and the heat map only.

## G. Word Clouds for the Seven G-T Technological Areas

We showed the word clouds for seven technological areas in the following seven figures, Figure 38 to Figure 44.

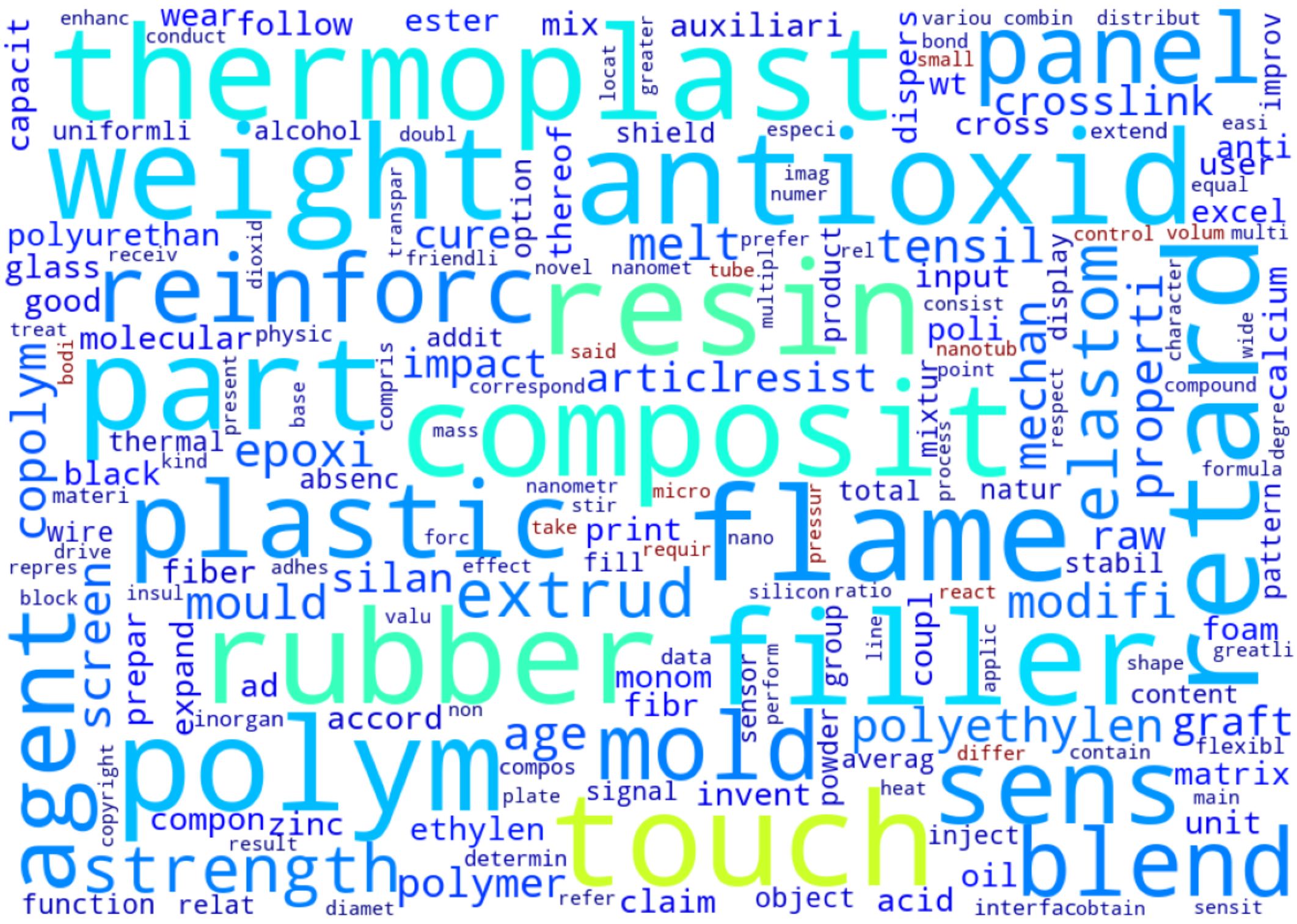

Figure 38: Word Cloud for (0) *composites*.

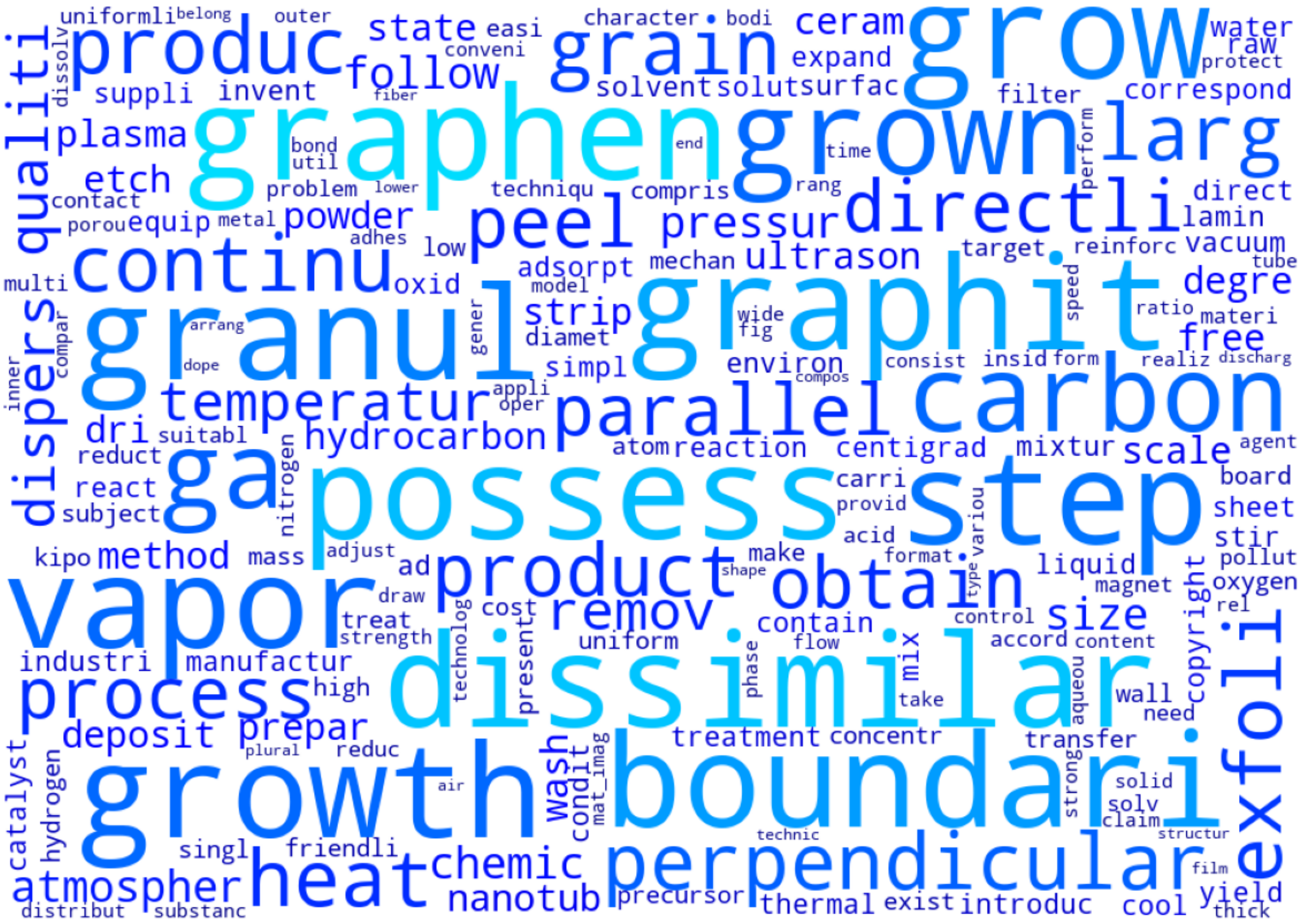

Figure 39: Word Cloud for (1) *synthesis*.

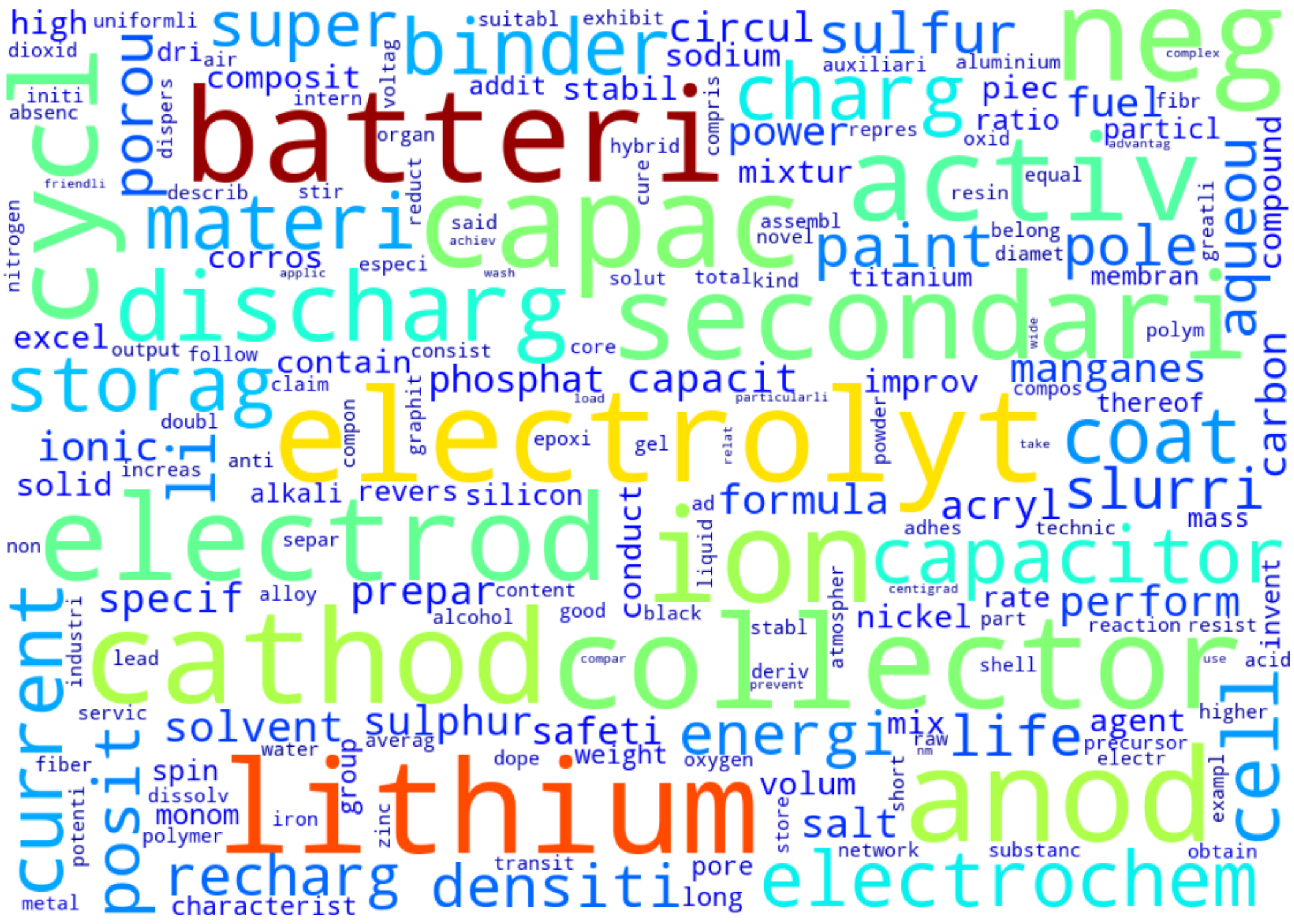

Figure 40: Word Cloud for (2) *batteries*.

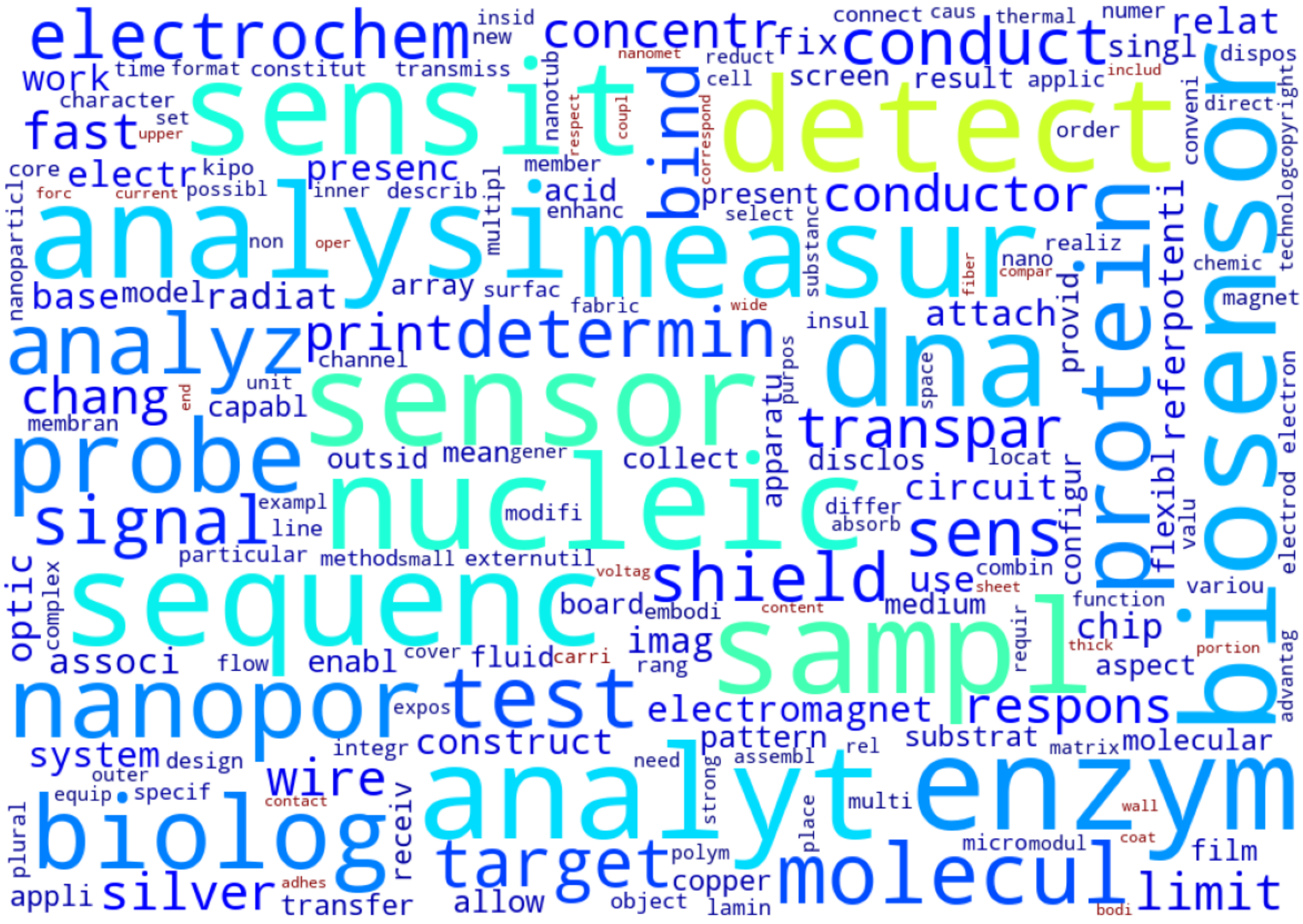

Figure 41: Word Cloud for (3) *sensors*.

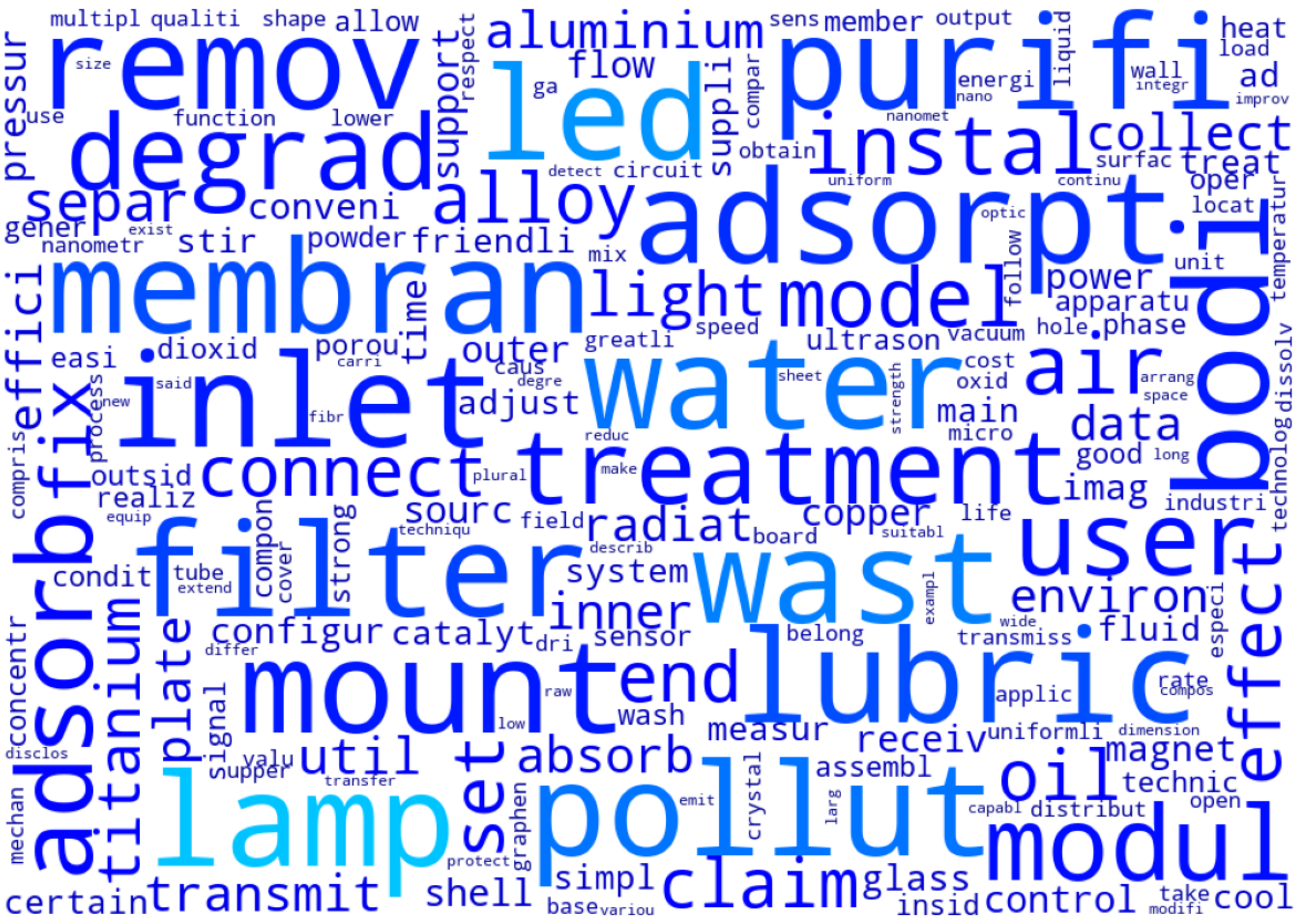

Figure 42: Word Cloud for (4) *water treatment*.

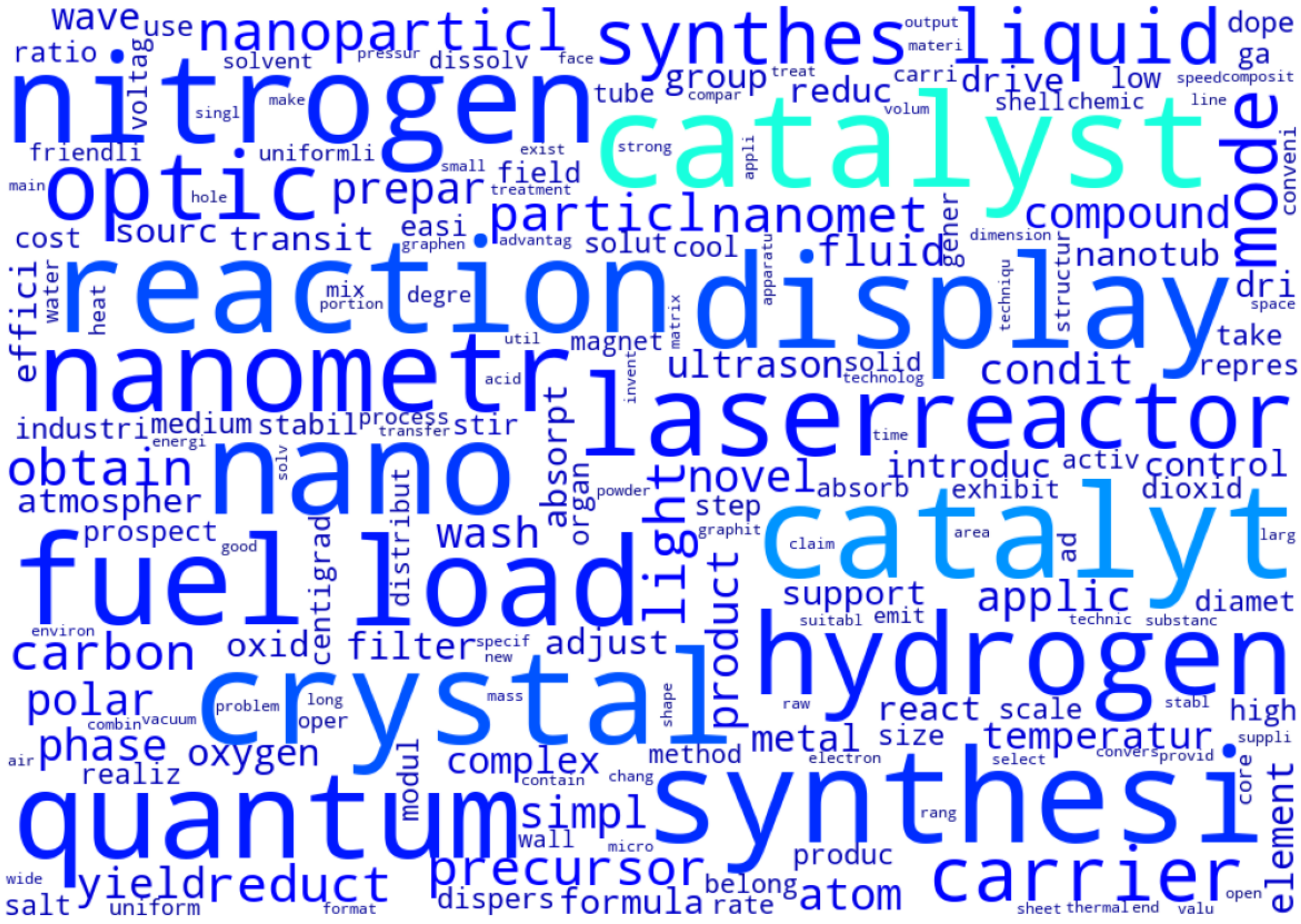

Figure 43: Word Cloud for (5) *catalyst*.

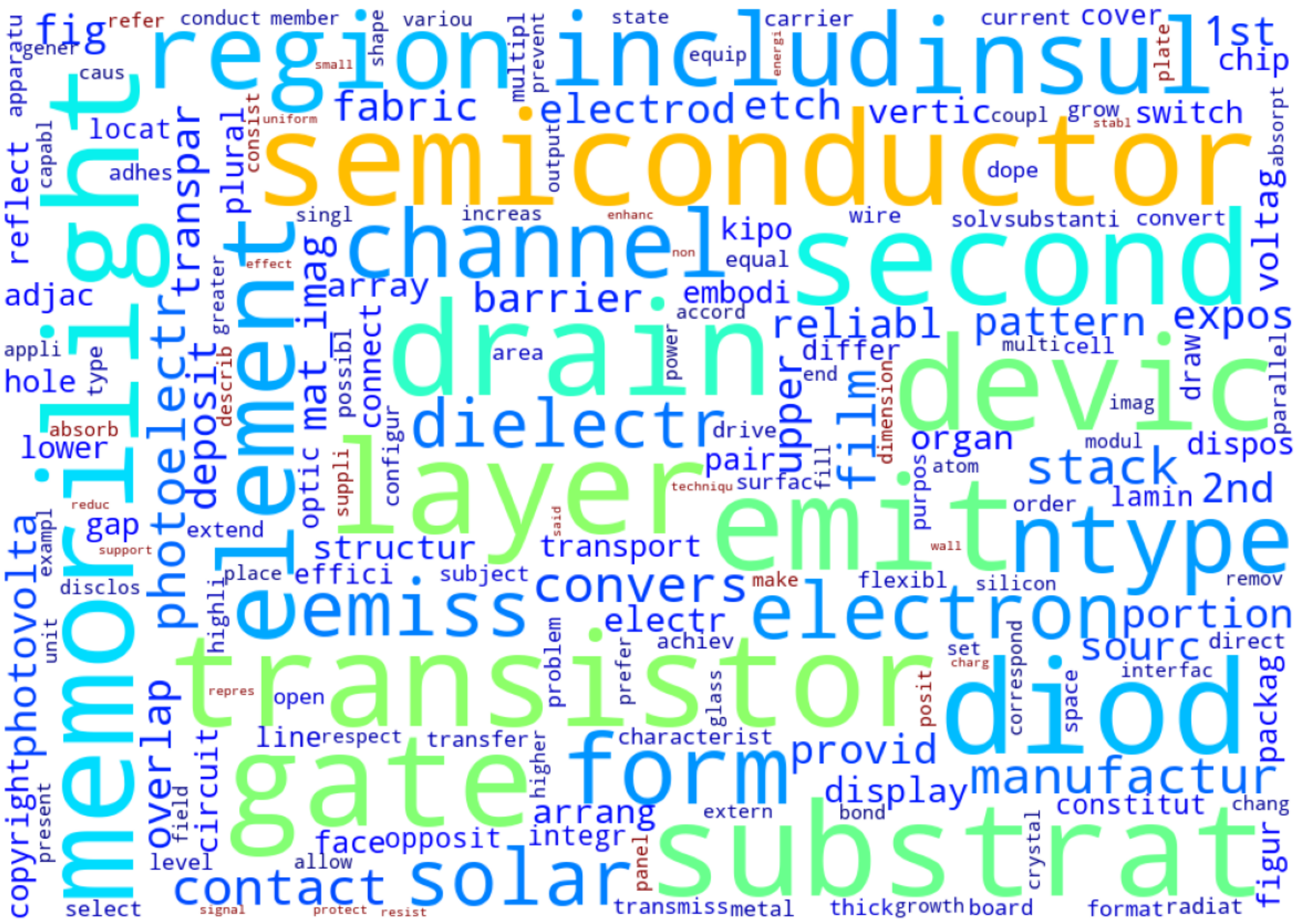

Figure 44: Word Cloud for (6) *devices*.